\documentclass[]{emulateapj}

\usepackage{color}
\usepackage{hyperref}

\newcommand{\OI}{\mbox{[\ion{O}{1}]}}
\newcommand{\CII}{\mbox{[\ion{C}{2}]}}

\newcommand{\lsun}{\mbox{L$_\odot$}}
\newcommand{\loh}{\mbox{$L_{\rm OH}$}}
\newcommand{\lwater}{\mbox{$L_{\rm H_2O}$}}
\newcommand{\loi}{\mbox{$L_{\rm [OI]}$}}
\newcommand{\lcont}{\mbox{$L_{\rm cont}$}}
\newcommand{\lbol}{\mbox{$L_{\rm bol}$}} 
\newcommand{\tbol}{\mbox{$T_{\rm bol}$}} 
\newcommand{\um}{$\mu$m} 
\newcommand\cmv{\mbox{cm$^{-3}$}}

\shorttitle{PACS observations of Taurus protostars}
\shortauthors{Lee. et al.}

\begin{document}


\title{Herschel Key Program, ``Dust, Ice, and Gas In Time" (DIGIT): 
the origin of molecular and atomic emission in low-mass protostars in Taurus
}


\author{Jeong-Eun Lee\altaffilmark{1,3}, Jinhee Lee\altaffilmark{2}, Seokho
Lee\altaffilmark{1}, Neal J. Evans II\altaffilmark{3},
Joel D. Green\altaffilmark{3}}
\altaffiltext{1}{Department of Astronomy and Space Science, Kyung Hee
University,
   Yongin-shi, Kyungki-do 449-701, Korea}
\email{jeongeun.lee@khu.ac.kr}
\altaffiltext{2}{Department of Physics and Astronomy, The University of
Georgia, Athens, GA 30602-2451, USA}
\altaffiltext{3}{Department of Astronomy, University of Texas at Austin,
2515 Speedway, Stop C1400, Austin, TX 78712-1205}

\begin{abstract}

Six low-mass embedded sources (L1489, L1551-IRS5, TMR1, TMC1-A, L1527, and TMC1) in Taurus have been observed with Herschel-PACS  to cover the full spectrum from 50 to 210 \um\ as part of the Herschel key program, ``Dust, Ice, and Gas In Time (DIGIT)". The relatively low intensity of the interstellar radiation field surrounding Taurus minimizes contamination of the \CII\ emission associated with the sources by diffuse emission from the cloud surface, allowing study of the \CII\ emission from the source.  In several sources, the \CII\ emission is distributed along the outflow, as is the \OI\ emission. The atomic line luminosities correlate well with each other, as do the molecular lines, but the atomic and molecular lines correlate poorly. The relative contribution of CO to the total gas cooling is constant at $\sim$30\%, while the cooling fraction by H$_2$O varies from source to source, suggesting different shock properties resulting in different photodissociation levels of H$_2$O. The gas with a power-law temperature distribution with a moderately high density can reproduce the observed CO fluxes, indicative of CO close to LTE. However, H$_2$O is mostly subthermally excited. L1551-IRS5 is the most luminous source ($L_{\rm bol}=24.5$ \lsun) and the \OI\ 63.1 \um\  line accounts for more than 70 \% of its FIR line luminosity, suggesting complete photodissociation of H$_2$O by a J-shock. In L1551-IRS5, the central velocity shifts of the \OI\ line, which exceed the wavelength calibration uncertainty ($\sim$70 km s$^{-1}$) of PACS, are consistent with the known red- and blue-shifted outflow direction.

\end{abstract}



\section{Introduction}

Jets and outflows are the most spectacular features of the star formation 
process. Although infall and accretion are the key processes
 in star formation,  
they are difficult to observe directly. Because models predict that 
jets and outflows correlate with accretion \citep{Shu87, Bontemps96},
they provide indirect probes of accretion. For example,
the average outflow force is stronger in Class 0 objects than Class I sources, 
indicating that  the mass accretion rate decreases as the central protostar
evolves \citep{Bontemps96, Curtis10}.
Analysis of outflows can also probe the energetics and chemistry that occur
when the jets carve out the envelope to construct bipolar cavities. 
Both high energy photons and shocks can heat the cavity walls farther from the 
central protostar than would be possible without the cavities \citep{Visser12}.
This bipolar cavity structure has been traced by the scattered emission at
short wavelengths (e.g., \citet{Tobin08}) and by maps of outflows with high
resolution (e.g., \citet{Wu09} at 1.3 mm).

Recent observations with instruments aboard the $Herschel$ Space Observatory
(HSO, \citet{Pilbratt10})
strongly suggest that the far-infrared (FIR) line emission is predominantly
produced by shocks
associated with jets and outflows \citep{Manoj13, Karska13} although energetic
photons produced by the accretion process also contribute to exciting 
the FIR line (especially CO line) emission (e.g., \citet{Visser12}).
In addition, according to the detailed models of \citet{Kristensen11}, \citet{Goi12}, \citet{Karska13}, and \citet{Lee13},
molecules with temperatures of $\sim$100-1000 K emit most of their
luminosity ($\sim 80 \%$) in the wavelength range of 50 to 200 $\mu$m, which is
covered by
the Photodetector Array Camera and Spectrometer (PACS; \citealt{Poglitsch10}).

Therefore, covering the SED from 50 to 210 \um, 
the PACS full SED range mode observations 
of a large number of embedded sources can provide
an excellent opportunity to study the dependence of the gas cooling budget on evolutionary phase.
However, kinematic constraints cannot usually be obtained from unresolved PACS lines ($\delta v \sim 50-100$ km/s).
Complementary kinematic
information comes from the 557 GHz H$_2$O 1$_{10}$-1$_{01}$ line survey
with the Heterodyne Instrument for the Far-Infrared on $Herschel$ (HIFI,
\citet{deGraauw10}) obtained by the $Herschel$ key programs, 
``Water in star forming regions with $Herschel$" (WISH) and
``Dust, Ice, and Gas In Time" (DIGIT) carried out 
toward the same sources that were observed with PACS \citep{Kristensen12,
Green13}.
According to \citet{Kristensen12}, the H$_2$O line emission is correlated with
other evolutionary indicators such as \lbol, \tbol, and envelope mass
as well as the indicators of outflow strength such as outflow momentum flux
calculated from the CO J=3-2 line and the line width of the 557 GHz water line.
While HIFI can resolve the 557 GHz water line into kinematic components in the
protostellar system, the PACS spectra are well suited to study the 
cooling of the heated gas because all of dominant cooling
lines are covered by the observations.

We present PACS spectra toward 6 embedded YSOs in Taurus, from the DIGIT program 
(PI: N. Evans, \citet{Green13}).
Because they are forming in the same environment, that variable can
be removed. In particular,  the Taurus cloud exists in a region of low
excitation by the interstellar radiation field, minimizing external radiation effects.
These effects can influence the emission, as seen in sources
in the Serpens \citep{Dionatos13}, Corona Aus \citep{Lindberg14},
and Ophiuchus (Je et al., in prep.) clouds.
Although all sources in our sample except L1527 are Class I sources, they span a substantial
range of \lbol\ (0.7 to 24.5 \lsun) and \tbol\ (67 to 226 K), as noted
in Table 1.

The PACS observations, data reduction, and flux measurement are explained in \S
2.
Spatial distributions of lines and their luminosities are described in \S 3, and
the analyses of FIR line fluxes with
rotation diagrams and the non-LTE LVG code, RADEX \citep{Tak07} are presented
in \S 4.
L1551-IRS5, which is the most luminous and peculiar source  among our samples,
is separately
discussed in \S 5, and we discuss the mechanisms for the FIR line emission in \S 6.
Finally, the summary of this work will be given in \S 7.

\section{Observations and flux measurement}

PACS provides a 5 $\times$ 5 array of $9.4\arcsec \times 9.4\arcsec$ spatial 
pixels
(hereafter ``spaxels") covering the spectral range from 50 to 210 \um\ with
$\lambda/{\Delta\lambda}\sim1000$--2500, divided into four segments, covering
$\lambda \sim50$--75, 70--105, 100--145, and 140--210 \um.
The PACS spatial resolution ranges from $\sim9\arcsec$ at the shortest
wavelengths (50 \um) to $\sim18\arcsec$ at the longest 210 \um\  (refer to PACS observer's manual (2010)\footnote{$\rm http://herschel.esac.esa.int/Docs/PACS/pdf/pacs\_om.pdf$
}).

\begin{figure}
\epsscale{0.8}
\plotone{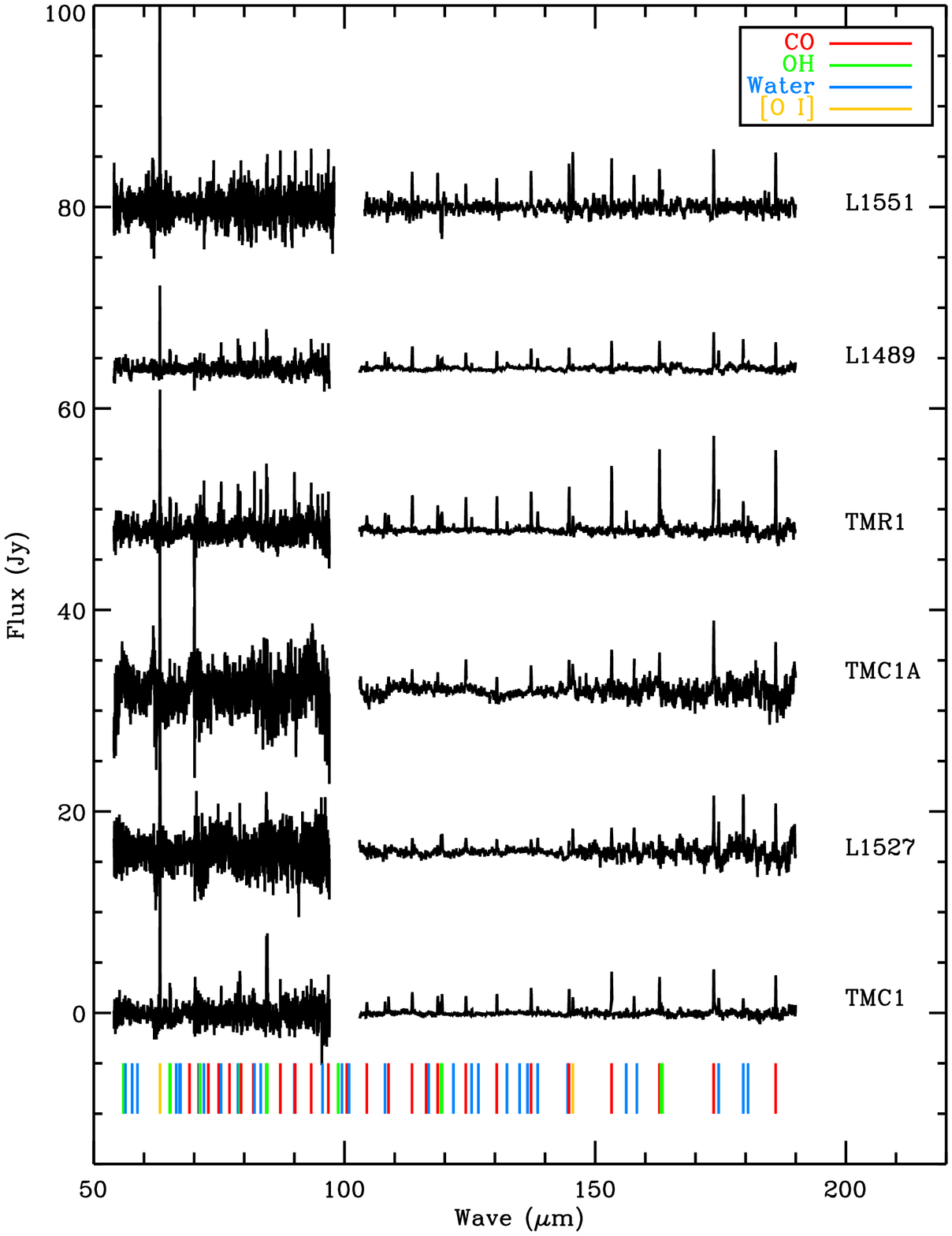}
\caption{PACS continuum subtracted spectra of 6 Taurus sources. 
The spectra are extracted from the central spaxel in order to show lines with better S/N ratios. However, for flux measurements, spectra extracted from multi-spaxels over which emission was detected (check the text).
From bottom to top, sources are arranged as their bolometric luminosities increase.
(The fluxes of TMC1-A, L1527, and TMC1 have been doubled to show their lines
better.)
}
\label{lines}
\end{figure}

\begin{figure}
\epsscale{1.0}
\plotone{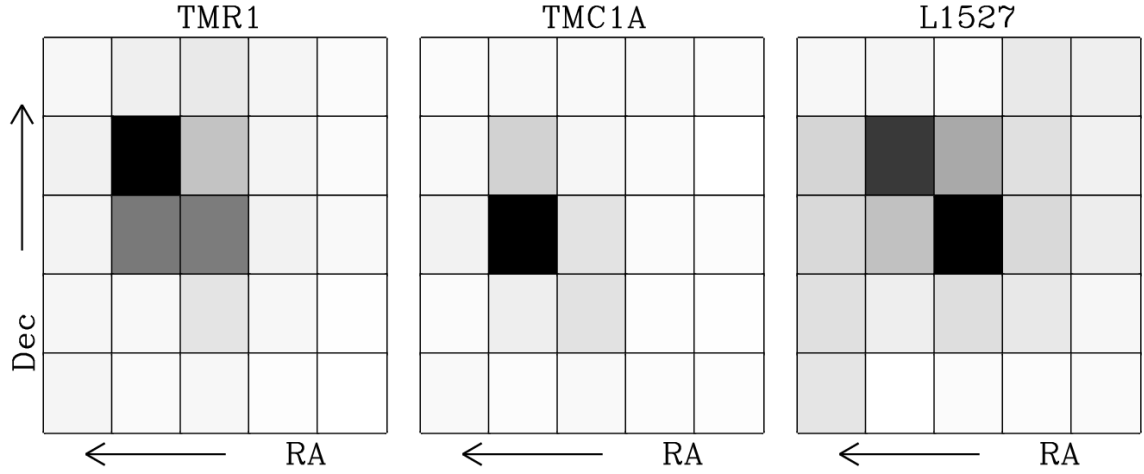}
\caption{5$\times$5 cube images of mispointed sources. 
Fluxes are integrated from 67.5 \um\ to 68.5 \um\ and decrease from black to white in the images. 
}
\label{cubeimages}
\end{figure}

Information on the Taurus sources covered by the DIGIT program are listed
in Table~\ref{sourceinfo}.
More detailed information on the PACS observations of our targets can be found
in \citet{Green13}: basic parameters of protostars (\lbol\ and \tbol),
the rotation diagrams for the total numbers of each molecule and its rotation temperature, 
and the correlations between various physical parameters
and line properties.
The PACS SEDs of these sources are presented in \citet{Green13}, so we present
only
the continuum subtracted line spectra in Fig.~\ref{lines}.
The basic data reduction procedures are explained in detail in \citet{Green13}.
Among our sources, TMR1, TMC1-A, and L1527 were mispointed (due to a known early mission issue).
TMC1-A is mispointed by one spaxel to the east, but the emission is well confined in 
this single off-center spaxel. 
TMR1 and L1527 both are shifted to the north-east from the central spaxel; the strongest continuum emission for 
L1527 is still at the central spaxel while the emission peak is shifted to the NE by one spaxel for TMR1.  
In TMR1 and L1527, the continuum emission distributes over a few spaxels due to the mispointing. 
Fig.~\ref{cubeimages} shows the PACS images at the 68 \um\  continuum for the mispointed sources.

We follow the line flux measurement method of \citet{Lee13} rather than that of
\citet{Green13}. In order to correct for the extended nature of line emission,
\citet{Green13} derived an equation for line flux as a function of wavelength
using strong CO lines and applied the equation to all species, except for [O I].
However, the emission lines of different species are distributed differently.
In addition, three of our sources are mispointed, so the line emission distributes 
over multiple spaxels.
In particular, the atomic lines are spatially distinct from the molecular lines in our
targets; the atomic line emission is extended beyond the PACS PSF, but
molecular line emission is mostly compact at similar wavelengths.
To solve this problem, we measured fluxes of the extended (or mispointed)
line emission by calculating the equivalent width over multiple spaxels, over which the line emission distributes, 
as explained in the Appendix of \citet{Lee13}.
However, the key concept of flux measurement in \citet{Lee13} is the same as that in
\citet{Green13}.
Both combine two data reductions done with two different Herschel Interactive
Processing Environment (HIPE) versions (6.1, circa 2010-2011 and 8.1, circa 2012-2013) 
to produce reliable
absolute flux calibration as well as the best S/N; for line fluxes, the equivalent width calculated from the HIPE 8.1 reduction (with a higher S/N) is multiplied by the local continuum extracted from the whole 25 spaxels of the HIPE 6.1 reduction (with more accurate absolute flux calibration).

We improved upon the reduction used in \citet{Green13} by applying 
correction factors to the blue and red spectra in order to align the different orders smoothly around 100 \um. We calculated the correction factors using the PACS 70 and 160 \um\ photometric data, measured with the same aperture size as the PACS field of view because the continuum spectra are extracted from all 25 spaxels.  The SEDs were convolved with the PACS wideband response functions to compare with actual photometric data. Correction factors were found separately for the blue and red spectra and listed in Table~\ref{factortable}.   

The flux of each transition of each source is listed in Table~\ref{fluxtable}.
The line detection criterion was S/N$\ge$3 from the Gaussian line fitting.
The errors presented in Table~\ref{fluxtable} are total flux errors calculated from the error propagation equation in the Appendix of \citet{Lee13}. However, the significance of line detection is decided by the rms noise of the baseline local to the line. In some sources, although the higher excitation lines are detected, the lower excitation CO lines are listed as non-detections because of their higher noise levels.  
In addition, some of the H$_2$O and CO lines detected by Karska et al. (2013) were not detected in our observations. This apparent discrepancy in line detection is caused by the two different observing modes; the rangescan mode used in this work accepts lower sensitivities to gain the full coverage of the PACS wavelength region, whereas the linescan mode from Karska et al. (2013) gains better spectral resolution and sensitivity over a very narrow wavelength range.  
In addition, the CO J=14-13 flux is systematically lower than the CO J=15-14 flux for all sources as listed in Table~\ref{fluxtable} 
because the level population of J=15 is greater than that of J=14 
at the gas temperature above 250 K, 
which is the case in our sources (see \S 4).

\section{Detected line transitions}

\subsection{Molecular and atomic lines}
As shown in Fig.~\ref{lines} and Table~\ref{fluxtable}, L1489, TMR1, and TMC1
have very rich emission line spectra.
In contrast to previous studies (e.g., \citet{Nisini02}, \citet{Karska13}), these sources do not have very high bolometric luminosities.
TMC1, especially, has \lbol\ less than 1 \lsun.
The most luminous source among our targets is L1551-IRS5, but it does not show
water lines except the $3_{03}-2_{12}$ 174.6 \um\ transition.
In addition, L1527, which has been known as a prominent outflow Class 0 source \citep{Davidson11},
shows neither high-J ($E_{\rm u}>$ 1500 K) CO transitions nor many water line transitions.

Toward all 6 sources,  \OI\
63 and 145 \um\  lines are detected.
The \CII\ 157 \um\  line was also detected in all sources except L1489.
These atomic lines have been considered as good tracers of the strength of the
interstellar
radiation field (ISRF) based on observations with the
$Infrared~Satellite~Observatory~(ISO)$ \citep{Kaufman99}.
However, it was difficult to separate the contribution from gas heated 
by the ISRF from the contribution by the source
itself because of the poor spatial resolutions of $ISO$.
Because $Herschel$ has much better spatial resolution than did $ISO$
(by a factor of $\sim 10$), most $Herschel$ observations have found that
the \OI\
emission distributes along the outflow direction
\citep{Karska13, Dionatos13, Lee13}, while \CII\ emission generally
shows no correlation with the source, indicating that it comes from the 
surface of the cloud.

\begin{figure}
\epsscale{1.0}
 \plottwo{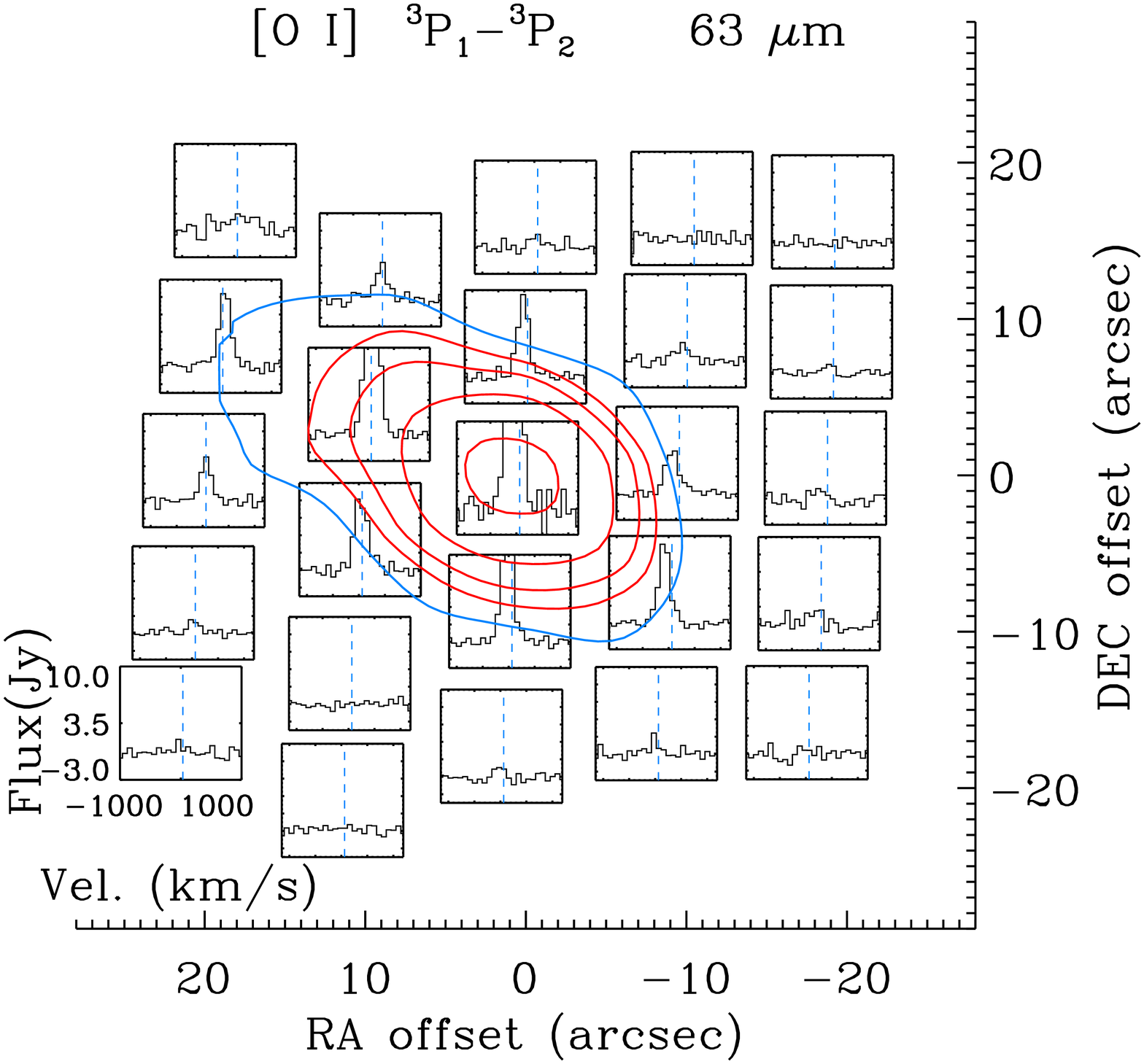}{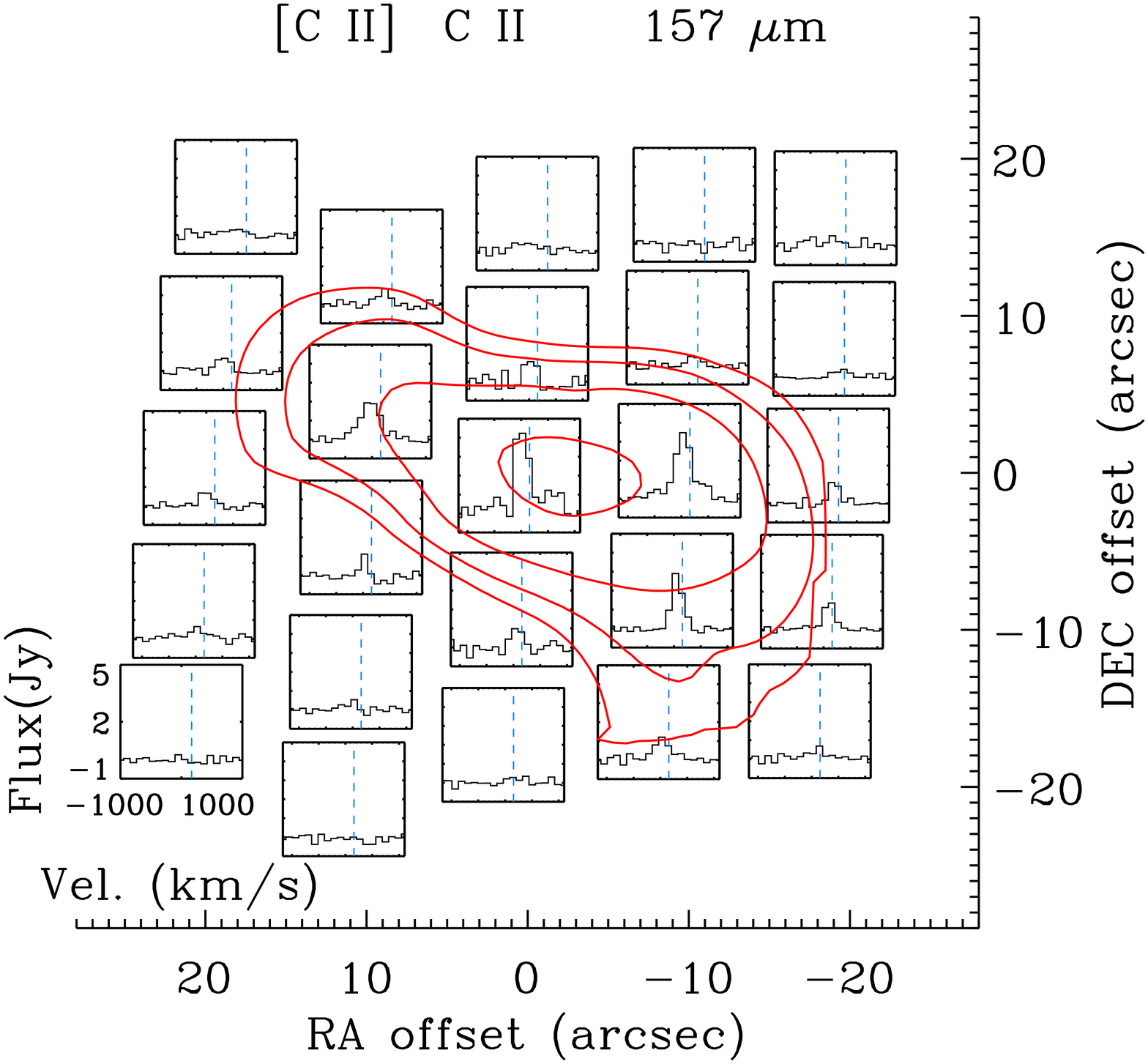}
\caption{Spectral maps of \OI\ 63.1 $\mu$m and \CII\ 157 $\mu$m for L1551-IRS5.
Contour levels are 0.1 (blue contour), 0.2, 0.3, 0.5, and 0.9 times of the peak flux for \OI\
while 0.3, 0.5, 0.7, and 0.9 of the peak flux for \CII\ .
We include one more blue contour in the \OI\ map to make a better comparison 
with the \CII\ map.
}
\label{l1551_OI_CII}
\end{figure}

\begin{figure}
 \plottwo{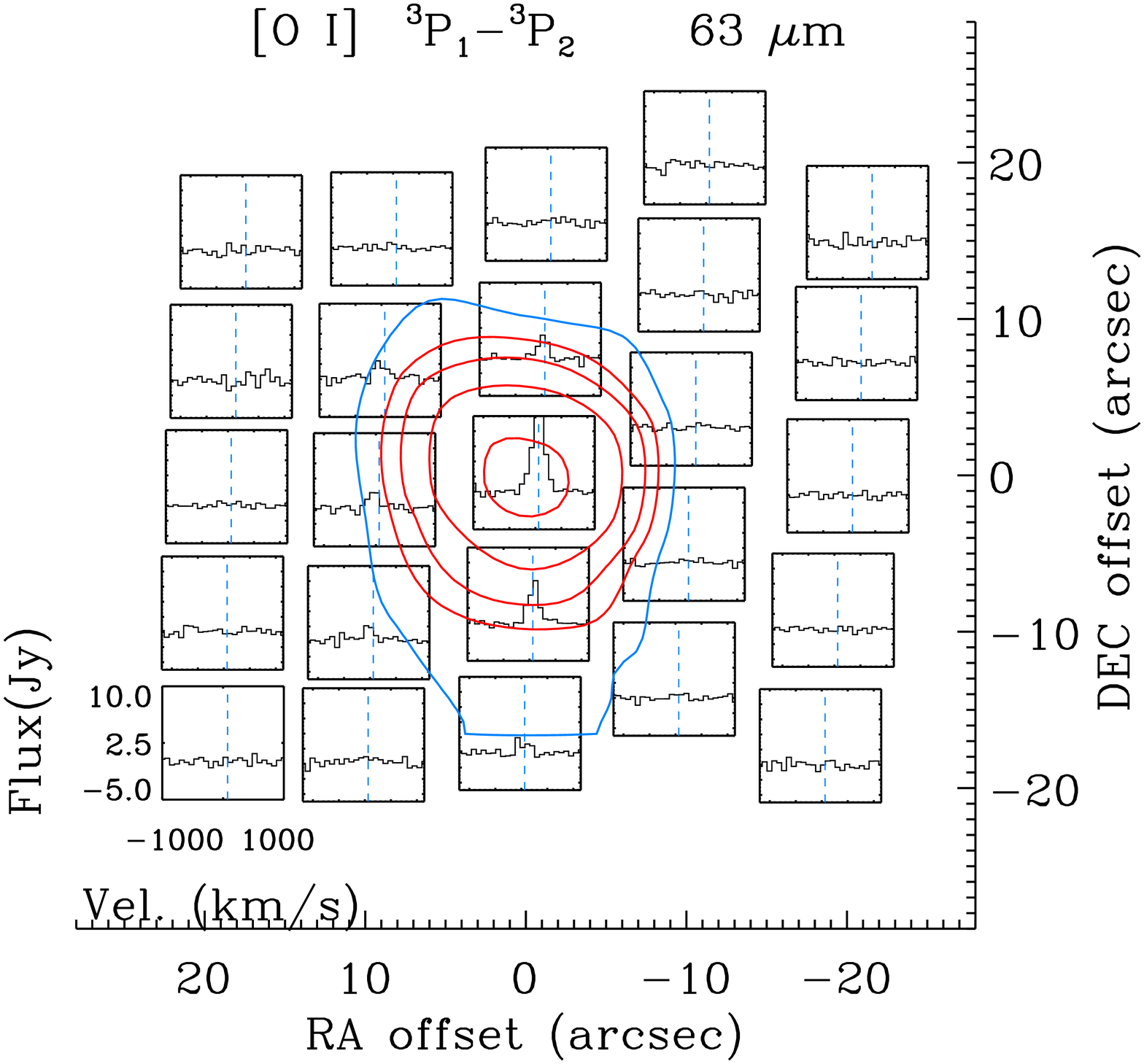}{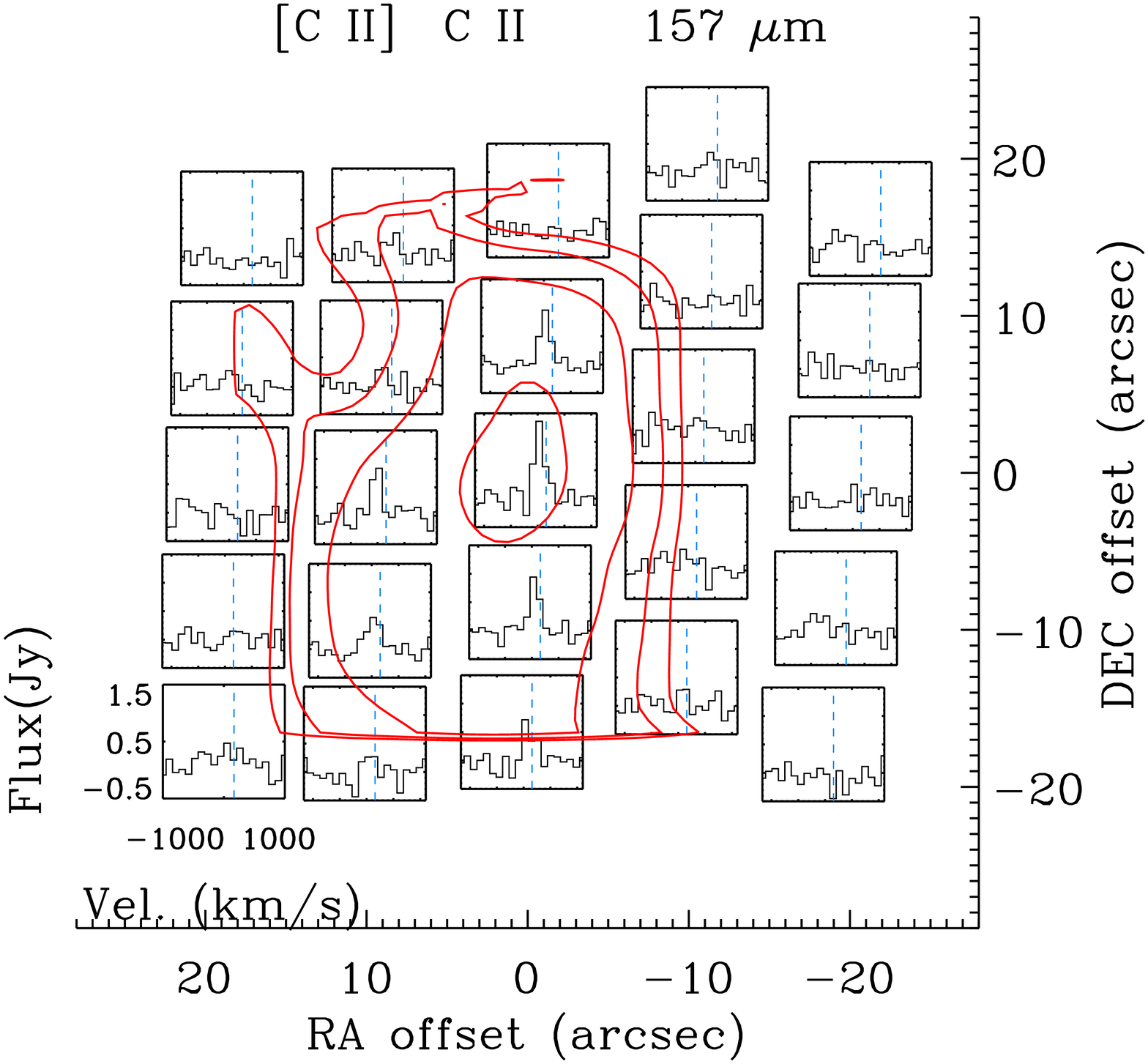}
\caption{Spectral maps of \OI\ 63.1 $\mu$m and \CII\ 157 $\mu$m for TMC1.
Contour levels are the same as Fig.~\ref{l1551_OI_CII}.
}
\label{tmc1_OI_CII}
\end{figure}

\begin{figure}
\plotone{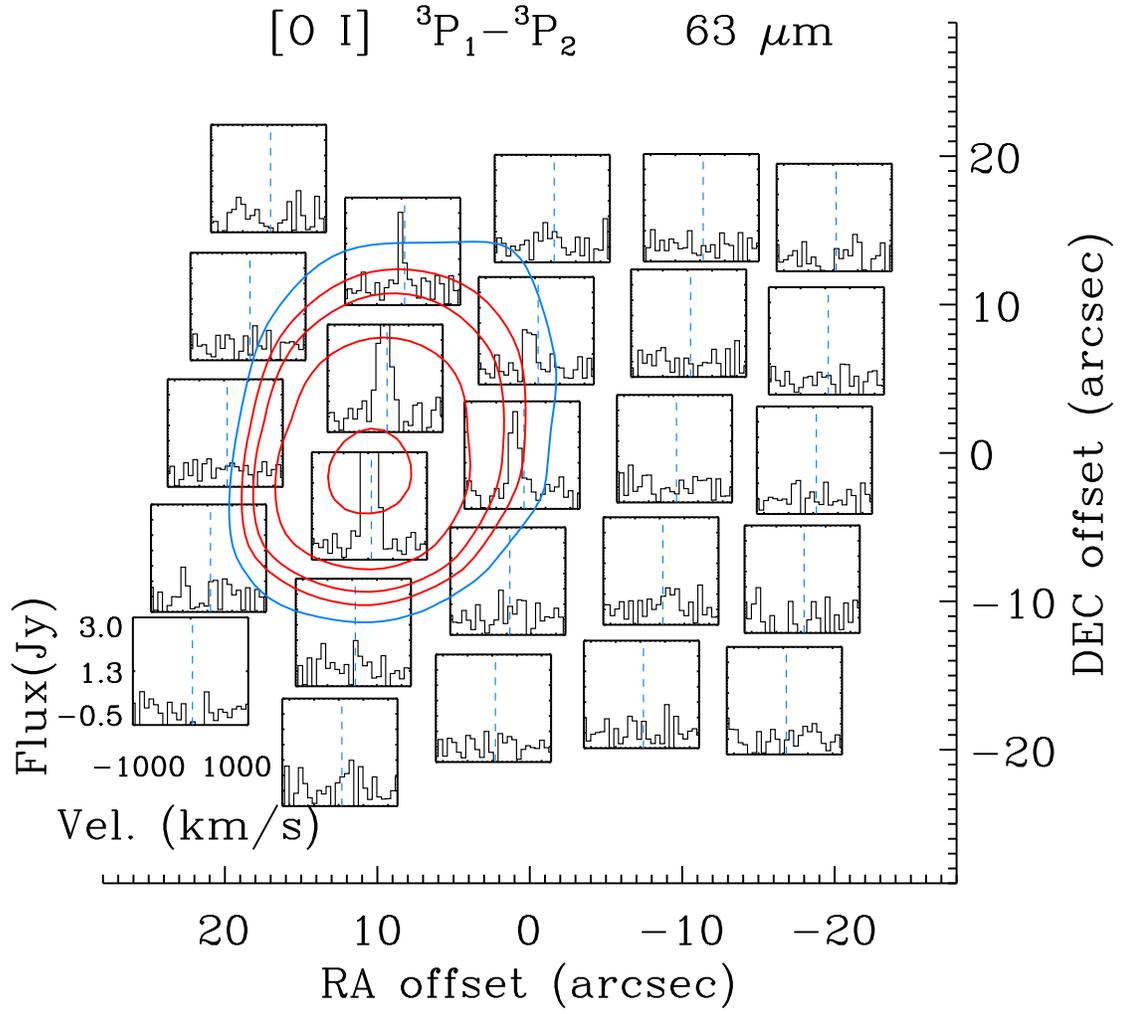}
\caption{Spectral map of \OI\ 63.1 $\mu$m for TMC1-A. 
Contour levels are the same as Fig.~\ref{l1551_OI_CII}a.
}
\label{tmc1a_oi}
\end{figure}

Atomic emission is also extended in our sources;
the \OI\ 63 \um\ emission is extended along the known outflow direction in all sources, 
but the \CII\ 157 \um\ line is too weak  to determine its extent in most sources. However, in L1551-IRS5 and TMC1, 
the \CII\ line is strong and extended like the \OI\ line (see Fig.~\ref{l1551_OI_CII} and~\ref{tmc1_OI_CII}) 
although their distributions differ somewhat.
In TMC1-A, which is mispointed to the east only by one spaxel, the \OI\ emission is elongated to the north, where the 
blueshifted outflow is located \citep{Hog98, Nara12}, as seen in Fig.~\ref{tmc1a_oi}.
The redshifted outflow located to the south in TMC1-A is very weak compared to 
the blueshifted component \citep{Hog98, Nara12}. 
Unlike the atomic emission, almost all molecular transitions are compact and peaked on the continuum
source (see Fig. 27 of \citet{Green13}); the mispointed sources do not seem to have extended emission structure
either since the displacement of emission does not change with wavelength.

No O and B type massive young stars exist near the Taurus cloud
 to illuminate the
large cloud structure. Therefore, these sources might be the best places to
study the \OI\ and \CII\ emission
directly related to protostars themselves.
Indeed, in the Taurus sources, the \CII\ emission is not diffuse and is 
very well correlated spatially with the \OI\ emission, 
which distributes along the outflow direction (see
Fig.~\ref{l1551_OI_CII} and~\ref{tmc1_OI_CII}) in L1551-IRS5 and TMC1.
\citet{Karska13}  also pointed out that the \CII\ emission is extended along
the outflow direction in TMC1, and \citet{Goi12} showed a similar trend in Serpens SMM1.
This might indicate that the main heating source of the atomic gas is the fast
J-type
shock, which can produce UV photons, and thus, dissociate H$_2$O and CO
resulting in high abundances of O and C$^+$ \citep{Snell04}.
Alternatively, the atomic line emission may be produced along the outflow
cavity wall, where
high energy photons can penetrate to dissociate molecules to O and C$^+$ \citep{Mamon88}.
In the latter case, a C-shock should still exist along the cavity wall to explain
the broad component of molecular emission detected by HIFI \citep{Kristensen12}.

\subsection{FIR line and continuum luminosities}

\begin{figure}
\epsscale{0.8}
 \plotone{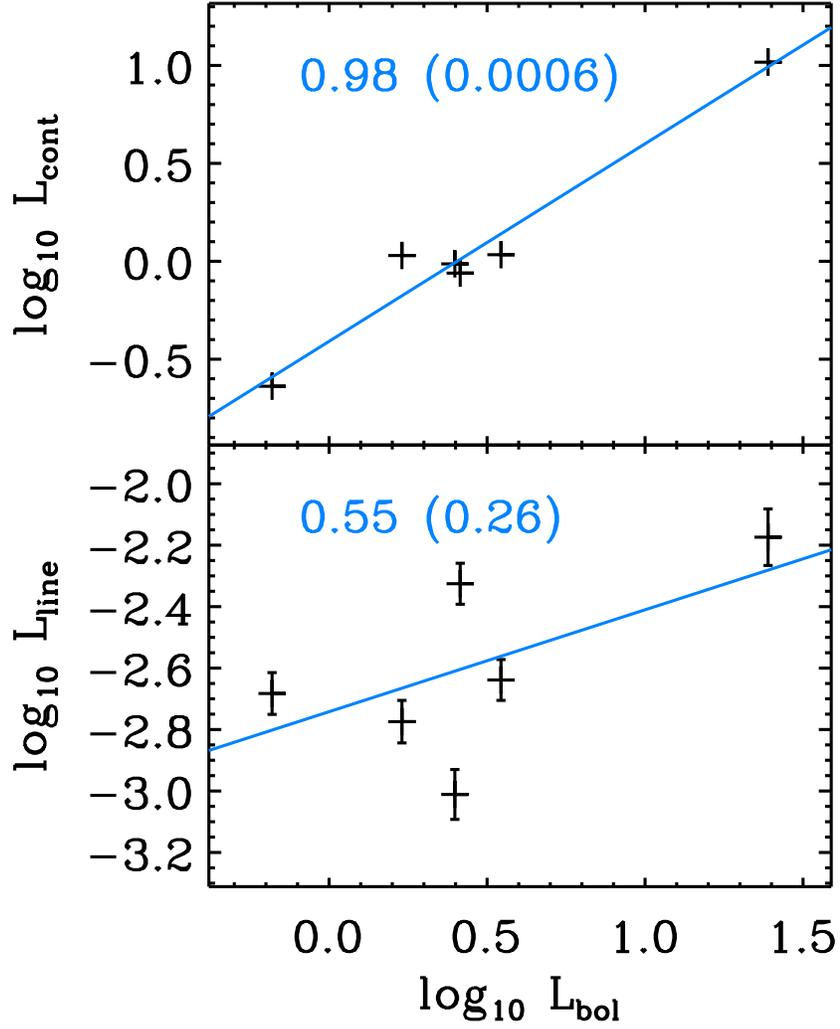}
\caption{Correlations of continuum luminosity (Top) and total FIR line luminosity (Bottom)
with bolometric luminosity .
Luminosities are in the logarithmic scale.
The FIR continuum luminosity has been calculated by integrating flux density from 55 \um\ to 190 \um\ with d=140 pc. The correlation coefficients with $p$-values in parentheses are marked in the upper left of each box.
}
\label{correlation1}
\end{figure}

The observed PACS line luminosity of each species for each source is listed in
Table~\ref{lumtable}.
The FIR continuum luminosity in the PACS range
is also listed. The total line luminosity is $\sim 0.1$ to 1 \% of the FIR
continuum luminosity. The dominant cooling channel for the protostellar
envelope is the FIR dust continuum emission.
The brightest source, L1551-IRS5 has the smallest percentage (0.06 \%) of line
luminosity
while the faintest source, TMC1 has the biggest percentage (1 \%) of line
luminosity. 
As seen in Fig.~\ref{correlation1}, the total FIR line luminosities ($\it L_{\rm line}$) have 
no correlation with \lbol\ (bottom panel), 
which indicates that the line excitation is not simply related to the current
source luminosity.
The Pearson correlation coefficient between $\it L_{\rm line}$ and \lbol\ is 0.53, and its $p$-value is 0.26. The $p$-value shows the fractional chance that the correlation is not significant. Therefore, in our samples, the correlation between $\it L_{\rm line}$ and \lbol\ is not statistically significant.
However, as expected for embedded sources, the FIR continuum luminosity ($\it L_{\rm cont}$) in the
PACS range tightly correlates with \lbol\ ($r$=0.98 with $p$-value=0.0006) (see the upper panel of Fig.~\ref{correlation1}); we find the
relation of $\lbol=2.55~\it L_{\rm cont}^{\rm 0.99}$. 
Therefore, the FIR continuum luminosity can be used as an accurate tracer of
\lbol\ for embedded protostars.
\citet{Dunham08} showed that \lbol\ can also be calculated from the flux at 70 \um\ 
($\lbol \propto F_{\rm 70}^{0.94}$).

\begin{figure}
\epsscale{1.0}
\plotone{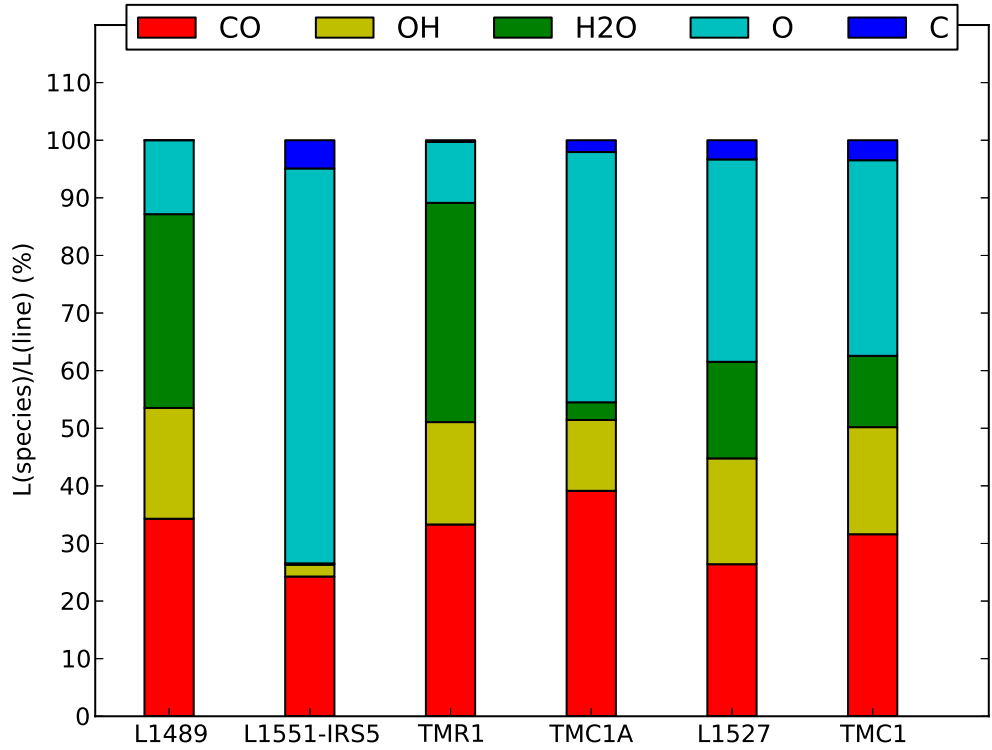}
\caption{Relative contribution of each species to the total FIR line cooling for each source. 
}
\label{lumratio}
\end{figure}

Fig.~\ref{lumratio} shows the relative contribution of each species to the
total FIR line cooling.
The CO luminosity is about 30\% of the total FIR line luminosity for all
sources.
Except L1551-IRS5, the sum of CO and OH luminosity and the sum of H$_2$O and
\OI\ luminosity
each contribute half of the total FIR line luminosity.
Except for 
L1551-IRS5, OH contributes about 20\% to the total FIR line
luminosity.

The full spectral scans from the DIGIT program provide a check on 
conclusions from programs that observed only selected transitions to
calculate the total line cooling.
\citet{Karska13} used the spectral mode of PACS to observe selected lines
toward low mass protostars, including all of our targets, 
but not L1551-IRS5.
In order to calculate the total CO luminosity in the PACS range, they
used rotational temperatures from a subset of lines within their linescans while they
used the ratios of twelve H$_2$O and  four OH line fluxes with respect to the
total H$_2$O and OH fluxes in
the full PACS range for NGC1333-IRAS4B and Serpens SMM1 to calculate the total
PACS line luminosity of H$_2$O and OH (Fig. 8 and Table 4 of \citet{Karska13}).
Their results are not completely consistent with our results, based on
the full spectral scans.  
For example, for TMR1, water contributes $\sim 40$ \% to the total FIR line
luminosity in our calculation, but only $\sim 20$ \% in \citet{Karska13}. 
In addition, for TMC1-A, there is
no contribution by water to the FIR line luminosity according to
\citet{Karska13} (because of no detection for the selected lines) while our observation
shows that $\sim 5$ \% of the FIR line luminosity is caused by water.
On the other hand, for L1527,  \citet{Karska13} overestimated the contribution
of water line to the
total FIR line luminosity ($\sim$30 \%) by a factor of 2 compared to our
calculation.
As a whole, we calculated the percentage difference between our results and Karska et al. (2013) 
for each species among five sources; the biggest difference appears in water ($\sim$60 \%). 
The average difference for CO, OH, and \OI\ are about 20 \%.
The average difference over four species for each source is about 65 \% for TMC1-A, 30 \% 
for L1527 and TMR1, and about 20 \% for L1489 and TMC1.
These differences suggest that the flux ratio between selected lines and the
full array of lines (especially for  H$_2$O) in specific sources is not applicable to all sources.

The obvious result from our analysis is that water is not the most dominant
coolant in most Taurus cloud  sources; only
in L1489 and TMR1, about 40\% of FIR line luminosity is attributed to water
lines while the water line
luminosity is less than 20\% of the total FIR line luminosity for the other 4
sources.
For L1551-IRS5, about 70 \% of the FIR line luminosity is contributed by \OI\
with little contribution
by H$_2$O and OH, indicative of nearly complete dissociation of H$_2$O and OH.
The differing contributions of each species to the total FIR line luminosity
suggests different detailed heating and chemistry in each source.

\section{Line Analysis}

\subsection{Rotation diagrams}

The simplest analysis of the rotational transitions of each molecule detected
with PACS is the rotation diagram, where lines are assumed optically thin, and
the same excitation temperature is applied to a range of level populations.
\citet{Green13} and \citet{Lee13} have described the construction of
rotation diagrams with DIGIT data in detail.
We plot the log of total number of molecules per degenerate sublevel
($\mathcal N/g\rm_J$) versus the energy in the upper state, expressed in
kelvin ($E\rm_u$).
As found in many other sources, the observed CO fluxes show a positive curvature in the rotation diagrams of four of our sources.
Therefore, for L1489, L1551-IRS5, TMR1, and TMC1, we fitted the CO rotation diagram with
two rotational temperatures with a break point at $\it E\rm_u\sim~$1800 K as is typical from
previous studies; the fitting results are not affected by moving
the break point by $\Delta$ J$\le$2 in either direction \citep{Green13}.
In L1551-IRS5, only two transitions with $\it E\rm_u \ge$ 1800 K were detected, 
and thus the hot component is poorly constrained.
For the other two sources, without detection of CO lines at ${\it E\rm_u}>1800$ K, 
we fitted only one temperature. 
The results are summarized in Table~\ref{rotco}.

For the rotation diagrams with two temperatures, high-$\it J$ CO lines (${\it
E\rm_u}\ge1800$ K) were fitted by a component $\ge$ 700 K (referred to as the
``hot" component) while low-$\it J$ CO fluxes
were fitted to 350$\sim$400 K (the ``warm" component).
The total numbers of emitting molecules, 
$\mathcal N\rm(CO)$, is always greater for the warm
component than the hot component, as found by 
Green et al. (2013) and Karska et al. (2013). 
The average ratio of $\mathcal N\rm(CO)$ between warm and hot components is about 2.5 
(with the standard deviation $\sigma$ of 0.2), which is close to the average ratio 
($\sim 2.1$ with $\sigma\sim 0.7$) in \citet{Karska13} but much smaller than the average ratio 
($\sim 8.5$ with $\sigma\sim 1.7$) in Green et al. (2013) when 
only Taurus sources in those papers  are included.
However, if all WISH and DIGIT sources are considered, the average ratios are $\sim 6$ and
$\sim 10$, respectively, indicating that Taurus sources have more 
hot CO components compared to YSOs in other star forming regions.
We note that our line fluxes are more consistent with \citet{Karska13} than \citet{Green13}.

The OH fluxes were fitted separately for lines in the 
$\Pi_{1/2}$ and $\Pi_{3/2}$
ladders, and the fluxes were fitted to $\it T\rm_{rot}$ of $\sim$100 K with
$\mathcal N\rm(OH)$ of $\sim10^{45}$.
The results are summarized in Table~\ref{rotoh}.
If we fit both ladders all together, the rotation temperature results in a value
between two temperatures fitted separately. 
We also fitted OH fluxes only at $\it E\rm_u <$ 400 K
and found higher rotation temperatures ($\sim 200$ K). 
These higher rotational temperatures were seen in low mass protostars in \citet{Karska14} 
although intermediate mass protostars show much lower rotational temperatures.
For L1551-IRS1, TMC1-A, and L1527, very few OH lines are detected so that
rotation diagrams are not useful.

\begin{figure}
\epsscale{1.0}
 \plotone{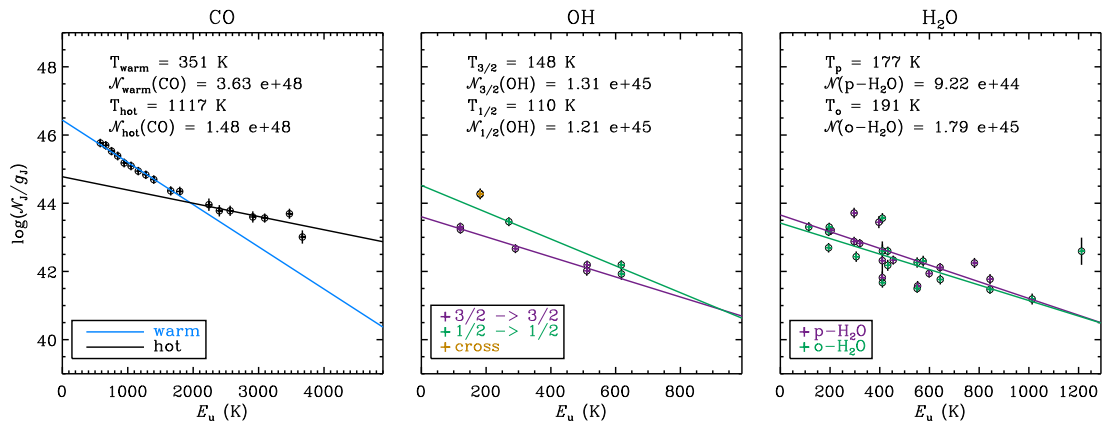}
\caption{Rotation diagrams of CO, OH, and H$_2$O for TMR1.
For CO, two temperatures (warm and hot) components are fitted separately. 
For OH, two ladder transitions ($\Pi_{1/2}$ and $\Pi_{3/2}$) are fitted separately,
and for H$_2$O, the ortho- and para-H$_2$O are fitted separately. 
The observed data are marked by open circles with error bars.}
\label{rotation}
\end{figure}

For water, we fitted o-H$_2$O (ortho-water)  and p-H$_2$O (para-water)
separately and derived $\it T\rm_{rot}$ of 100--300 K with $\mathcal N\rm(H_2O)$ of
10$^{44}$--$10^{45}$ (See Table~\ref{roth2o}).
L1551-IRS1 and TMC1-A have only one and two water line detected, respectively,
so that rotation diagrams are not useful.
The ortho-to-para ratio (OPR) derived from the total numbers of H$_2$O 
molecules varies from
source to source but is never greater than 2.5, indicative 
of a low temperature formation mechanism. 
In temperatures higher than 50 K, the equilibrium value of the OPR is 3
\citep{Mumma87}.
For all sources except for L1527, the OPR is certainly smaller than 2.5 
even if the uncertainties of the derived column densities are considered. 
A ratio of 2.5 suggests
formation below $\sim30$ K \citep{Mumma87}, much lower than their
current rotational temperature. 
The water that we observe might have formed on 
cold grain surfaces and evaporated when the grains were heated by energetic photons.
The water ice can be also sputtered by shock \citep{Neufeld14} or photodesorbed by UV
photons (Lee et al. submitted).
The timescale after evaporation (sputtering or photodesorption) should not be long enough to equilibrate the
ratio with the gas temperature.
Fig.~\ref{rotation} presents the CO, H$_2$O, and OH rotation diagrams of TMR1
as examples
since TMR1 shows the richest line emission among our sources.
The rotation diagrams for other sources can be found in the online supplementary material.

\subsection{LVG models}

The assumptions of the rotation diagram analysis 
(optically thin lines and a single rotational temperature)
may not hold for all PACS lines. 
Even if a single rotational temperature fits the data, it may not
be equal to the true kinetic temperature of  the gas.
In addition, the rotation diagram cannot provide direct
information on the density of the associated gas.
Therefore, we adopted the non-LTE LVG code, RADEX \citep{Tak07} to model the
intensities of the detected lines.
The three parameters ($\it T\rm_K$, $\it n\rm{(H_2)}$, and $\it
N\rm(molecule)$/$\Delta \it v$) are used to determine the level population.
We used the LAMDA database \citep{Schoier05} for molecular data.
Because we do not know the actual solid angle of the source or the 
linewidth, we cannot convert intensity to line flux. Therefore, we
applyed a scaling factor to models that gives the minimum $\chi^2$.
The scaling factor is the product of emitting area and line width.

In the rotation diagrams, the CO gas seems to have two temperature components
of 300$-$400 K and 700$-$1100 K in our samples (Table~\ref{rotco}).
However, \citet{Neufeld12} suggested that one gas component with a very high
temperature ($>1000$ K) with a low $\it n\rm{(H_2)}$ of $10^4\sim10^5$ cm$^{-3}$
could generate a positively curved CO rotation diagram.
They also pointed out that gas with a power-law temperature distribution can
account for the observed curvature in rotation diagrams.

\subsubsection{CO}

In our RADEX models of CO, we tested the case of a single gas component and also gas
with a power-law temperature distribution ($dN\propto T^{-p} dT$).
The ranges of physical conditions tested were 50 K $\le \it T
\rm_K \le$ 5000 K, 10$^3 \rm cm^{-3} \le \it n\rm{(H_2)} \le 10^{13} cm^{-3}$, and
10$^{13} \rm cm^{-2}/km$ $\rm s^{-1} \le \it N\rm(CO)/\Delta\it v \le\rm
10^{20} cm^{-2}/km$ $\rm s^{-1}$ for the one component model.
The temperature range for the power-law model is 10 to 5000 K, and the total
$N\rm(CO)/\Delta\it v$ is fixed to 10$^{15} \rm cm^{-2}/km$ $\rm s^{-1}$,  which 
satisfies the optically thin assumption and reduces the number of  free parameters.
The density range was 10$^3 \rm cm^{-3} \le \it n\rm{(H_2)} \le 10^{13} cm^{-3}$, and
the power index range was $0<p<4$. 
We calculated the collisional rates up to 5000 K by following \citet{Neufeld12} and \citet{Schoier05}.
The best-fit models are summarized in Table~\ref{lvgco}.

For the one component model, these sources prefer a 
sub-thermal solution with $n \le 10^5$ cm$^{-3}$ and T=5000 K, which is the highest temperature for 
available collision rates. 
Therefore, the temperature of this model is not constrained. 
If a higher temperature is available, it would fit the observations better with a lower density.
 
Therefore, if a C-shock is responsible for these CO fluxes, the shock velocity must be $\ge 40$ km s$^{-1}$ 
in order to provide $T_K \ge 5000$ K \citep{Flower10}.
The density derived from this model is the post-shock density. Therefore, if we consider the enhancement factor of density by shocks, 
the pre-shock density is smaller than the derived values by an order of magnitude (for C-shocks) or more (for J-shocks).
The densities for the one-component models are mostly $\le 10^5$ \cmv, 
which would imply pre-shock densities $\le 10^4$ \cmv. If these are shocks
into the cavity walls, they would need to be far from the source to get
such low densities, but we observe compact emission. The alternative is 
low density molecular gas {\it within} the outflow cavity rather than on the
walls.

For the power-law model, our sources require a 
higher density ($1.0\times 10^5$ to  $3.2\times 10^6$ cm$^{-3}$) with $p\sim 1-3$.
The pre-shock density for the best-fit models of TMR1 and TMC1 can be found at $\sim 1000$ AU for their envelope models
\citep{Kristensen12} if a C-shock is responsible for the CO emission.
However, the pre-shock density for the best-fit model for L1489 can be found
at the radius greater than 1000 AU or within the outflow cavity.

\begin{figure}
\epsscale{0.7}
\plotone{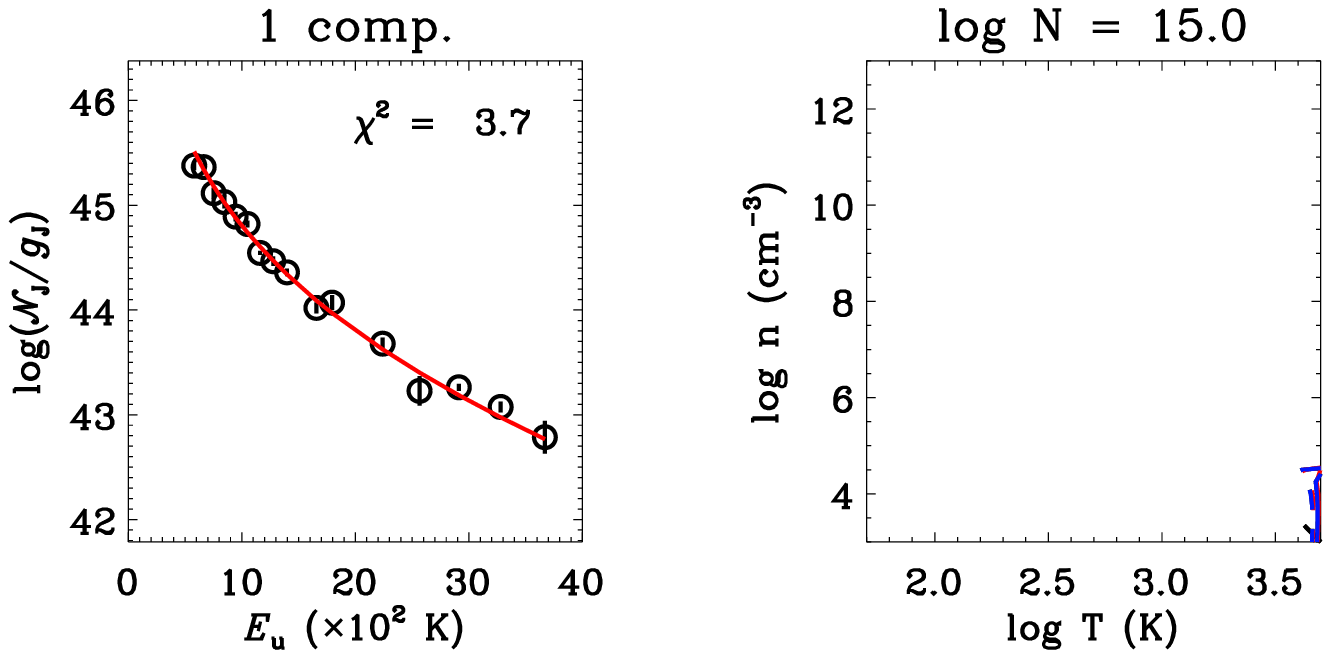}
\plotone{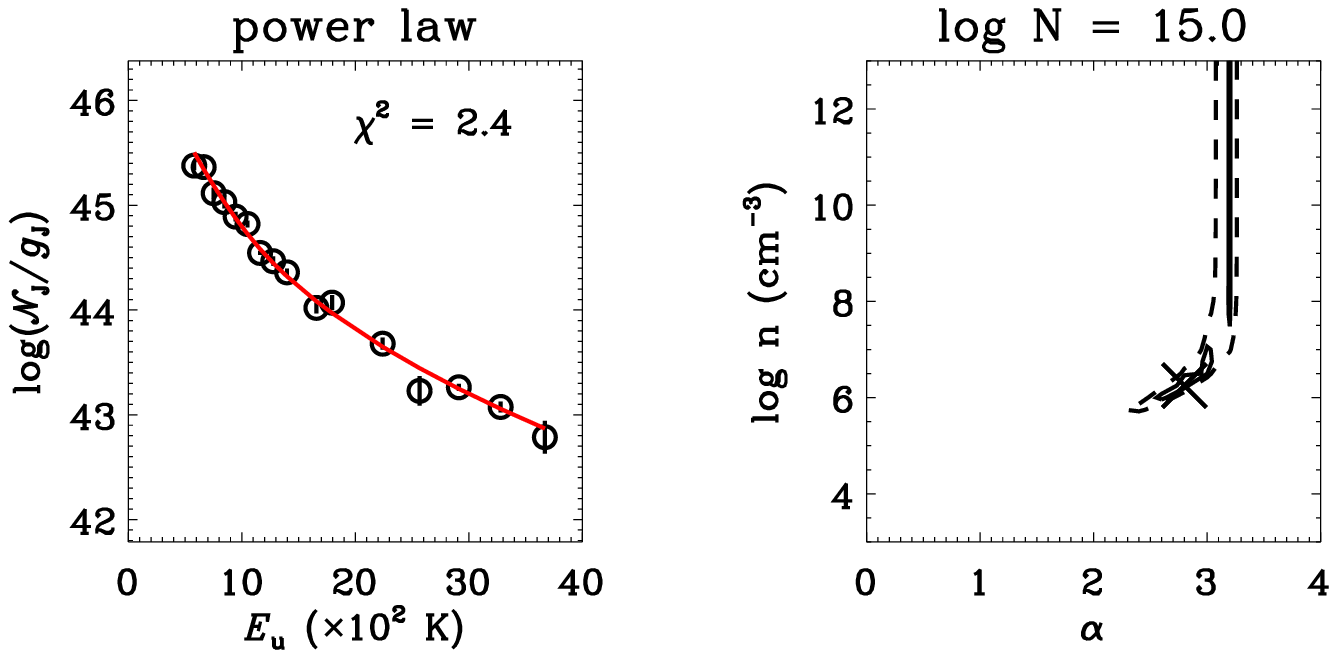}
\caption{Best-fit RADEX models for CO lines (top panels for the one-component model and 
bottom panels for the power-law temperature distribution model) in TMC1.
In rotation diagrams (left panels), circles with error bars are observed data, and the solid lines represent
the best-fit models. 
The physical conditions for the models are summarized in Table~\ref{lvgco}.
The reduced $\chi^2$ is presented as contours in the ($n$, $T$) space at the best-fit column density
(right panels). The contours are 1.2 (solid line) and 1.5 (dashed line) times the best $\chi^2$ value 
(cross), which is presented in the rotation diagrams.
For the one component model (top right), the black contours represent $\chi^2$ for the best-fit model, 
but blue and red contours show $\chi^2$ for the models with lower and higher column densities, 
respectively, than the best-fit column density by a factor of 10. However, all these models collapse to 
the sub-thermal solution with a low density and the highest temperature available, so those different
color contours are not distinguishable in the plot.
}
\label{tmc1_lvgco}
\end{figure}

\begin{figure}
\epsscale{1.0}
 \plotone{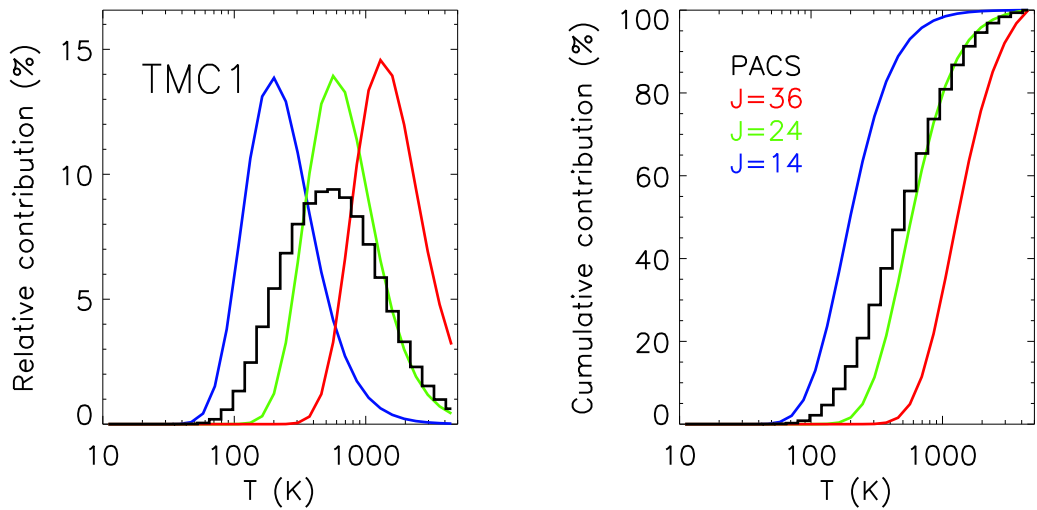}
\caption{The relative (left) and cumulative (right) contribution of different temperatures to the total CO flux
at the PACS wavelength range (black histogram) and fluxes of J=14-13 (blue line), J=24-23 (green line),
and J=36-35 (red line) in the best-fit power law model for TMC1. 
Most ($\sim$70 \%) of the total CO flux comes from the gas with T$<$1000 K.}
\label{tmc1_pow_cont}
\end{figure}

The best CO models of TMC1 are presented in Fig.~\ref{tmc1_lvgco}. Figures for
the RADEX models for other sources can be found in the online material.
For the CO models, only line flux measurement errors are considered because we
are the most
interested  in fitting the shape of the CO rotation diagrams. (For other molecules, 
we consider the total errors, which are dominated by the
continuum calibration errors. The continuum calibration errors are 
assumed to be 20 \% of the continuum
fluxes.)
For TMC1 and TMR1, the power-law model fits the CO observations better than the single temperature model, although all other sources show similar reduced $\chi^2$ for the power-law and one component models (see Table~\ref{lvgco}). 
Fig.~\ref{tmc1_pow_cont} shows the relative (left) and cumulative (right) 
contributions of different temperatures
to the total flux for the best-fit power law temperature model of TMC1. 
Most of flux comes from the gas with T$<$1000 K for TMC1 and TMR1, where the best-fit density is greater than 10$^6 \rm cm^{-3}$ with $p$=2.8.

\begin{figure}
\epsscale{0.8}
\plotone{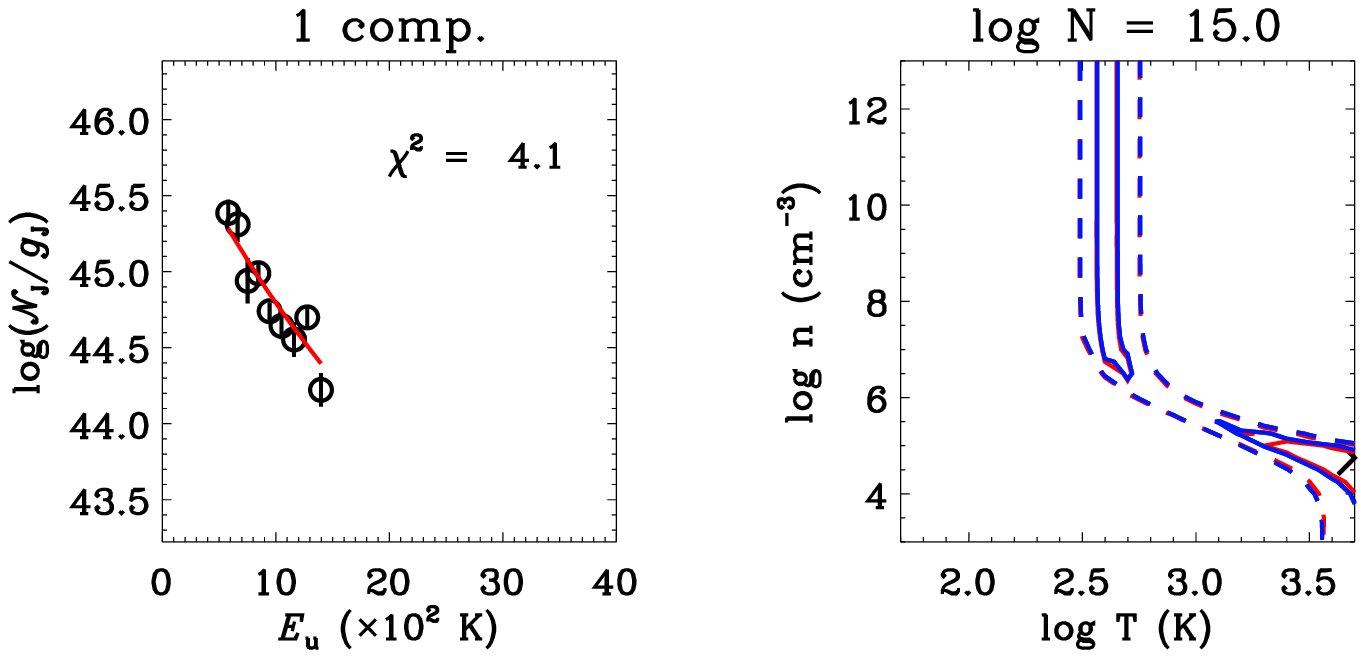}
\plotone{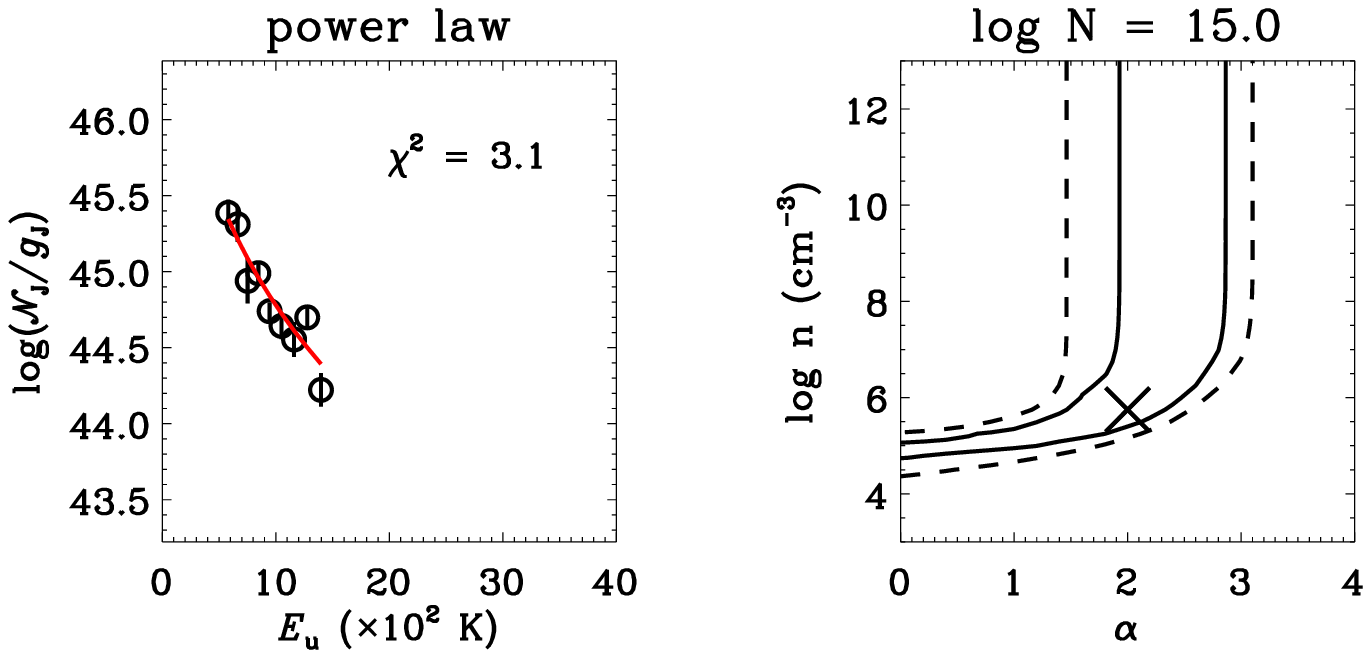}
\caption{The same as Fig.~\ref{tmc1_lvgco} except for L1527}
\label{L1527_lvgco}
\end{figure}

On the other hand, L1489 is best fitted with the sub-thermal solution 
for both the one-component model and the power-law 
model because its rotation diagram shows less curvature than TMC1 and TMR1. 
For L1527 and TMC1-A, the fits for the two models 
are not very different; as seen in the $\chi^2$ plot (Fig.~\ref{L1527_lvgco}), both thermal 
and sub-thermal solutions are possible both in the one-component model 
and in the power-law model 
although the best-fit model has been selected to have the lowest $\chi^2$ for Table~\ref{lvgco}.
This unconstrained result is caused by the lack of lines at $E_{\rm u}$ above 1800 K in the two sources.
L1551-IRS5 shows a similar result to L1527 and TMC1-A.
Generally speaking, the single-temperature,
 non-LTE models tend to favor the highest kinetic
temperatures in the grid, so they are not well constrained.
The power law models, on the other hand, result in a better fit,
with a reasonably well-constrained power-law index, but the densities
are often relatively unconstrained on the high side, suggesting that 
the excitation is close to thermal.

\subsubsection{H$_2$O}

We also modeled the H$_2$O fluxes with RADEX considering both one component and
 power-law temperature distribution as was done for CO.
The OPRs found in the rotational diagrams have been adopted.
The $\chi^2$ and physical properties are similar for the best-fit 
models with different OPRs between 1 and 3.

\begin{figure}
\epsscale{0.8}
 \plotone{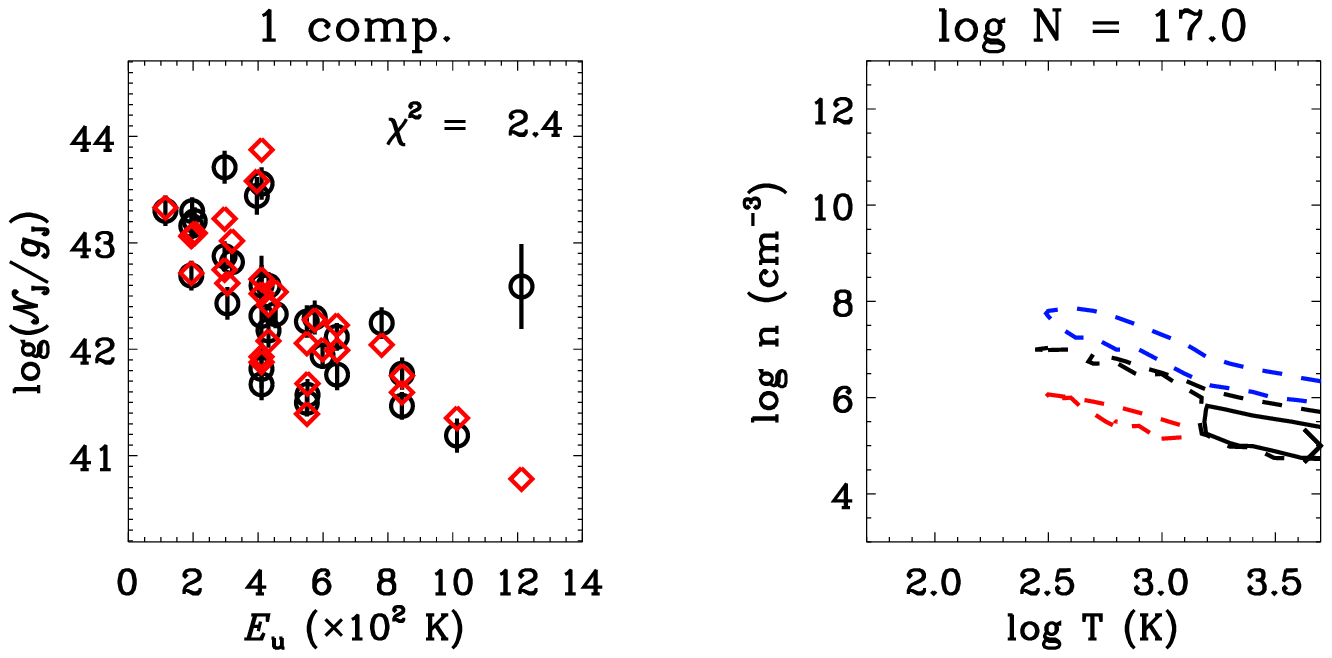}
 \plotone{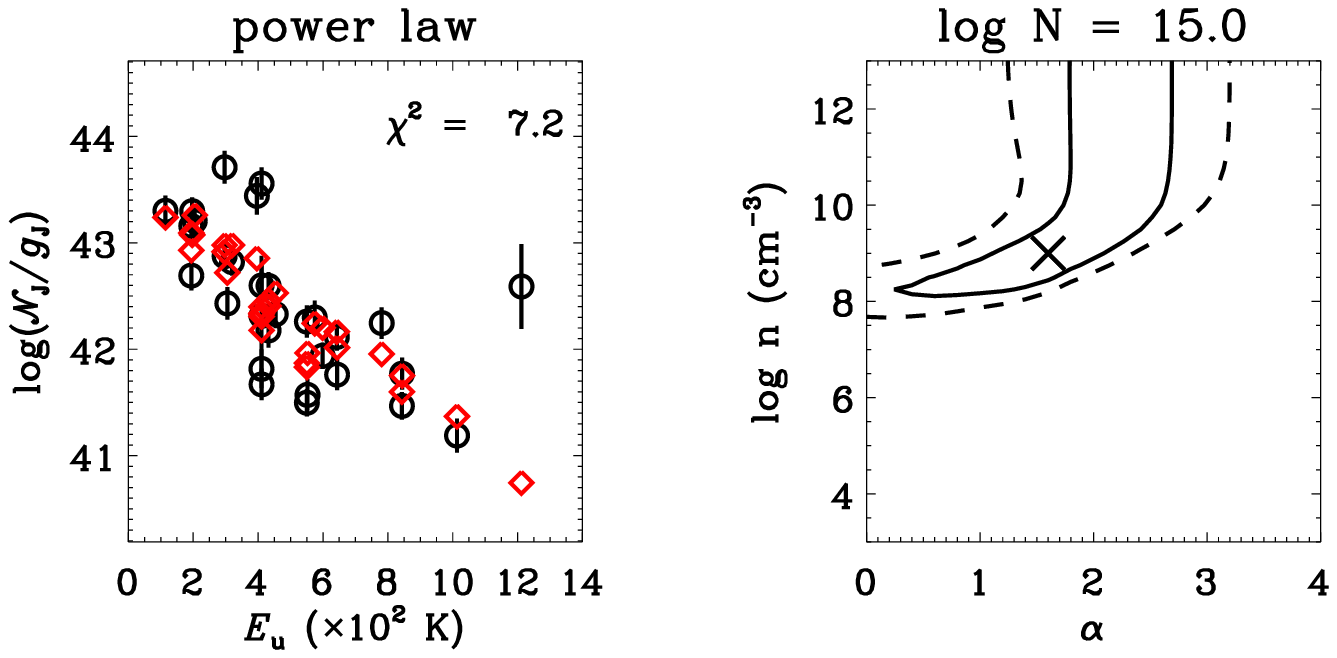}
\caption{Best-fit RADEX models for H$_2$O lines (top panels for the one-component model and 
bottom panels for the power-law temperature distribution model) in TMR1.
The physical conditions for the models are summarized in Table~\ref{lvgh2o}.
Diamonds without error bars indicate model points. Contours for $\chi^2$ are the same as in 
Fig.~\ref{tmc1_lvgco}.
}
\label{fig_lvgh2o}
\end{figure}

For the one component model, we explored the same physical range as CO. 
For the power-law model, we used the temperature range of 10 K to 5000 K as CO, but the total $N\rm(H_2O)/\Delta\it v$ ranges from $10^{13}$ to $10^{15}$ cm$^{-2}\rm/km $ s$^{-1}$.
The results for our best models are summarized in Table~\ref{lvgh2o}, and
Fig.~\ref{fig_lvgh2o} presents
the best-models for TMR1, which are the richest in water lines among
our sources.

The one component model fits observations better than the power-law model for water, unlike CO. 
The power-law model is meaningful only when lines are optically thin 
because we do not have information on how the gas components with different temperatures 
distribute along the line of sight and the projected space. 
For the best-fit models, this condition is satisfied by CO lines, but half of water lines are 
optically thick in half of our sources. 
Therefore, the power-law temperature distribution 
in the 1-D physical structure cannot provide a good model for water lines.
The gas temperatures of the best-fit one component models for H$_2$O are much greater than
 the rotational temperatures derived from its rotation diagrams, i.e., water is
sub-thermally excited because of its high critical density.
The densities ($10^4 \sim 10^7$ cm$^{-3}$) of our best-fit models are much
lower than the critical densities of many of water lines at the PACS range, which can
reach up to $10^{11}$ cm$^{-3}$  \citep{Herczeg11}.

The best-fit models for CO (power-law model) and H$_2$O (one component model) 
indicate that CO and H$_2$O are not excited by the same gas component; models
that fit the CO do not fit the H$_2$O.
CO seems to be excited by the warm gas mostly 
(at least  for TMC1 and TMR1),
but H$_2$O is excited by the hot and dense gas. 
However, the 1-D power-law temperature model cannot deal with appropriately
optically thick lines, and most water lines are optically thick.

\subsubsection{OH}

It has been suggested 
that IR pumping could have an effect on the population of
OH \citep{Wampfler13}, so we included FIR radiation in our model.
We considered FIR radiation by including blackbody radiation from the inner
boundary of the envelope on top of the cosmic background radiation, which is
implemented in RADEX.
The temperature at the inner boundary of the envelope is assumed to be 250 K.
In order to calculate the dilution of the IR radiation from the central source
to the given density, we assumed that the envelope has a power-law density
structure with $n=n_{0}\times(\frac{r}{r_0})^{-p}$
($r_0$ = inner radius at AU, $n_0$ = density at $r_0$ cm$^{-3}$), and $r_0$,
$n_0$, and $\it p$ were adopted from \citet{Kristensen12}.
For more details of our inclusion of the central IR source, refer to
\citet{Lee13}.
We modeled the range of 50 K $\le \it T \rm_K
\le$250 K, 10$^3 \rm cm^{-3} \le \it n\rm{(H_2)} \le$ (maximum density of the
envelope model), and $\rm10^{10} cm^{-2}/km$ $\rm s^{-1} \le \it
N\rm(OH)/\Delta\it v \le\rm 3.2\times10^{18} cm^{-2}/km$ $\rm s^{-1}$.

\begin{figure}
\epsscale{1.0}
\plottwo{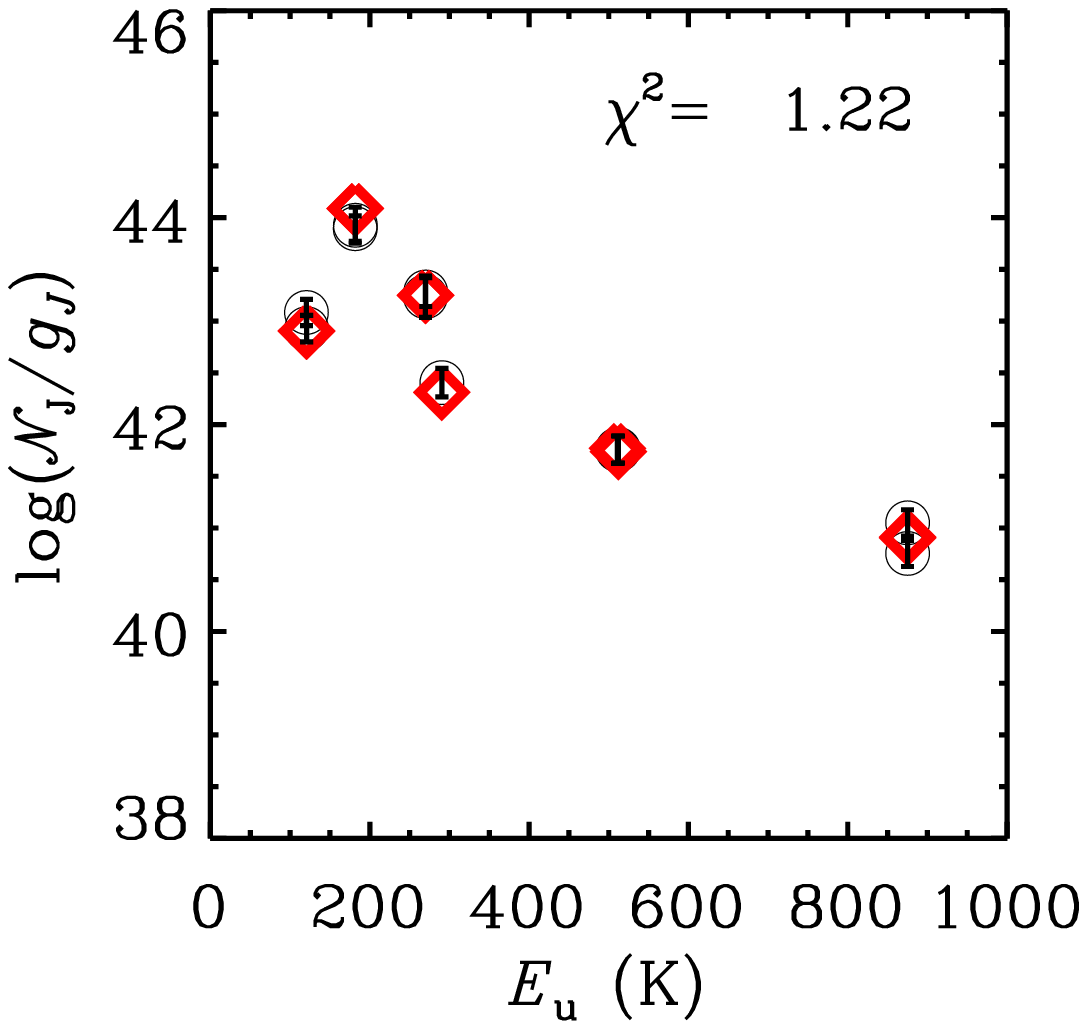}{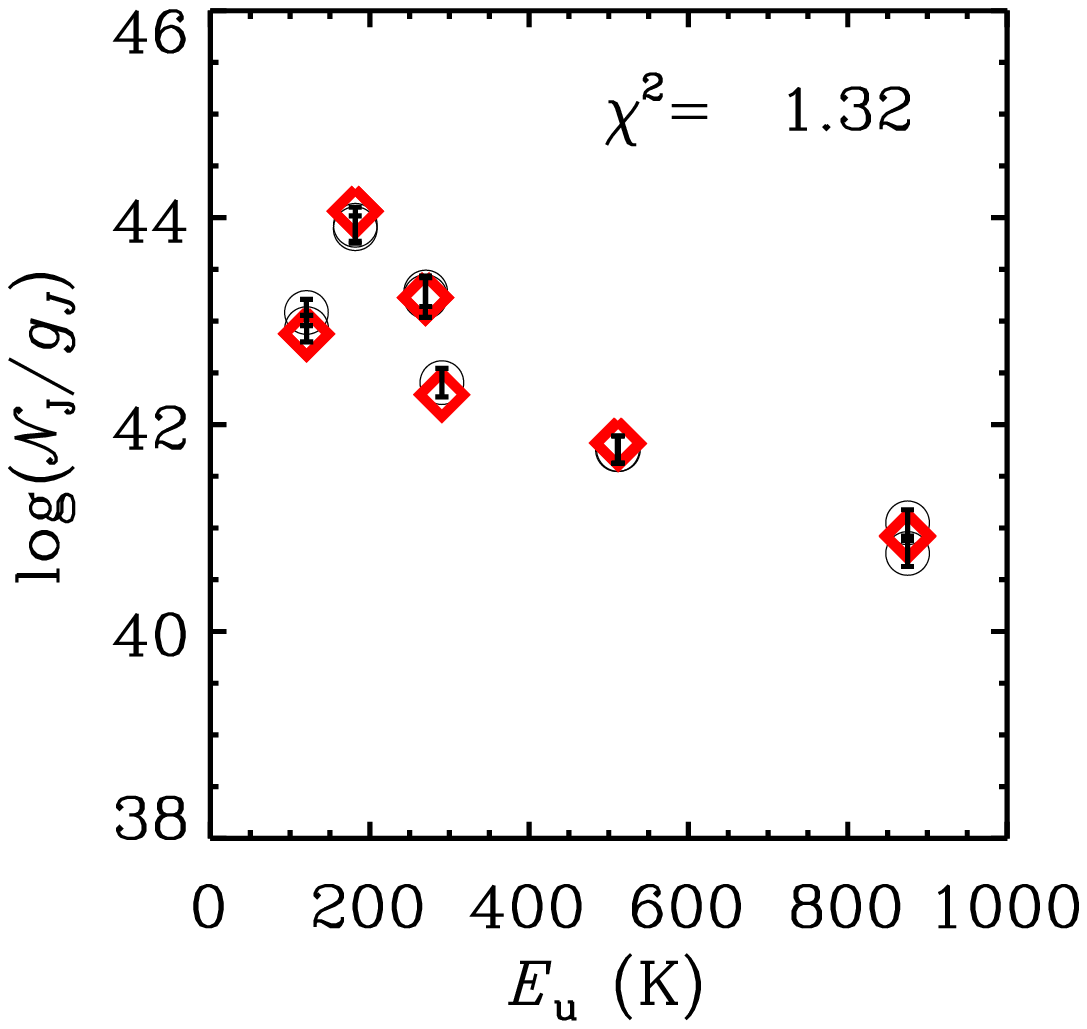}
\caption{Best-fit RADEX models for OH lines in TMC1; (Left) the best-fit LVG
model with the IR-pumping effect and (right) the best-fit LVG model without the
IR-pumping effect.
The physical conditions for the models are summarized in Table~\ref{lvgoh}.
Symbols are the same as in Fig.~\ref{fig_lvgh2o}.
}
\label{tmc1_lvgoh}
\end{figure}

We considered fits with and without the IR pumping effect only from the central IR source.
The OH fluxes of all the sources could be explained by either model; IR pumping
effect is not necessary to reproduce the observed OH fluxes
(see Fig.~\ref{tmc1_lvgoh} and Table~\ref{lvgoh}).
\citet{Lee13} showed that the IR pumping is important only for the OH line with
the highest energy
level of 875 K (at 55.9 \um), which was not detected toward these sources except for TMC1.
For TMC1 where the highest energy level OH line has been detected, the
consideration of IR pumping lowers the best-fit OH column density by a factor
of 10, compared to the model without the IR pumping effect.
However, a caveat exists in the modeling of OH fluxes; the collision rates for
OH are limited to $T_{\rm K} < 250 K$ in the current molecular data. As in CO and H$_2$O,
OH might be also excited by a much hotter gas.
In addition, unlike \citet{Wampfler13}, our model did not consider the IR radiation
from the dust component mixed with gas in situ.
The lower energy level transitions are mostly optically thick, so 
the external IR photons do not strongly affect those lines.

We conclude that ad-hoc models with a single temperature or with a temperature
power law can fit some molecules but not others. Clearly a self-consistent
physically motivated model including shocks and radiation is needed to make
further progress.

\section{L1551-IRS5}

L1551-IRS5  stands out among our targets as the most luminous in both
continuum and lines.  This source shows
very dry emission without any significant water line emission but with very
strong \OI\ and \CII\ emission.
Similar characteristics (with mostly CO and [O I] emission without much 
water emission) has been shown in more luminous protostars \citep{Karska14}.
The OH absorption line at 119 $\mu$m has been detected toward
L1551-IRS5.
\citet{Wampfler13} found the same absorption line feature toward several
intermediate mass protostars (AFGL 490, Vela IRS 17, and Vela IRS 19),
and one low mass protostar, the Class 0 source,  NGC 1333  IRS 2A 
\citep{Wampfler10}.  \citet{Karska14} also found the OH absorption feature, 
even in transitions from higher energy levels, toward more luminous protostars.

L1551-IRS5 is a deeply embedded protostellar binary system
\citep{Rodriguez98},
and X-ray emission associated with HH 154 was discovered
 \citep{Favata02, Bally03, Favata06}.
The presence of X-ray sources in the vicinity of the protostar \citep{Bonito04,
Favata06,Schneider11} suggests that high velocity jets ($v > 500\sim1000$ km
s$^{-1}$) might have induced
the X-ray emission, indicative of very  fast shocks
associated with L1551-IRS5.

\begin{figure}
\epsscale{0.7}
 \plotone{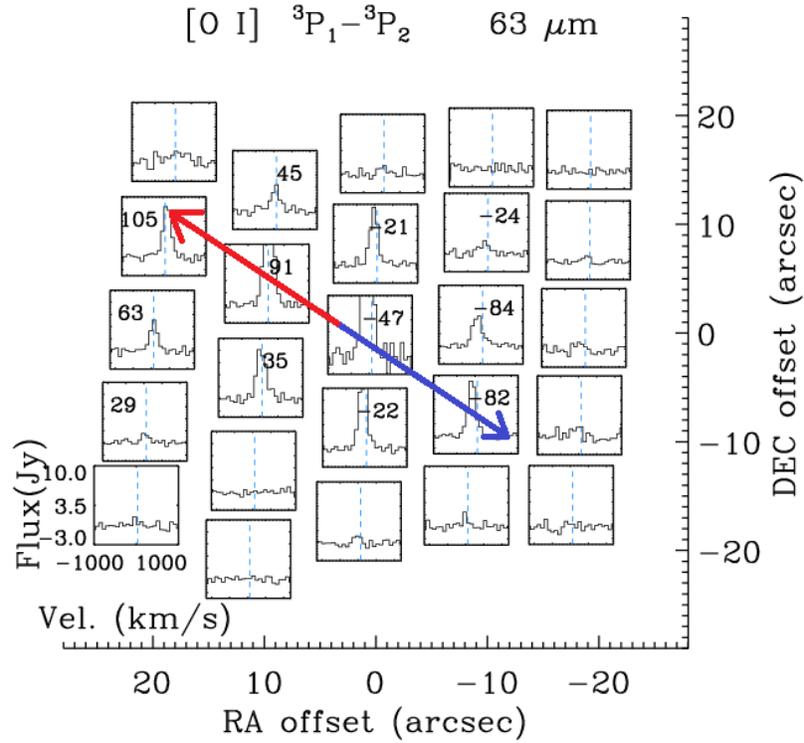}
 \caption{The central velocity shift of the \OI\ 63.1 $\mu$m line in
L1551-IRS5.
 The central velocity is noted in the top of each spectrum.
The marked red and blue arrows indicate the outflow direction for the
red-shifted and blue-shifted components, respectively.
}
\label{l1551_OI_vel_map}
\end{figure}

Although the PACS spectra do not have sufficient spectral resolution to resolve
the detailed kinematics in most sources, 
very fast outflows can make detectable velocity shifts.
Fig.~\ref{l1551_OI_vel_map} shows the \OI\ 63 $\mu$m line
spectrum in each spaxel.
The \OI\ emission distributes along the known outflow direction
\citep{Irena06}.
The number marked in the top of each box indicates the central velocity
of the \OI\ line derived from the Gaussian profile fitting.
The known source velocity ($v_{\rm LSR}$) is +7.2 km s$^{-1}$.
Therefore, 
this map shows that the central velocity shifts up to $\pm 100$  km s$^{-1}$ approximately, 
with the redshifted emission in
the NE and the blueshifted emission in the SW, which is consistent with
the known outflow feature \citep{Irena06}.
Therefore, the velocity structure indicates that the \OI\ emission is tracing
the jet/wind
in L1551-IRS5.

\begin{figure}
\epsscale{1.0}
\plotone{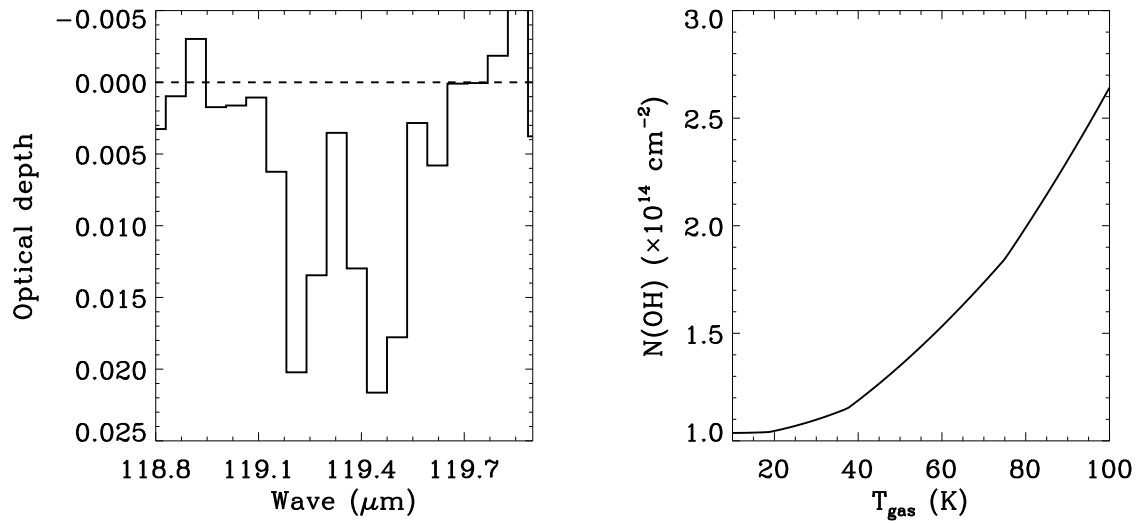}
\caption{(left) The optical depth of the OH 119 $\mu$m line in L1551-IRS5 and
(right)
the OH column density, calculated from the optical depth profile, as a function
of the kinetic temperature.
}
\label{l1551_oh119}
\end{figure}

The OH absorption profile is shown in Fig.~\ref{l1551_oh119} (left) as the
optical depth versus wavelength.
OH molecules in the dense envelope can produce this absorption
feature due to the high FIR radiation field from the central source.
We calculate the column density of OH based on the LTE assumption using the
equation below.

\begin{equation}
\int \tau_{\lambda} d\lambda = \frac {c}{\nu^2} \frac{c^2}{8\pi\nu^2} A_{ul}
[\frac{g_u}{g_l} x_l-x_u] N, ~~~
x_j  = \frac{g_j}{Q} e^{(-E_j/KT)},
\end{equation}

\noindent where $\tau$ is the observed optical depth, $N$ is the total column
density,  $Q(T)$ is the partition function , $T$ is the gas temperature, $c$ is
the speed of light, and $g_u$ ($g_l$) and $E_u$ ($E_l$) is the statistical
weight and the energy of  upper (lower) level, respectively. $A_{ul}$ and $\nu$
 are the Einstein coefficient and frequency, respectively, for this transition.

The column density needed to produce the absorption as a function of
temperature is presented in
Fig.~\ref{l1551_oh119} (right).
If the cold envelope (T$<$20 K) produces this absorption feature, then the
lower limit of
the OH column density in the cold envelope is about $1 \times 10^{14}$
cm$^{-2}$.

\section{Discussion}

\subsection{FIR Line luminosities}

Our 6 sources show no significant trend
in FIR line luminosity ($\it L_{\rm line}$) with either \lbol\ ($r$= 0.53, $p$-value=0.26) or
\tbol\ ($r$=0.04, $p$-value=0.94).
\citet{Karska13} found a significant correlation ($r$=0.71, $p$-value=0.0009) between $\it L_{\rm line}$ and \lbol\ at the 3-$\sigma$ level with 18 sources. Therefore, the small sample in this work may obscure the actual correlation. 

\begin{figure}
\epsscale{0.7}
 \plotone{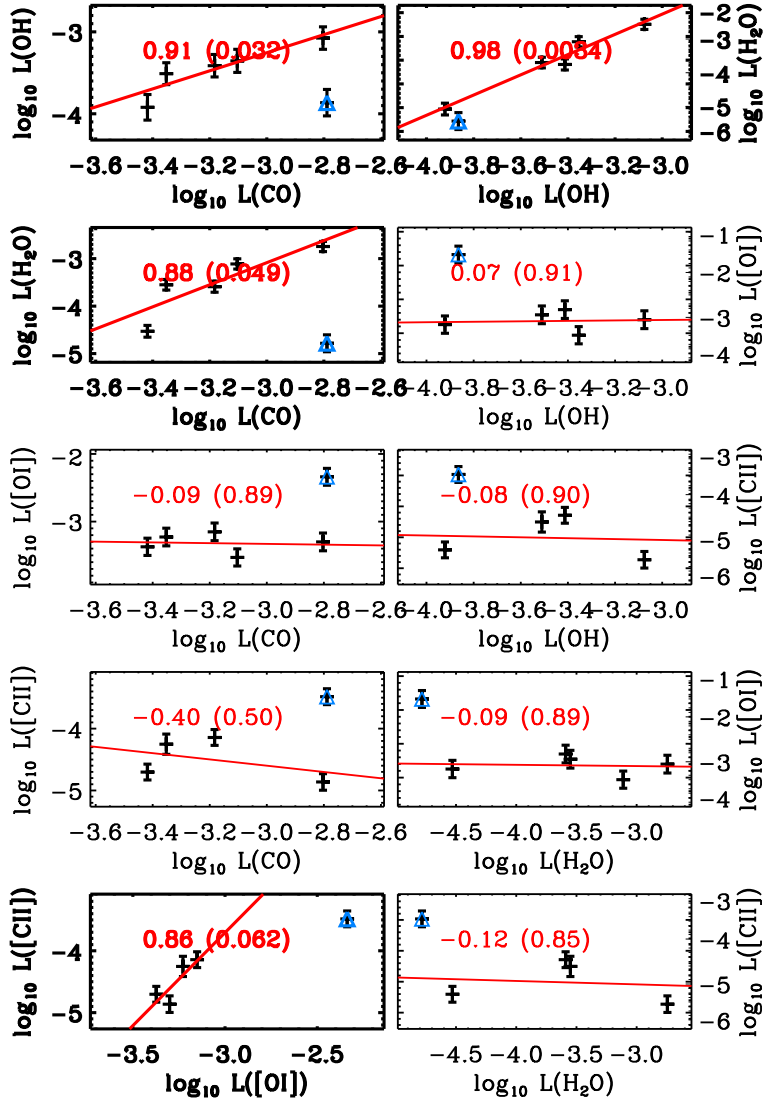}
\caption{Correlations among line luminosities of different species.
Luminosities are in the logarithmic scale. The blue triangle in each box
indicates values for L1551-IRS5, which is not included in the correlation
calculation.
The boxes with thick lines represent species with some correlation.
The red numbers in the upper left corner indicate the Pearson correlation coefficient with its $p$-value in parentheses. 
}
\label{correlation2}
\end{figure}

Our targets include one Class 0 and five Class I sources in the same
environment.
The Class 0 object, L1527, shows no special features compared to five other
Class I sources.
However, the most luminous source, L1551-IRS5 shows very different line
luminosities compared to
other less luminous sources; its FIR line luminosity is dominated by the
\OI\ lines. 
Thus we show correlation coefficients and fits without L1551-IRS5 in
Figure~\ref{correlation2} and Table~\ref{correl}.

The Pearson correlation coefficients at 3-$\sigma$ (99.7$\%$ confidence) and 2-$\sigma$ (95.4$\%$ confidence) levels for the sample number of 5 are 0.98  and 0.89, respectively.
Therefore, if we consider only the five less luminous sources, 
molecular line luminosities correlate with each other at the 2-$\sigma$ level, but no correlation with atomic line luminosities is
seen (Fig.~\ref{correlation2}, Table~\ref{correl}).
However, the [O I] luminosity is correlated with the [C II] luminosity,
indicative of the same heating mechanism.
This correlation between [O I] and [C II] luminosities is consistent with the
similar spatial distribution of
the [O I] and [C II] line intensities shown in L1551-IRS5 and TMC1
(Fig.~\ref{l1551_OI_CII},
Fig.~\ref{tmc1_OI_CII}).
Therefore, the [C II] emission is possibly associated with the jet shock or the
outflow cavity walls heated
by the UV photons from the central source.

\subsection{Origins of FIR line emission}

As shown \citet{Green13} and \citet{Karska13}, the CO rotation diagrams
for YSOs are commonly fitted by two components (warm and hot).
\citet{Manoj13} showed that the curvature in the CO rotation diagram
could be explained either by a hot gas with low densities,which
possibly  exists inside the outflow cavity, or a gas with a power-law
temperature distribution.
Data on DIGIT sources, including L1448-MM \citep{Lee13} and 
GSS30-IRS1 (Je et al. in prep), seem to prefer the latter solution.

Visser et al. (2012) combined the PDR and a C-shock along the outflow cavity
walls to fit the PACS CO line fluxes of
several embedded protostars; they attributed the warm CO ($T\rm_{rot} =  300$ K)
and the hot CO component of $T\rm_{rot} \sim 1000$ K, 
respectively to the PDR and C-shock.
The 300 K component appears in the PACS CO rotation diagram for almost all
embedded YSOs regardless of their luminosities, however, \citet{Manoj13} and
\citet{Karska13} argued that the PDR has a minor
contribution to the CO emission in the PACS range.

Surprisingly, according to our analysis, the relative contribution of CO to the total gas cooling is kept approximately constant. 
The relative contribution of CO to the total gas cooling in the PACS range is about 30\% within errors in
our 6 sources (see Fig.~\ref{lumratio}) as well as in L1448-MM \citep{Lee13} and GSS30-IRS1 (Je et al. in prep.).
Somehow, the relative contribution of CO to the total gas cooling is kept
approximately constant.

\begin{figure}
\epsscale{1.0}
\plotone{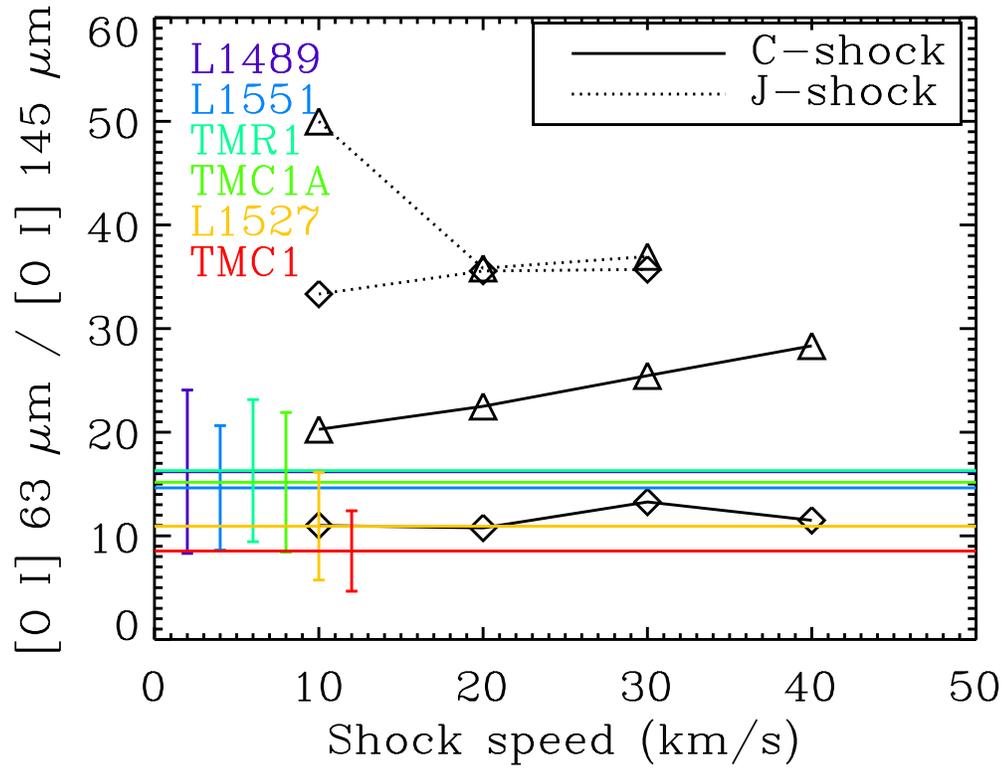}
\caption{The comparison between the shock models by \citet{Flower10} and the
observations in the flux ratio between the \OI\ 63 and 145 $\mu$m.
The black lines and symbols represent the model results and the color lines
show the observed values in  our targets. Solid lines and dashed lines
represent C- and J-shock models, respectively. Diamonds refer
to $2\times10^4$ cm$^{-3}$ and triangles to $2\times10^5$ cm$^{-3}$ of hydrogen
density.
}
\label{oi63_145}
\end{figure}

\begin{figure}
\epsscale{0.8}
 \plotone{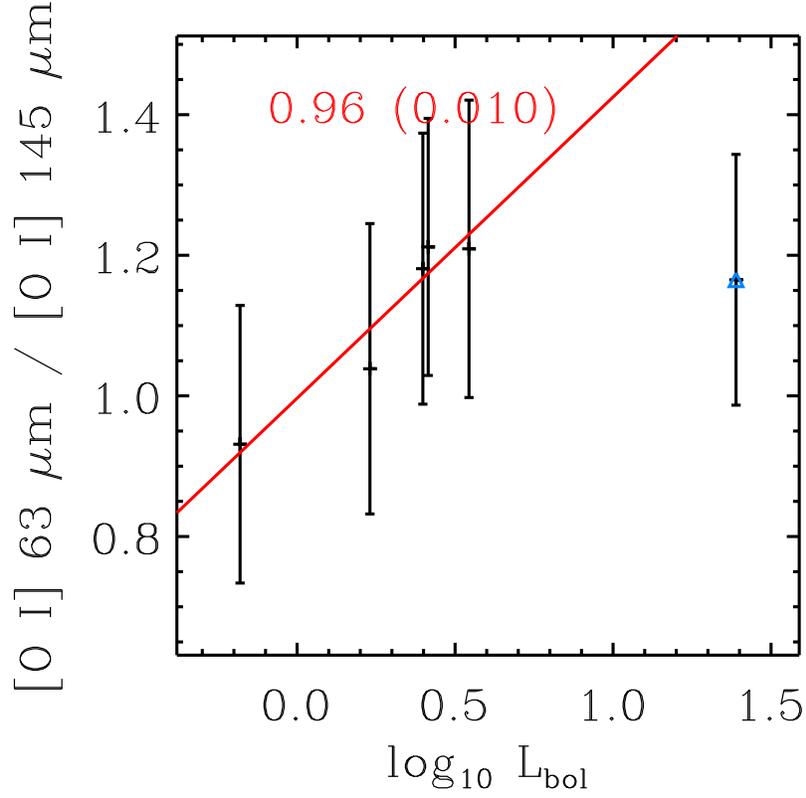}
\caption{Correlation between flux ratio of \OI\ 63 $\mu$m and 145 $\mu$m and
\lbol.
The correlation was calculated by excluding L1551-IRS5, which is marked as a
blue triangle. The red number at the upper left corner is the correlation coefficient with its $p$-value inside parentheses.
}
\label{correlation4}
\end{figure}

We compared the line fluxes of our targets with the model fluxes for shocks calculated by
\citet{Flower10}.
The \OI\ 63/145 \um\ ratio of all sources is consistent with the C-shock model (Fig.~\ref{oi63_145}).
According to \citet{Kristensen12}, our sources have 
$n(\rm H_2)\sim 2\times10^5$ cm$^{-3}$  at the radius of 1000 AU.
Fig.~\ref{correlation4} shows the \OI\ 63/145 \um\ ratio as a function of
\lbol; the ratio increases with \lbol.
In C-shock models, the ratio increases with shock velocity, at least at the
density of $2\times 10^5$ cm$^{-3}$ as seen in Fig.~\ref{oi63_145}. Therefore,
a higher \lbol\ may result in a C-shock with a higher velocity.

\begin{figure}
\epsscale{0.8}
 \plotone{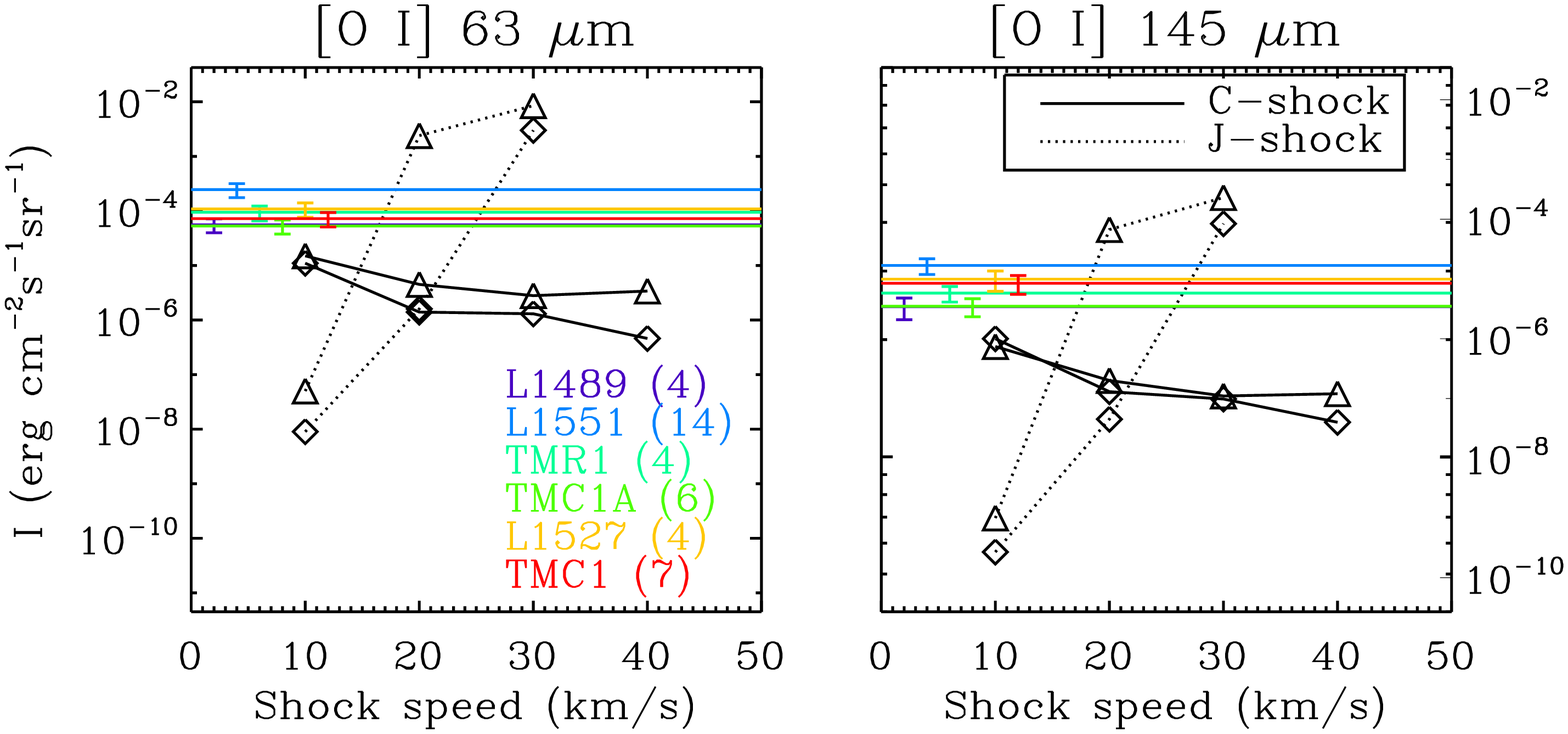}
\caption{The average intensity of [O I] lines over emitting area for each source. 
The numbers in parentheses are the number of spaxels where the [O I] 63 \um\ line was detected.
Symbols and colors are the same as in Fig.~\ref{oi63_145}.}
\label{abs_oi}
\end{figure}

The actual intensity of [O I] lines can be calculated with an assumption of emitting area, which 
we adopt as the total size of spaxels where the [O I] emission is detected. 
The average [O I] 63 \um\ intensity is 5.6$\times10^{-5}$, 2.5$\times10^{-4}$, 9.4$\times10^{-5}$, 
5.3$\times10^{-5}$, 1.1$\times10^{-4}$, and 7.2$\times10^{-5}$ erg $\rm cm^{-2}s^{-1}sr^{-1}$  
for L1489, L1551-IRS5, TMR1, TMC1-A, L1527, and TMC1, respectively.
When compared with the shock models of \citet{Flower10} (Fig.~\ref{abs_oi}), 
the average intensity is consistent more with J-shock models ($15<v<25$ km s$^{-1}$) for all sources.
However, if a C-shock in an irradiated environment (such as outflow cavity walls)
is considered, the less luminous five sources might be fitted with a lower shock velocity.
If a J-shock was indeed important, the [O I] line ratio would indicate that 
the 63 \um\ line was optically thick.
However, the CO line fluxes are not well-described by either J- or C-shock models as a whole.

\begin{figure}
\epsscale{1.0}
\plottwo{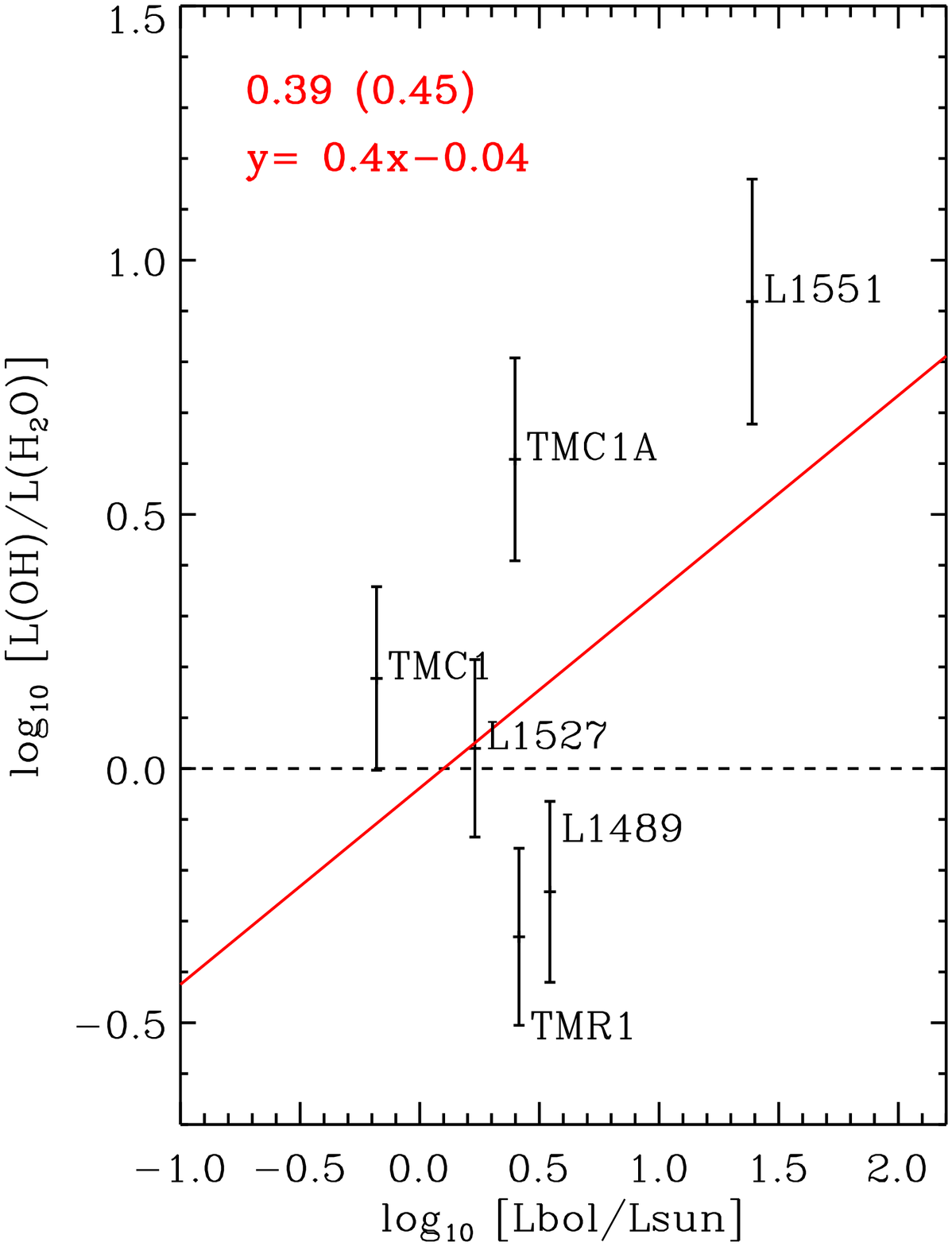}{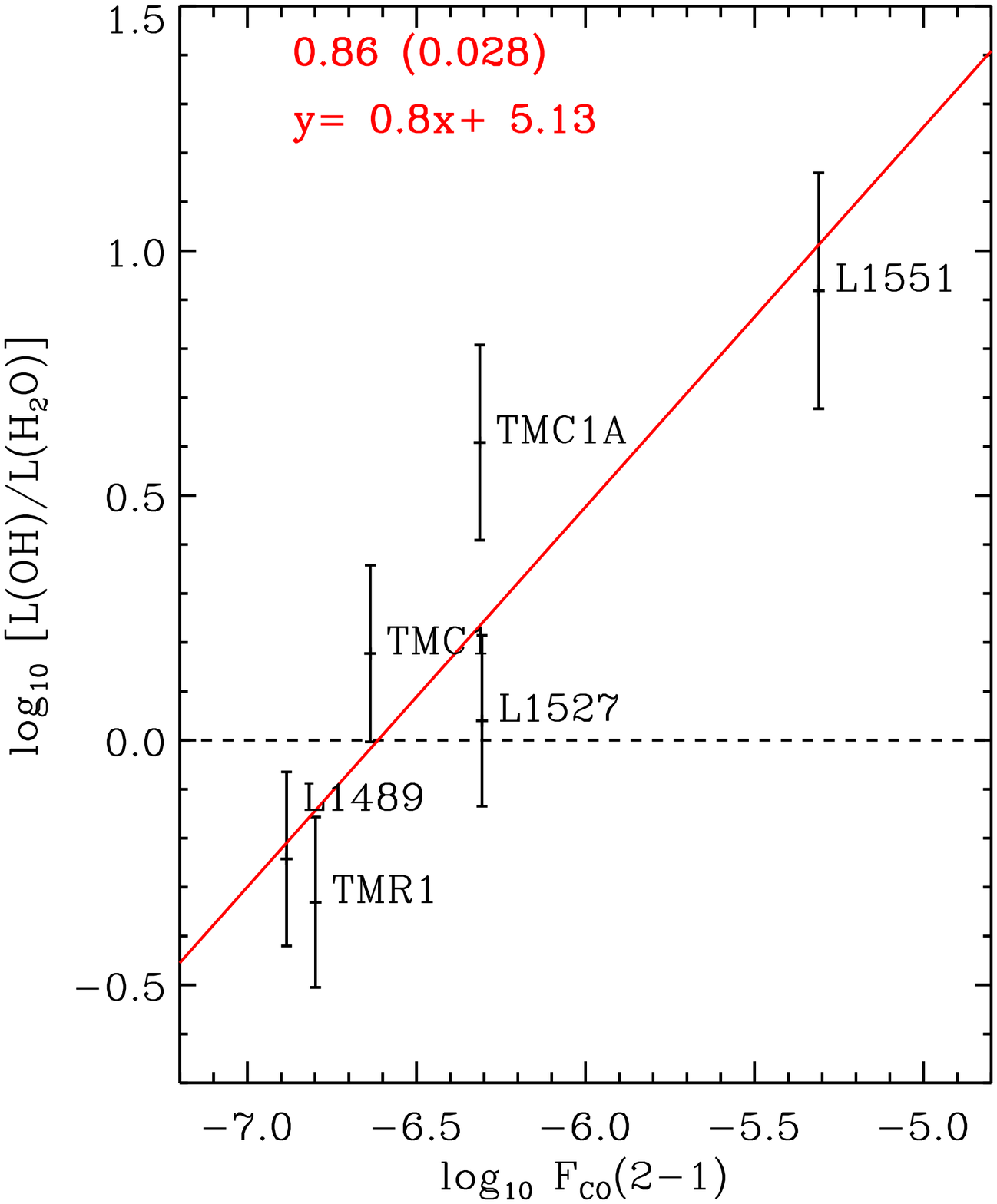}
\caption{The luminosity ratios of OH relative to that of H$_2$O with respect to \lbol\  and $\it F\rm_{CO}$(2-1).
The red numbers in the upper left corner indicate the correlation coefficients with their $p$-values inside parentheses.
The equations given inside boxes present the fitting result of red lines.
$\it F_{\rm CO}$(2-1) are adopted from Kang et al. (in prep.).
}
\label{oh_ratio}
\end{figure}

\begin{figure}
\epsscale{1.0}
\plottwo{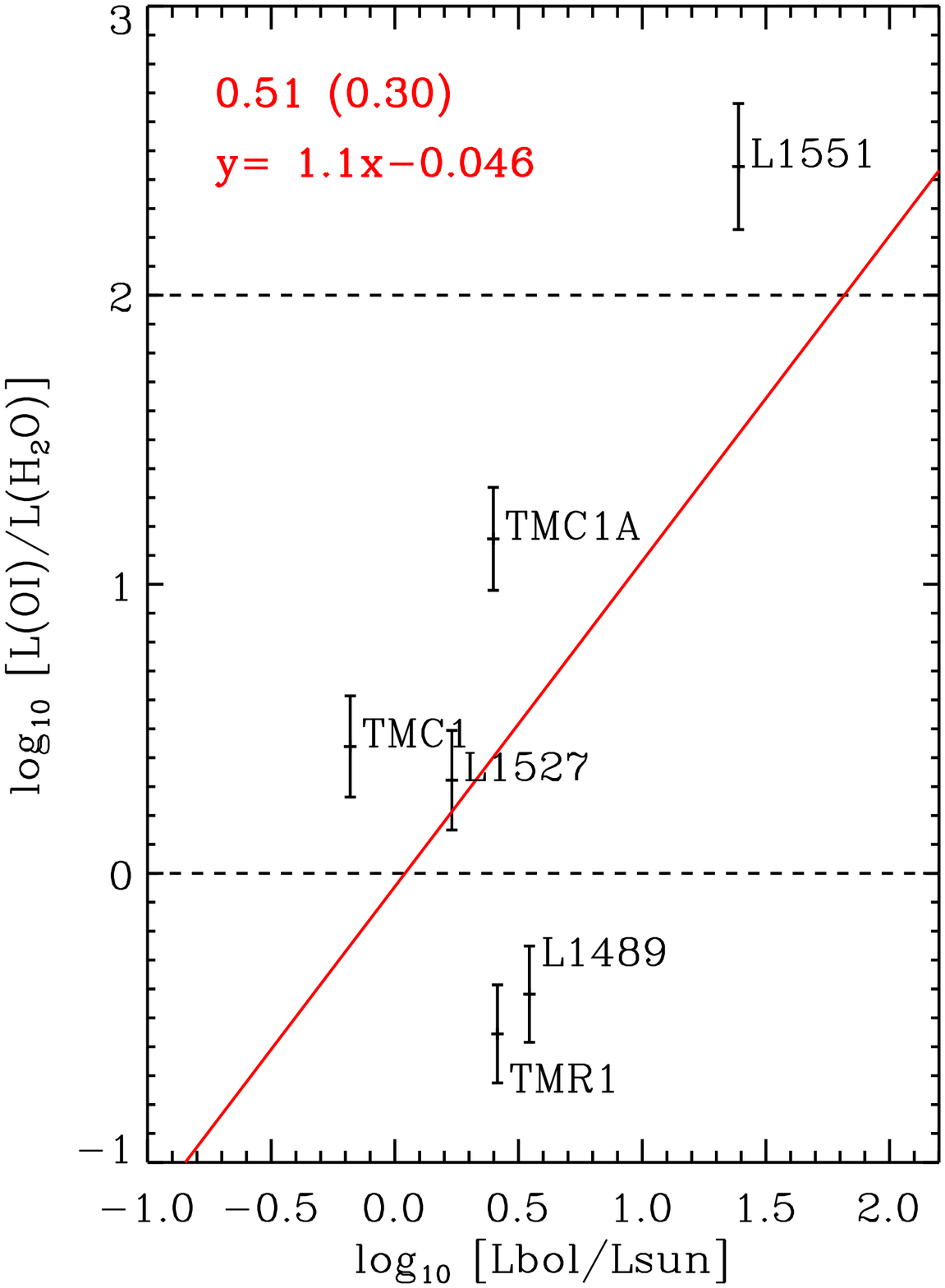}{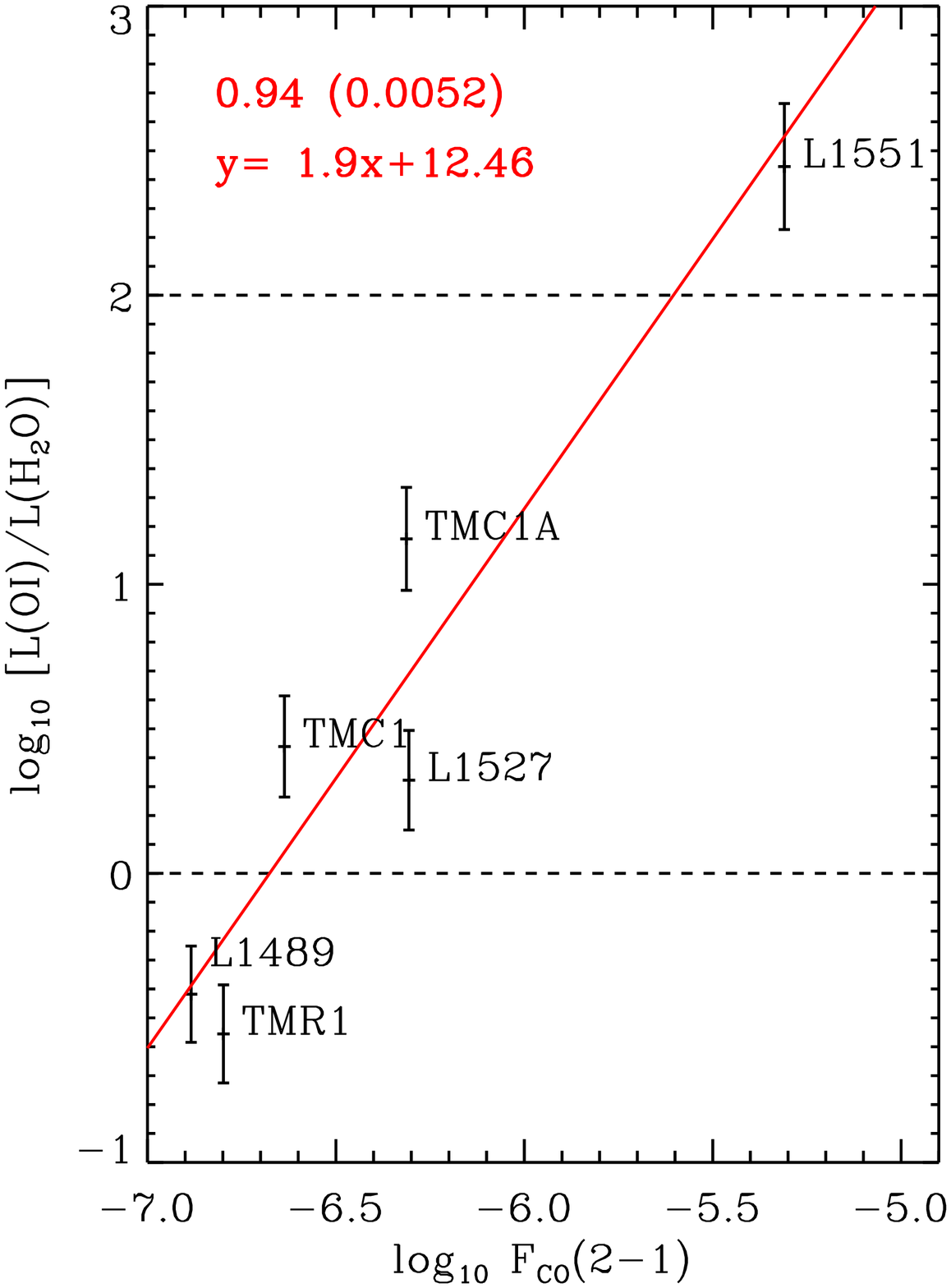}
\caption{The luminosity ratios of  \OI\ relative to that of H$_2$O with respect to \lbol\ and $\it F_{\rm CO}$(2-1).
The numbers are the same as in Fig.~\ref{oh_ratio}.
}
\label{oi_ratio}
\end{figure}

Fig.~\ref{oh_ratio} and Fig.~\ref{oi_ratio} present the luminosity ratio of OH and \OI, 
respectively, relative to that of H$_2$O.
According to \citet{Lindberg14}, the high flux ratios of OH/H$_2$O and
\OI/H$_2$O might be
caused by the PDR process, which dissociates H$_2$O to OH and O, but
their analysis was based on one specific line of each species.
Although not very distinct in OH/H$_2$O luminosity ratio (Fig.~\ref{oh_ratio}), 
our sources are readily divided into three groups in the \OI/H$_2$O luminosity ratio 
(Fig.~\ref{oi_ratio}); \loi/\lwater $<$ 1 in L1489 and TMR1, between 1 and a few 10s 
(more conservatively, 100) for TMC1, L1527, and TMC1-A, and above 100 for L1551-IRS5.
The group with \loi/\lwater ratios greater than 1 is likely affected by photodissociation of water.
A portion of CO must be dissociated in the same gas component since the spatial
distribution of the \CII\
emission is consistent with that of \OI\ (at least for L1551-IRS5 and TMC1),
and their luminosities have a correlation.

However, the origin of the high-energy photons does not need to be the central
or external (proto-)stars.
For instance, L1551-IRS5 is known to have shock-induced X-ray emission \citep{Favata02, Favata06}.
Therefore, a dissociative J-shock can also increase those OH/H$_2$ and \OI/H$_2$O luminosity ratios.
However, the low \lbol\  of TMC1, for example, indicates a low mass accretion 
rate, and thus a very energetic shock capable of dissociating H$_2$O, is not
expected.
Indeed, as seen in Fig.~\ref{oh_ratio} (right) and Fig.~\ref{oi_ratio} (right), 
the outflow momentum flux ($F\rm_{CO}$) of L1551-IRS5 
is much larger than the other sources. 
Therefore, the two groups with the ratios lower than 100 may be explained by
C-shocks.
The middle group with the ratios greater than 1 and smaller than
$\sim$100 may have
been influenced by UV photons produced by the accretion to the central
protostar in addition to C-shocks, i.e.,
irradiated C-shocks may be important for the middle group.
Therefore, the ratio of \loi/\lwater\ may characterize the heating
mechanisms in the YSO envelopes; $\sim$100 for the division between J- and
C-shock and 1 for the division between an irradiated C-shock and C-shock without
irradiation.

One more important point from Fig.~\ref{oh_ratio} and Fig.~\ref{oi_ratio} is that 
these ratios of \loi/\lwater\ and \loh/\lwater\ are better correlated with  $F\rm_{CO}$ ($r$=0.94 with $p$-value=0.0052 and $r$=0.86 with $p$-value=0.028) 
than \lbol\ ($r$=0.51 with $p$-value=0.30 and $r$=0.39 with $p$-value=0.45). 
$F\rm_{CO}$ traces the integrated activity over the entire lifetime of the YSO, but \lbol\
is associated more with the current accretion process. 
Therefore, it is interesting to find a tighter correlation of $F\rm_{CO}$ 
with the currently 
shocked gas probed by the FIR \OI, OH, and H$_2$O observations.
According to this result, the FIR line luminosity ratio between \OI (or OH) and H$_2$O should 
reflect the time averaged outflow properties.

According to the PDR model by Lee et al. (subm.), CO can have an abundance
greater than $10^{-5}$ at $n\rm(H_2)$ of $10^6$  to $10^8$  cm$^{-3}$  if
the gas temperature is greater than 300 K. These conditions may arise at the cavity wall
surface (${\it A\rm_V} < 0.5$ mag) when the ratio of UV flux ($G_0$; in the
unit of the average interstellar radiation field of $1.6\times10^{-3}$ erg
cm$^{-2}$ s$^{-1}$) to density ($n$), $G_0/n$  is $\sim10^{-3}$ cm$^3$.
Certainly, the gas in the cavity wall has been shocked and must have high
velocities as
observed with HIFI \citep{Kristensen12}. Therefore, the kinematical, physical,
and
chemical conditions along the outflow cavity wall cannot be explained solely
either by PDR
or shock.
Further investigation with more sophisticated models for irradiated shocks is
necessary.

\section{Summary}

The DIGIT observations of 6 Taurus sources with the PACS SED range mode have
been analyzed.

1.  The FIR continuum luminosity (\lcont) has a very strong correlation with \lbol,
but the richness of line emission ($L_{\rm line}$) is not related to \lbol. 
The main difference between sources is the amount of water emission, which is relatively low 
in the most luminous sources. L1489, TMR1, and
TMC1, which are the faintest sources among our targets, are rich in water lines
while L1551-IRS5 and L1527, which are known as strong outflow sources, do not
show much water emission. 

2.  Molecular emission is compact in these sources, but atomic emission such as \OI\ and \CII\
tends to be extended  along the known outflow directions.

3. The relative line luminosity of each species varies by source, although CO consistently 
contributes $\sim$30\% to the total FIR line cooling.

4.  Similar to previous studies, the rotation diagram of CO shows is well-characterized 
by two temperature components of $\sim$350 K and $\ge700$ K. 
The rotation diagrams of OH and H$_2$O can be fitted by a single comopnent 
between 100 K and 200 K.

5. The non-LTE LVG models of CO indicate that models with temperature
gradients are the most plausible.
In order to fit our H$_2$O fluxes, hot gas with kinetic temperatures above 1000
K, which is much higher than the excitation temperatures derived from the
rotation diagrams, is required.
The OH fluxes can be fitted by models either with or without the IR-pumping effect  from the central source.

6. Among our sources, L1551-IRS5 seems to have different physical and chemical
conditions.  Approximately around 70\% of its FIR line luminosity appears in two \OI\ lines, with
almost no accompanying water emission.
This strongly suggests that the very fast J-shock is working in L1551-IRS5 to
dissociate
water to atomic oxygen. In addition, the \OI\ emission is distributed along the
outflow direction
with the consistent shift of the central velocity,
probably induced by jet-shocks.
In L1551-IRS5, the OH 119 \um\ line shows absorption, which might be produced
by the OH column density of $\ge 1\times 10^{14}$ cm$^{-2}$.

7. The FIR line luminosities of each molecular species correlate with each other among
the less luminous five sources at the confidence greater than 95\%. 
The \OI\ and \CII\ line luminosities correlated with each other 
(94\% confidence)
but not with other molecular luminosities.

8. It is difficult to designate one heating mechanism to explain all line
emission in the PACS range. It is unrealistic to assume that
only the PDR or shock alone heats
the gas in the embedded sources. Rather the combination of the two processes,
i.e., irradiated shocks must produce the FIR line emission for most sources.
However, the shock properties, and thus, the resulting chemistry must be very
different in L1551-IRS5 from the other five sources; a dissociative J-shock
likely controls the energetics and its chemistry in L1551-IRS5.

9. The line luminosity ratio, $\it L\rm_{[OI]}$/$\it L\rm_{H_2O}$ can be used to
characterize the heating mechanisms in the YSO envelopes; $\sim$100 for
the division between J- and C-shock and 1 for the division between irradiated
C-shock and
normal C-shock.

\acknowledgments

We thank A. Karska and the anonymous referee, whose comments led to improvements 
in the paper.
J.-E. L. is very grateful to the department of Astronomy, University of Texas
at Austin for the hospitality
provided to her from August 2013 to July 2014. J.-E. L. was supported by LG
Yonam Foundation Oversea Research Professor Program in 2013.
Support for this work, part of the Herschel Open Time Key Project
Program, was provided by NASA through an award
issued by the Jet Propulsion Laboratory, California Institute of
Technology.
This research was supported by the Basic
Science Research Program through the National Research Foundation
of Korea (NRF) funded by the Ministry of Education of
the Korean government (grant No. NRF-2012R1A1A2044689)
and the 2013 Sabbatical Leave Program of Kyung Hee Unviersity
(KHU-20131724, 20131727).

\clearpage

\clearpage

\clearpage

\clearpage

\clearpage


\begin{deluxetable}{lrrrr}
\tablewidth{0pt}
\tablecaption{Source information}
\scriptsize
\tablehead{
\colhead{Name} 		& \colhead{RA}	 	& \colhead{Dec}	 &
\colhead{\tbol}\tablenotemark{a}(K)		& \colhead{\lbol}\tablenotemark{a}(\lsun)}
\startdata
L1489 	&	04:04:42.9 &	+26:18:56.3	&	226	&	3.5 \\
L1551-IRS5 & 	04:31:34.1 &	+18:08:04.9	&	106	&	24.5 \\
TMR1 	&	04:39:13.9 & 	+25:53:20.6	&	140	&	2.6 \\
TMC1-A 	&	04:39:35.0 & 	+25:41:45.5	&	164	&	2.5 \\
L1527 	&	04:39:53.9 & 	+26:03:09.8	&	67	&	1.7 \\
TMC1 	&	04:41:12.7 & 	+25:46:35.9	&	171	&	0.66 
\enddata
\tablenotetext{a}{values from \citet{Green13}}
\label{sourceinfo}
\end{deluxetable}

\begin{deluxetable}{lcccccc}
\tablewidth{0pt}
\tablecaption{Correction factors}
\tablehead{
\colhead{  } & \colhead{L1489} & \colhead{L1551-IRS5} & \colhead{TMR1} & \colhead{TMC1-A} & \colhead{L1527} & \colhead{TMC1}  }
\startdata
Blue ($<$ 100 \um) & 1.0	&	0.891\tablenotemark{a}, 0.979\tablenotemark{b} & 1.076	&	1.054 &	1.028	&	0.960\\
Red  ($>$ 100 \um) & 0.939	&	1.007\tablenotemark{c}, 1.057\tablenotemark{d} & 0.904	& 0.985 & 0.928 & 0.827
\enddata

\tablenotetext{a}{for $\lambda<$72 \um}
\tablenotetext{b}{for 72 \um$<\lambda<$100 \um}
\tablenotetext{c}{for 100 \um$<\lambda<$142 \um}
\tablenotetext{d}{for $\lambda>$142 \um}
\label{factortable}
\end{deluxetable}

\clearpage

\begin{deluxetable}{llllcccccc}
\tablewidth{0pt}
\tabletypesize{\footnotesize}
\tablecaption{Line Fluxes (10$^{-18}$ W m$^{-2}$)}
\scriptsize
\tablehead{
\colhead{Species} 		& \colhead{Transition}	 	& \colhead{E$_u$ (K)}	 &
\colhead{$\lambda$ ($\mu$m)} &
\colhead{L1489} 		& \colhead{L1551-IRS5} 			& \colhead{TMR1}	&
\colhead{TMC1-A} 	&
\colhead{L1527}		& \colhead{TMC1}   }
\startdata
CO& 36-35&       3668.78&       72.84& --& --&           31$\pm$          14& --& --&           18$\pm$           8\\
& 35-34&       3471.27&       74.89& --& --&          133$\pm$          45& --& --& --\\
& 34-33&       3279.15&       77.06&           38$\pm$          14& --& --& --& --&           28$\pm$           8\\
& 33-32&       3092.45&       79.36&           25$\pm$          10& --&           77$\pm$          23& --& --& --\\
& 32-31&       2911.15&       81.81&           37$\pm$          16& --&           72$\pm$          27& --& --&           33$\pm$           9\\
& 30-29&       2564.83&       87.19&           52$\pm$          23&          126$\pm$          55&           82$\pm$          28& --& --&           23$\pm$           9\\
& 29-28&       2399.82&       90.16& --& --&           70$\pm$          26& --& --& --\\
& 28-27&       2240.24&       93.35& --&          106$\pm$          36&           90$\pm$          37& --& --&           47$\pm$          14\\
& 25-24&       1794.23&       104.45&           65$\pm$          23& --&          130$\pm$          41&           50$\pm$          17& --&           69$\pm$          22\\
& 24-23&       1656.47&       108.76&           77$\pm$          22& --&          111$\pm$          35&           55$\pm$          24& --&           51$\pm$          15\\
& 22-21&       1397.38&       118.58&           87$\pm$          25&          332$\pm$          97&          158$\pm$          45&           51$\pm$          16&           53$\pm$          20&           73$\pm$          21\\
& 21-20&       1276.05&       124.19&          112$\pm$          32&          255$\pm$          75&          175$\pm$          50&           82$\pm$          23&          129$\pm$          41&           75$\pm$          22\\
& 20-19&       1160.20&       130.37&          117$\pm$          33&          268$\pm$          92&          178$\pm$          51&           52$\pm$          19&           73$\pm$          28&           71$\pm$          20\\
& 19-18&       1049.84&       137.20&          121$\pm$          34&          259$\pm$          73&          197$\pm$          58&           62$\pm$          19&           70$\pm$          24&          106$\pm$          31\\
& 18-17&       944.97&       144.78&          123$\pm$          34&          282$\pm$          80&          186$\pm$          55&           62$\pm$          17&           67$\pm$          21&           95$\pm$          29\\
& 17-16&       845.59&       153.27&          108$\pm$          33&          246$\pm$          72&          226$\pm$          67&           56$\pm$          19&           91$\pm$          30&          100$\pm$          31\\
& 16-15&       751.72&       162.81&          113$\pm$          32&          192$\pm$          63&          232$\pm$          66&           43$\pm$          13&           60$\pm$          26&           90$\pm$          27\\
& 15-14&       663.35&       173.63&          131$\pm$          37&          383$\pm$         115&          251$\pm$          73&           69$\pm$          20&          103$\pm$          40&          117$\pm$          33\\
& 14-13&       580.49&       186.00&           98$\pm$          28&          248$\pm$          74&          209$\pm$          60&           47$\pm$          15&           88$\pm$          30&           86$\pm$          25\\ 

\\

OH& $\frac{1}{2}$,$\frac{9}{2}$-$\frac{1}{2}$,$\frac{7}{2}$ &       875.10&       55.89& --& --& --& --& --&           18$\pm$           5\\
& &       875.10&       55.95& --& --& --& --& --&           36$\pm$          10\\
& $\frac{3}{2}$,$\frac{9}{2}$-$\frac{3}{2}$,$\frac{7}{2}$&       512.10&       65.13&          140$\pm$          40& --&          257$\pm$          75& --& --&           94$\pm$          28\\
& &       510.90&       65.28&           81$\pm$          23& --&          170$\pm$          56& --& --&           93$\pm$          28\\
& $\frac{1}{2}$,$\frac{7}{2}$-$\frac{1}{2}$,$\frac{5}{2}$ &       617.60&       71.17&           46$\pm$          14\tablenotemark{b}& --&          150$\pm$          46&           74$\pm$          25\tablenotemark{a}& --& --\\
& &       617.90&       71.22&           46$\pm$          14& --&           80$\pm$          32&           74$\pm$          25& --& --\\
&  $\frac{1}{2}$,$\frac{1}{2}$-$\frac{3}{2}$,$\frac{3}{2}$ &       181.90&       79.12&           75$\pm$          24& --&          147$\pm$          51&           51$\pm$          21&          119$\pm$          37\tablenotemark{a}&           64$\pm$          25\\
& &       181.70&       79.18&           74$\pm$          28& --&          139$\pm$          46& --&          119$\pm$          37&           59$\pm$          16\\
&  $\frac{3}{2}$,$\frac{7}{2}$-$\frac{3}{2}$,$\frac{5}{2}$ &       290.50&       84.60&           71$\pm$          27&          122$\pm$          39&          193$\pm$          58& --& --&          105$\pm$          33\\
& $\frac{3}{2}$,$\frac{5}{2}$-$\frac{3}{2}$,$\frac{3}{2}$ &       120.70&       119.23&           73$\pm$          22& -260$\pm$77 &           99$\pm$          28& --&          113$\pm$          32&           49$\pm$          14\\
&  &       120.50&       119.44&           78$\pm$          23& -408$\pm$116&          118$\pm$          33& --&          160$\pm$          51&           70$\pm$          20\\
& $\frac{1}{2}$,$\frac{3}{2}$-$\frac{1}{2}$,$\frac{1}{2}$ &       270.20&       163.12&           23$\pm$           9& --&           38$\pm$          11& --& --&           25$\pm$           8\\
&  &       269.80&       163.40&           24$\pm$          10&          104$\pm$          45& --& --& --&           22$\pm$          10\\

\\
p-H$_2$O& 4$_{31}$--3$_{22}$&       552.30&       56.33&           94$\pm$          33& --&           73$\pm$          24& --& --& --\\
& 4$_{22}$--3$_{13}$&       454.30&       57.64& --& --&          106$\pm$          32& --& --& --\\
& 3$_{31}$--2$_{20}$&       410.40&       67.09&           30$\pm$          11& --&           70$\pm$          29& --& --& --\\
& 5$_{24}$--4$_{13}$&       598.80&       71.07&           53$\pm$          17& --&           75$\pm$          21& --& --& --\\
& 7$_{17}$--6$_{06}$&       843.80&       71.54&           39$\pm$          12& --&          123$\pm$          42& --& --& --\\
& 6$_{15}$--5$_{24}$&       781.10&       78.93& --& --&          111$\pm$          37& --& --&           18$\pm$           6\\
& 6$_{06}$--5$_{15}$&       642.70&       83.28& --& --&          123$\pm$          36& --& --&           17$\pm$           7\\
&3$_{22}$--2$_{11}$&       296.80&       89.99&           36$\pm$          14& --&          173$\pm$          55& --& --&           24$\pm$           7\\
& 4$_{04}$--3$_{13}$&       319.50&       125.35&           54$\pm$          17& --&           69$\pm$          20& --&           49$\pm$          17&           27$\pm$           8\\
& 3$_{31}$--3$_{22}$&       410.40&       126.71& --& --&            7$\pm$           9& --& --& --\\
& 3$_{13}$--2$_{02}$&       204.70&       138.53&           78$\pm$          24& --&           85$\pm$          25& --&           59$\pm$          20&           23$\pm$           7\\
& 4$_{13}$--3$_{22}$&       396.40&       144.52&           27$\pm$          11& --&           48$\pm$          19& --& --& --\\
& 3$_{22}$--3$_{13}$&       296.80&       156.19& --& --&          102$\pm$          36& --& --& --\\
& 3$_{31}$--4$_{04}$&       410.40&       158.31& --& --& --&           20$\pm$           7& --& --\\

\\

o-H$_2$O& 10$_{29}$--10$_{110}$&       1861.30&       55.84&           48$\pm$          16& --& --& --& --& --\\
& 4$_{32}$--3$_{21}$&       550.40&       58.70&           56$\pm$          20& --&          168$\pm$          49& --& --&           20$\pm$           6\\
& 7$_{16}$--6$_{25}$&       1013.20&       66.09&           57$\pm$          19& --&           84$\pm$          31& --& --& --\\
& 3$_{30}$--2$_{21}$&       410.70&       66.44&           62$\pm$          23& --&          155$\pm$          52& --& --&           20$\pm$           6\\
& 3$_{30}$--3$_{03}$&       410.70&       67.27& --& --&           81$\pm$          28& --& --& --\\
& 7$_{07}$--6$_{16}$&       843.50&       71.95&           65$\pm$          22& --&          179$\pm$          53& --& --&           22$\pm$           6\\
&3$_{21}$--2$_{12}$ &       305.30&       75.38&           89$\pm$          25& --&          211$\pm$          74& --& --&           54$\pm$          18\\
& 4$_{23}$--3$_{12}$&       432.20&       78.74&           83$\pm$          25& --&          211$\pm$          78& --& --&           36$\pm$          12\\
& 6$_{16}$--5$_{05}$&       643.50&       82.03&           77$\pm$          31& --&          174$\pm$          58& --&           40$\pm$          14&           41$\pm$          13\\
& 2$_{21}$--1$_{10}$&       194.10&       108.07&           92$\pm$          26& --&          148$\pm$          47& --&           67$\pm$          20&           40$\pm$          12\\
& 7$_{34}$--6$_{43}$&       1212.00&       116.78& --& --&           31$\pm$          28& --& --& --\\
& 4$_{32}$--4$_{23}$&       550.40&       121.72& --& --&           41$\pm$          14& --& --& --\\
& 4$_{23}$--4$_{14}$&       432.20&       132.41&           30$\pm$           9& --&           54$\pm$          16& --&           54$\pm$          19& --\\
& 5$_{14}$--5$_{05}$&       574.70&       134.94& --& --&           31$\pm$          11& --& --& --\\
& 3$_{30}$--3$_{21}$&       410.70&       136.50&           25$\pm$           9& --&           34$\pm$          15& --& --& --\\
& 3$_{03}$--2$_{12}$&       196.80&       174.63&           76$\pm$          25&           27$\pm$          11&          102$\pm$          30&           28$\pm$          12&           79$\pm$          24&           40$\pm$          16\\
& 2$_{12}$--1$_{01}$&       114.40&       179.53&          103$\pm$          30& --&           79$\pm$          25& --&          117$\pm$          35&           38$\pm$          15\\
& 2$_{21}$--2$_{12}$&       194.10&       180.49& --& --&           31$\pm$           8& --& --& --\\

\\

 $[O I]$&$^3$P$_1$--$^3$P$_2$	&       227.71&       63.18&          462$\pm$         132&         7147$\pm$        2051&          784$\pm$         237&          661$\pm$         192&          900$\pm$         266&         1049$\pm$         309\\
 $[O I]$&$^3$P$_0$--$^3$P$_1$&       326.58&       145.53&           28$\pm$          11&          488$\pm$         143&           48$\pm$          14&           43$\pm$          14&           82$\pm$          30&          123$\pm$          42\\
$[C II]$&$^2$P$_\frac{3}{2}$--$^2$P$_\frac{1}{2}$ &       91.21&       157.741& --&          545$\pm$         161&           22$\pm$           7&           32$\pm$           9&           93$\pm$          35&          119$\pm$          35\\

\enddata

\tablenotetext{a}{Unresolved doublet lines. Fluxes were divided by 2.}
\label{fluxtable}
\end{deluxetable}

\begin{deluxetable}{llcccccc}
\tablewidth{0pt}
\tabletypesize{\footnotesize}
\tablecaption{Luminosities}
\scriptsize
\tablehead{
\colhead{ } & \colhead{Species} & \colhead{L1489}	& \colhead{L1551-IRS5}	 &
\colhead{TMR1}	& \colhead{TMC1-A}   &\colhead{L1527}   &\colhead{TMC1}   }
\startdata
	 				  &CO		&7.9$\pm$2.4&16.3$\pm$5.0 &15.7$\pm$4.9	&3.8$\pm$1.3	&4.4$\pm$1.6
&6.6$\pm$2.0\\
Lines				  &OH		&4.4$\pm$1.4&1.4$\pm$0.5  &8.4$\pm$2.7	&1.2$\pm$0.4
&3.1$\pm$1.0  &3.9$\pm$1.2\\
(10$^{-4}$ \it L$_{\sun}$)&H$_2$O	&7.7$\pm$1.9&0.2$\pm$0.07 &18.0$\pm$4.4
&0.3$\pm$0.08    &2.8$\pm$0.7  &2.6$\pm$0.7 	\\
					  &$[$O I$]$&3.0$\pm$0.9&45.9$\pm$13.2&5.0$\pm$1.5  &4.2$\pm$1.3
&5.9$\pm$1.8  &7.1$\pm$2.1\\
					  &$[$C II$]$&		 -- &3.3$\pm$1.0  &0.1$\pm$0.04 &0.2$\pm$0.05	&0.6$\pm$0.2
 &	0.7$\pm$0.2	\\
\hline
Total FIR Line\tablenotemark{a} ($\it L_{\rm line}$) ($10^{-4}$ \it L$_{\sun}$)& &	23.0		&	67.0	&	47.3	&	9.8		&	16.8	&	20.8 \\
\hline
FIR Continuum\tablenotemark{b} ($\it L_{\rm cont}$) (\it L$_{\sun}$)& &	1.1		&	10.4	&	0.9	&	1.0		&	1.1	&	0.2 
\enddata
\tablenotetext{a}{$\it L_{\rm line}=\it L_{\rm CO}+\it L_{\rm OH}+\it L_{\rm H_2O}+\it L_{\rm[OI]}+\it L_{\rm[CII]} $}
\tablenotetext{b}{$\it L_{\rm cont}=\rm{4\pi}\it d^2 \int_{\rm PACS}\it f_{\nu}\,\it d\nu, d=$140 pc. 
We integrated the flux density from 55 \um\ to 190 \um\ for $L_{\rm cont}$.
}
\label{lumtable}
\end{deluxetable}

\clearpage

\begin{deluxetable}{llcc}
\tablewidth{0pt}
\tablecaption{Results of the CO rotation diagram}
\tablehead{
\colhead{ Source } &
\colhead{Component} &
\colhead{$\it T_{\rm rot}$ (K)} &
\colhead{$\mathcal N\rm(CO)$ ( $\times$10$^{47}$)}}
\startdata
L1489     & warm                  &    372$\pm$34        &  18.6$\pm$5.2     \\
              & hot                     &    754$\pm$446  &    6.2$\pm$14.7    \\
L1551-IRS5     & warm                  &    394$\pm$59        &  42.0$\pm$16.0    \\
              & hot                    &   2198$\pm$8274      &  458.9$\pm$1860.5    \\
TMR1     & warm                  &    351$\pm$30        &  36.3$\pm$10.0     \\
              & hot                      &    1117$\pm$371       &    14.8$\pm$13.0    \\
TMC1-A     & warm                  &    442$\pm$53      &  8.5$\pm$2.6    \\
L1527     & warm                  &    361$\pm$60       &  13.8$\pm$6.4      \\
TMC1     & warm                  &    363$\pm$32       &  15.6$\pm$4.4      \\
              & hot                      &    777$\pm$197      &    5.0$\pm$4.8    
\enddata
\label{rotco}
\end{deluxetable}

\begin{deluxetable}{llcc}
\tablewidth{0pt}
\tablecaption{Results of the OH rotation diagram}
\tablehead{
\colhead{ Source } &
\colhead{Component} &
\colhead{$\it T_{\rm rot}$ (K)} &
\colhead{$\mathcal N\rm(OH)$ $(\times$10$^{44}$)}}
\startdata
L1489    &$^2\Pi_{3/2}$-$^2\Pi_{3/2}$         &    132$\pm$13     & 8.7$\pm$2.4  \\
    &$^2\Pi_{1/2}$-$^2\Pi_{1/2}$         &    96$\pm$10    & 7.5$\pm$4.3    \\
TMR1    &$^2\Pi_{3/2}$-$^2\Pi_{3/2}$         &    148$\pm$17     & 13.1$\pm$3.4  \\
    &$^2\Pi_{1/2}$-$^2\Pi_{1/2}$         &    110$\pm$14    & 12.1$\pm$7.0    \\
TMC1    &$^2\Pi_{3/2}$-$^2\Pi_{3/2}$         &    136$\pm$14     & 7.4$\pm$2.0 \\
    &$^2\Pi_{1/2}$-$^2\Pi_{1/2}$         &    111$\pm$7    & 7.7$\pm$3.0     
\enddata
\label{rotoh}
\end{deluxetable}

\begin{deluxetable}{llcc}
\tablewidth{0pt}
\tablecaption{Results of the H$_2$O rotation diagram}
\tablehead{
\colhead{ Source } &
\colhead{Component} &
\colhead{$\it T_{\rm rot}$ (K)} &
\colhead{$\mathcal N\rm(H_2$O) ( $\times$10$^{44}$)}}
\startdata
L1489    & ortho                    &    328$\pm$22    & 7.4$\pm$1.1 \\
    & para                    &    155$\pm$14    & 5.0$\pm$1.5 \\
TMR1    & ortho                    &    191$\pm$12    &17.9$\pm$3.1  \\
    & para                    &    177$\pm$15    & 9.2$\pm$2.3 \\
L1527    & ortho                    &    119$\pm$11    &16.6$\pm$4.4 \\
    & para                    &    131$\pm$75    & 7.0$\pm$8.1 \\
TMC1    & ortho                    &    140$\pm$10    &2.8$\pm$0.7 \\
    & para                    &    210$\pm$31    & 2.2$\pm$0.7 
\enddata
\label{roth2o}
\end{deluxetable}

\begin{deluxetable}{llcccccc}
\tablewidth{0pt}
\tabletypesize{\footnotesize}
\tablecaption{The best-fit LVG models of CO}
\scriptsize
\tablehead{
\colhead{ } 		& \colhead{ } 		&
\colhead{L1489} 		& \colhead{L1551-IRS5} 			& \colhead{TMR1}	&
\colhead{TMC1-A} 	&
\colhead{L1527}		& \colhead{TMC1}   }
\startdata

&$\it T$ (K)					&	5.0(3)\tablenotemark{a}\tablenotemark{b}		&	5.0(3)\tablenotemark{b}&	5.0(3)\tablenotemark{b}&	4.0(3)&	5.0(3)\tablenotemark{b}	&	5.0(3)\tablenotemark{b}\\
1 comp.	& $\it n\rm(H_2)$ (cm$^{-3}$)	&	3.2(4)	&	3.2(4)	&
1.8(4)	&	1.0(5)	&	5.6(4)	&	1.0(3)	\\
&$\it N\rm(CO)$ (cm$^{-2}$/km $\rm s^{-1}$)	&	3.2(16) 				&	1.0(15)	&	1.0(15)	&
1.0(15)	&	1.0(15)	&	1.0(15)	\\
&scale factor\tablenotemark{c} & 1.2(0) & 1.0(2) & 1.4(2) & 1.1(1) & 1.8(1) & 1.5(3)\\
& $\chi^2$						&	2.5					&	6.6	&	3.7 	&	5.1	&	4.1	&	3.7	\\
\hline
& $\it n\rm(H_2)$ (cm$^{-3}$) &	1.0(5)				&	3.2(5)	&	3.2(6)	&	5.6(5) &	5.6(5)	&	1.8(6)	\\
power law 			&		$\it p$  			&	1.0						&	1.8		&	2.8		&	1.6		&	2.0		&	2.8 \\
&scale factor& 1.0(2) & 1.3(3) & 1.4(4) & 1.2(2) & 6.5(2) & 7.5(3) \\
		& $\chi^2$				&	2.6					&	5.3	&	2.1	&	4.0 	&	3.1	&	2.4	
\enddata

\tablenotetext{a}{a(b) $\equiv$ a$\times$10$^{b}$}
\tablenotetext{b}{This temperautre is the highest temperature in the collisional rate.}
\tablenotetext{c}{scale factor = AREA(arcsec$^2$) $\times$ FWHM (km/s)} 
\label{lvgco}
\end{deluxetable}

\begin{deluxetable}{llcccc}
\tablewidth{0pt}
\tabletypesize{\footnotesize}
\tablecaption{The best-fit LVG models of H$_2$O}
\scriptsize
\tablehead{
\colhead{ } & \colhead{}                        &  \colhead{ L1489}             & \colhead{TMR1} &  \colhead{L1527}  & \colhead{TMC1}   }
\startdata
& OPR                                          &   1.5$\pm$0.5                & 1.9$\pm$0.6  &   2.4$\pm$2.8  &    1.3$\pm$0.5  \\
\hline
& $\it T$ (K)                                   &       1.0(3)\tablenotemark{a}                    &    5.0(3)        &      2.0(3)         &       1.6(3)    \\
1 comp. & $\it n\rm(H_2)$ (cm$^{-3}$)           &       1.8(7) &    1.0(5)      &      3.2(4)      &       3.2(5)  \\
& $\it N\rm(H_2$O) (cm$^{-2}$/km $\rm s^{-1}$)  &       3.2(16)                 &    1.0(17)     &      3.2(16)     &       3.2(16) \\
&scale factor\tablenotemark{b}& 1.8(-2) & 2.1(-1) & 9.9(-1) & 8.2(-2)\\
& $\chi^2$                                      &         6.4                   &    2.4         &      4.0         &       1.6      \\
\hline
                  & $\it n\rm(H_2)$ (cm$^{-3}$) &       1.0(12)                 &    1.0(9)      &       3.2(7)      &       1.8(8)  \\
power law         &               $\it p$       &       2.0                     &    1.6         &       0.2         &       1.2      \\
 & $\it N\rm(H_2$O) (cm$^{-2}$/km $\rm s^{-1}$) &       1.0(13)                 &    1.0(15)     &       1.0(15)     &       1.0(13)   \\
 &scale factor&1.7(1) & 2.9(-1) & 7.4(-2) & 4.3(0)\\
                & $\chi^2$                      &       8.6                     &    7.2         &       4.9         &       1.1    
\enddata
\tablenotetext{a}{a(b) $\equiv$ a$\times$10$^{b}$}
\tablenotetext{b}{scale factor = AREA(arcsec$^2$) $\times$ FWHM (km/s)}
\label{lvgh2o}
\end{deluxetable}

\begin{deluxetable}{llcccccc}
\tablewidth{0pt}
\tabletypesize{\footnotesize}
\tablecaption{The best-fit LVG models of OH}
\scriptsize
\tablehead{
\colhead{ } & \colhead{ } 	&	\colhead{L1489}		&	\colhead{TMR1}	&
	\colhead{TMC1}	}
\startdata
& $\it T$ (K)							&	150						&	150		&	75		\\
including FIR & $\it n\rm(H_2$) (cm$^{-3}$)	&	2.5(8)\tablenotemark{a}	&	4.0(8)&	6.0(7)	\\
&$\it N\rm(OH)$ (cm$^{-2}$/km $\rm s^{-1}$)			&	1.0(15)					&	1.8(15)	&	1.8(17)\\
&scale factor\tablenotemark{b}& 1.7(-1) & 2.1(-1) & 3.1(-1) \\
& $\chi^2$								&	1.0					&	0.6	&	1.2	\\
\hline
& $\it T$ (K)							&	175						&	175		&	75			\\
not incl. FIR	&$\it n\rm (H_2$) (cm$^{-3}$)	&	2.2(8)					&	3.7(8)	&	6.0(7)	\\
&$\it N\rm(OH)$ (cm$^{-2}$/km $\rm s^{-1}$)			&	1.8(15)					&	1.8(15)	&	3.2(18)	\\
&scale factor&	1.1(-1) & 1.8(-1) & 2.9(-1) \\
& $\chi^2$								&	1.1				&	0.7	&	1.3		
\enddata
\tablenotetext{a}{a(b) $\equiv$ a$\times$10$^{b}$}
\tablenotetext{b}{scale factor = AREA(arcsec$^2$) $\times$ FWHM (km/s)}
\label{lvgoh}
\end{deluxetable}

\begin{deluxetable}{lccccccc}
\tablewidth{0pt}
\tabletypesize{\footnotesize}
\tablecaption{Pearson's correlation coefficients among line luminosites}
\scriptsize
\tablehead{
\colhead{ } 		 & \colhead{$\it L\rm_{CO}$}	 	& \colhead{$\it L\rm_{OH}$}	 &
\colhead{$\it L\rm_{H_2O}$} &
\colhead{$\it L\rm_{[OI]}$} 		& \colhead{$\it L\rm_{[CII]}$\tablenotemark{a}
}
   }
\startdata
{$\it L\rm_{CO}$}  & 								$-$  & +0.91(0.032\tablenotemark{b})	& 	 +0.88(0.049) & 	-0.09(0.89)  & 	-0.40(0.50)	 \\
{$\it L\rm_{OH}$}  & 								    &	 $-$	 &	 +0.98(0.0034) & 	+0.07(0.91)  & 	-0.08(0.90) \\
{$\it L\rm_{H_2O}$} & 								 &   		&		$-$ & 	-0.09(0.89)   &	 -0.12(0.85)   \\
{$\it L\rm_{O I}$} &     								&  			 & 			&		$-$  &	 +0.86(0.062) \\
{$\it L\rm_{C II}$} &									&			&			&			&	$-$		 
\enddata

\tablenotetext{a}{$\rm L$1489 was not detected in \CII.}
\tablenotetext{b}{$p$-value}
\label{correl}
\end{deluxetable}

\clearpage

\appendix
\section{Online Material}

\begin{figure}
\plotone{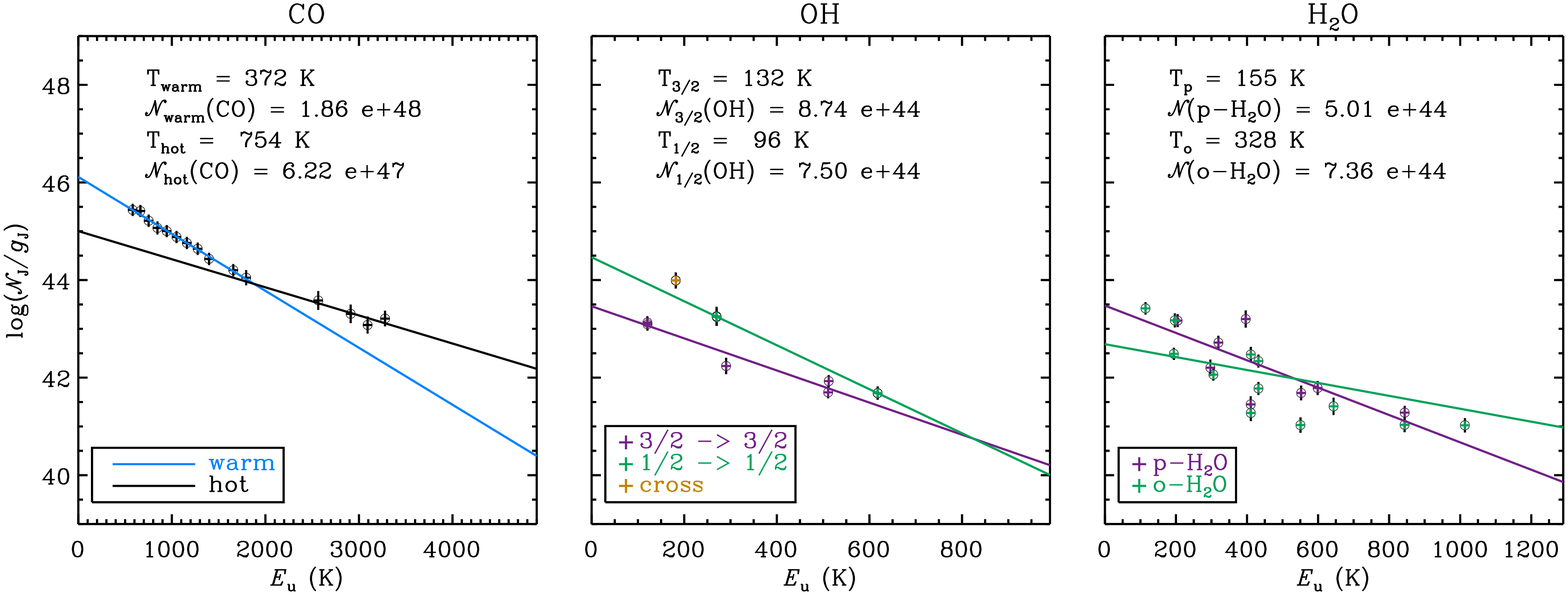}
\caption{The same as Fig.~\ref{rotation} but for L1489}
\label{rotation_l1489}
\end{figure}

\begin{figure}
\plotone{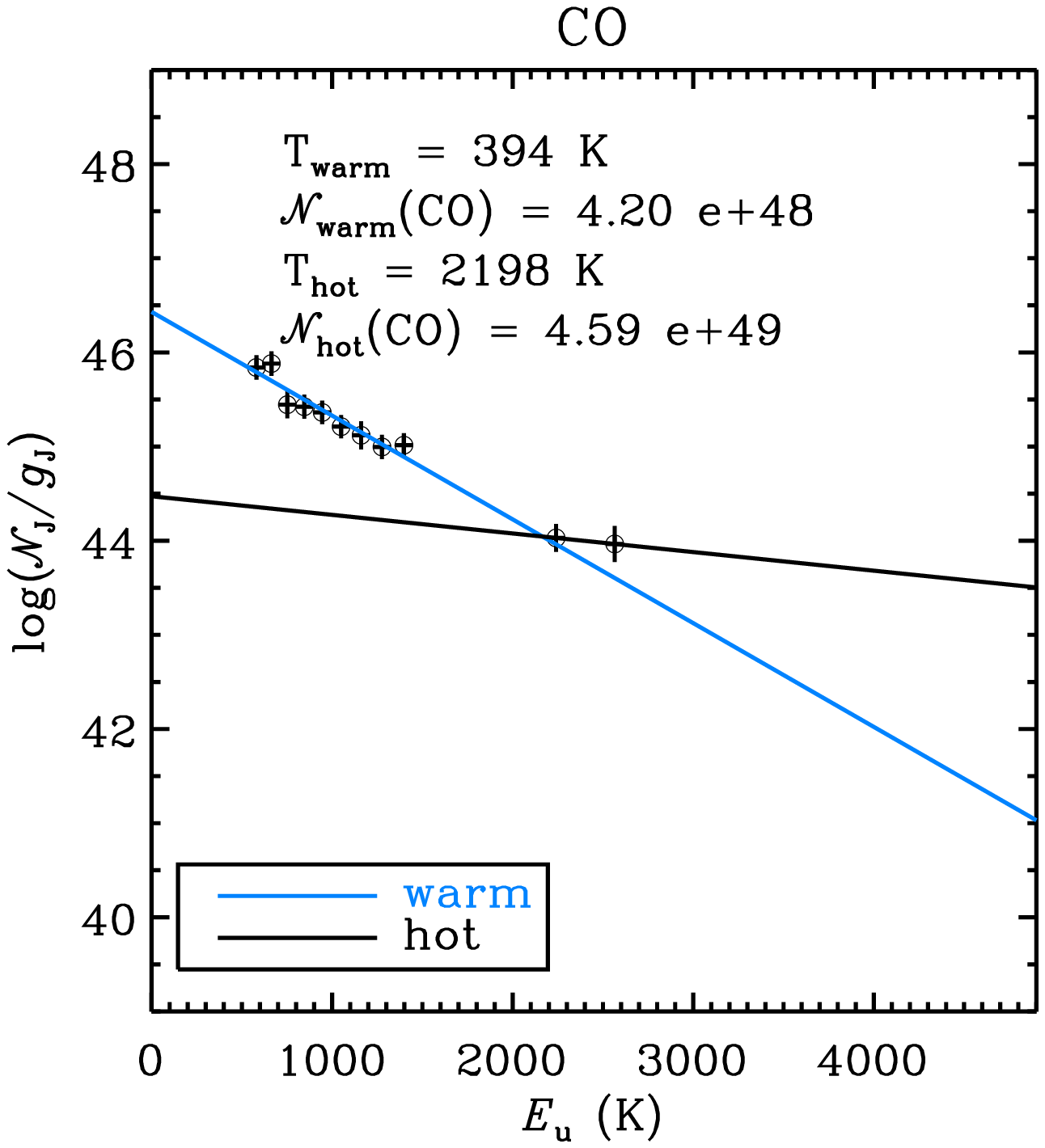}
\caption{The same as Fig.~\ref{rotation} but for L1551-IRS5}
\label{rotation_l1551}
\end{figure}

\begin{figure}
\plotone{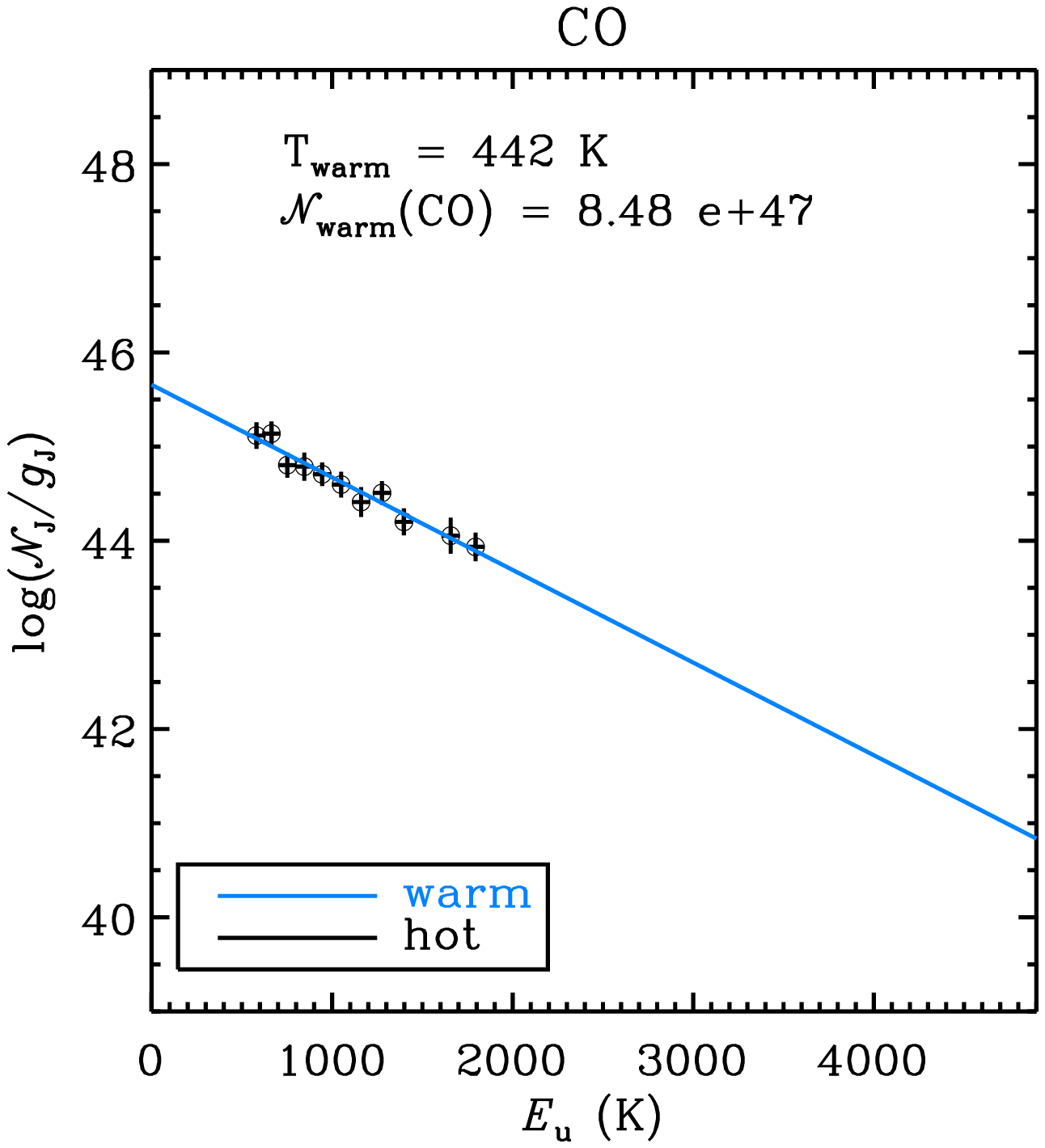}
\caption{The same as Fig.~\ref{rotation} but for TMC1-A}
\label{rotation_tmc1a}
\end{figure}

\begin{figure}
\plotone{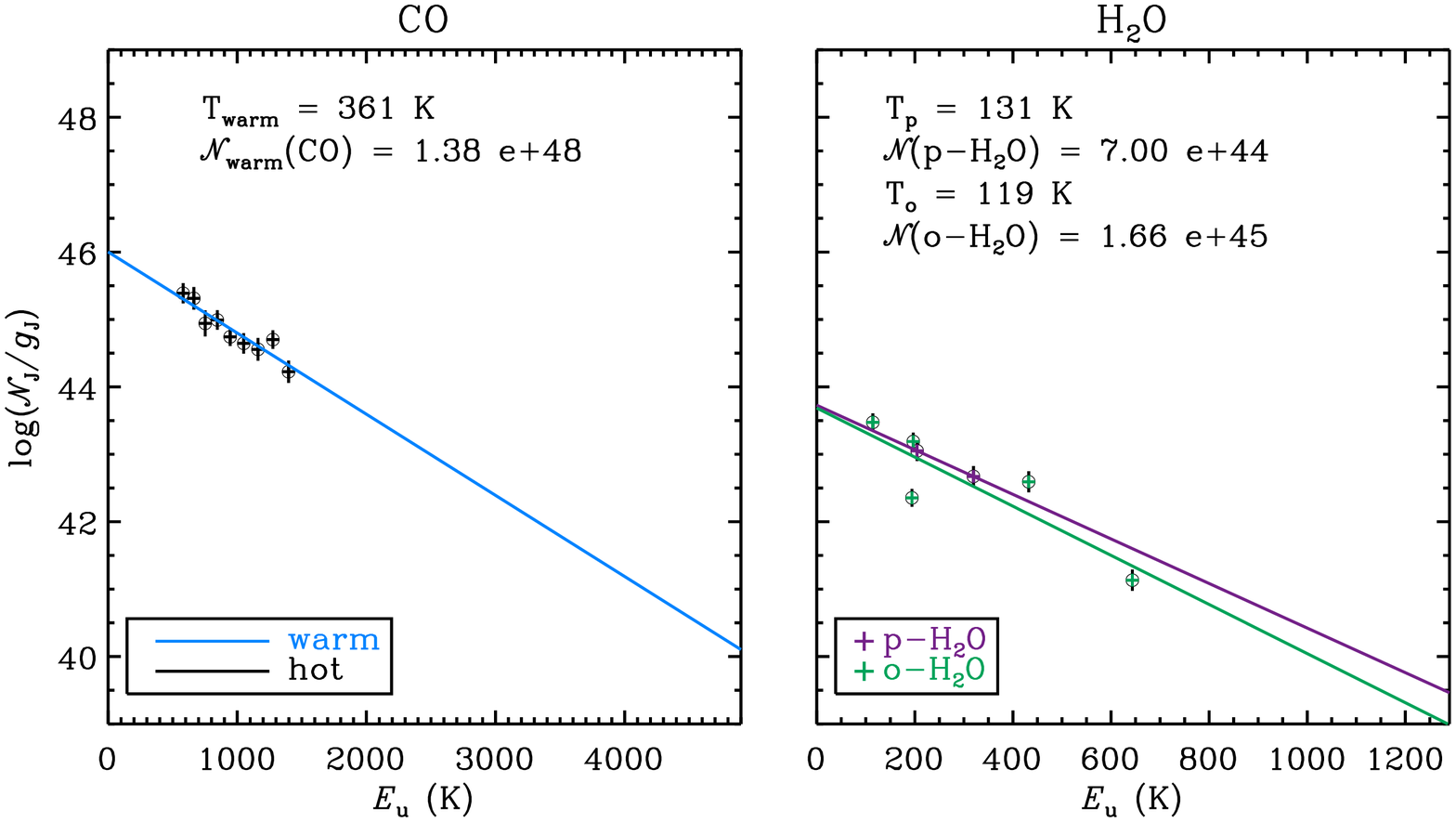}
\caption{The same as Fig.~\ref{rotation} but for L1527}
\label{rotation_l1527}
\end{figure}

\begin{figure}
\plotone{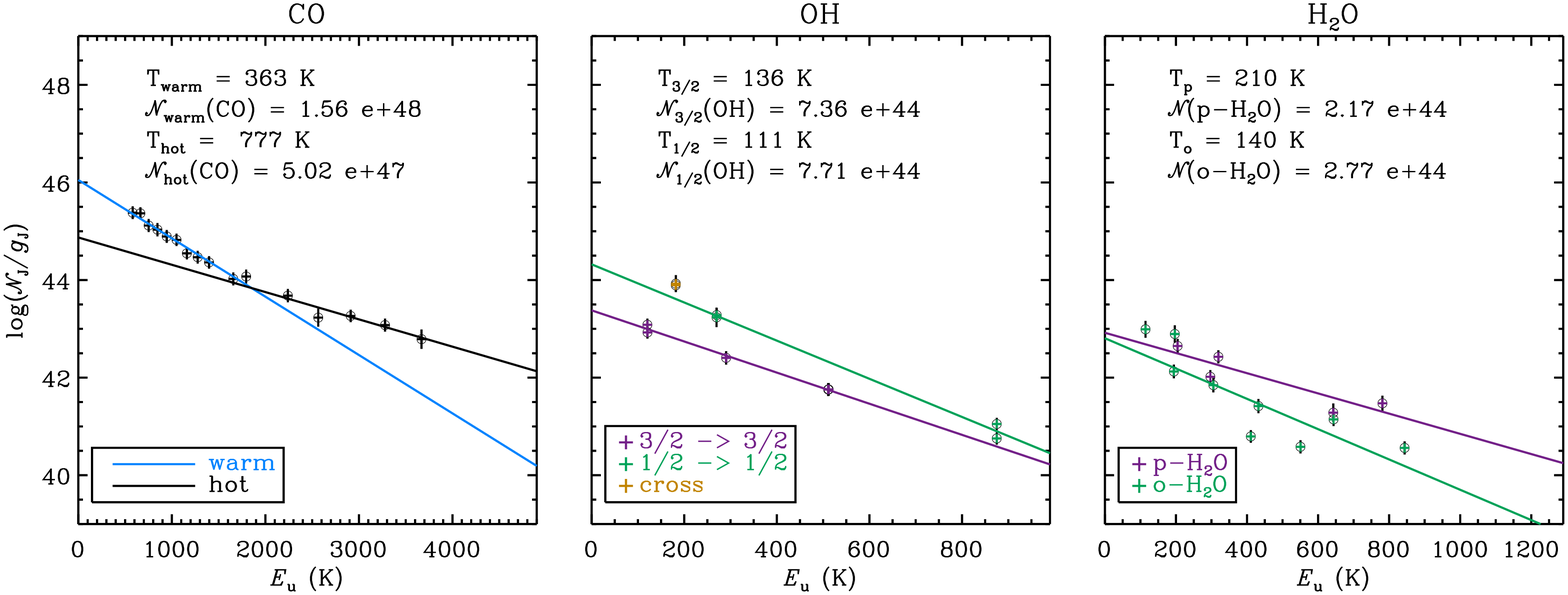}
\caption{The same as Fig.~\ref{rotation} but for TMC1}
\label{rotation_tmc1}
\end{figure}

\begin{figure}
\epsscale{0.8}
\plotone{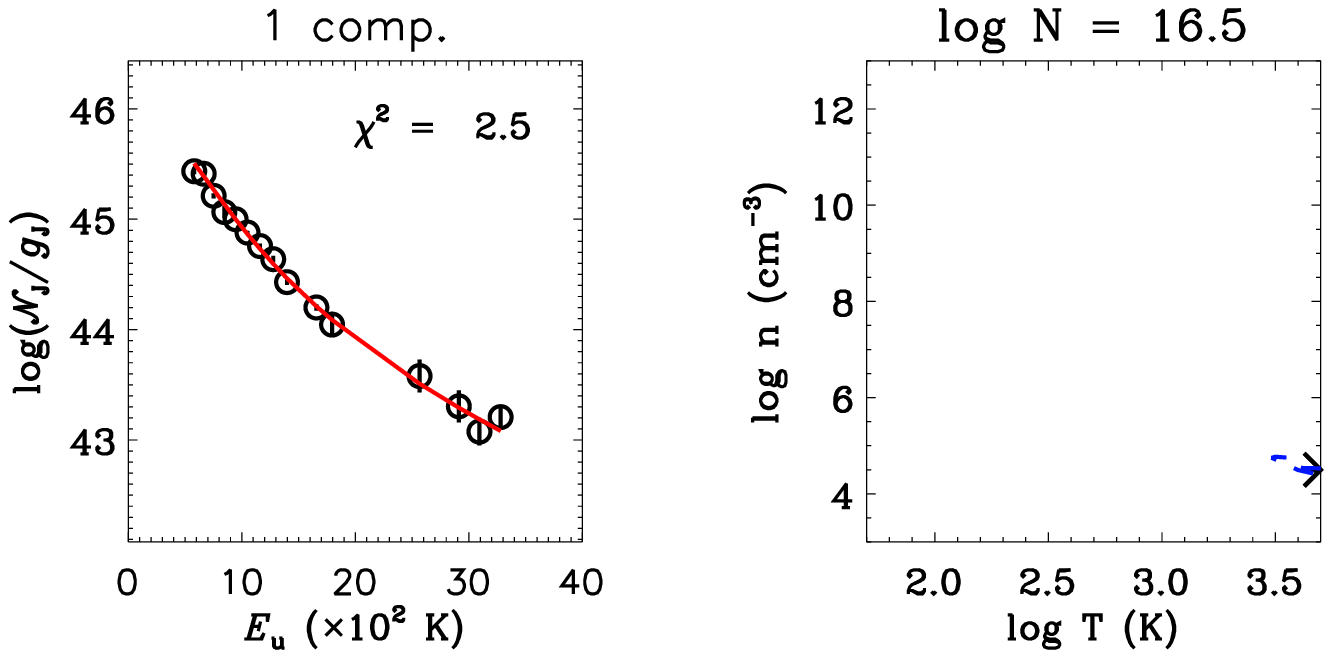}
\plotone{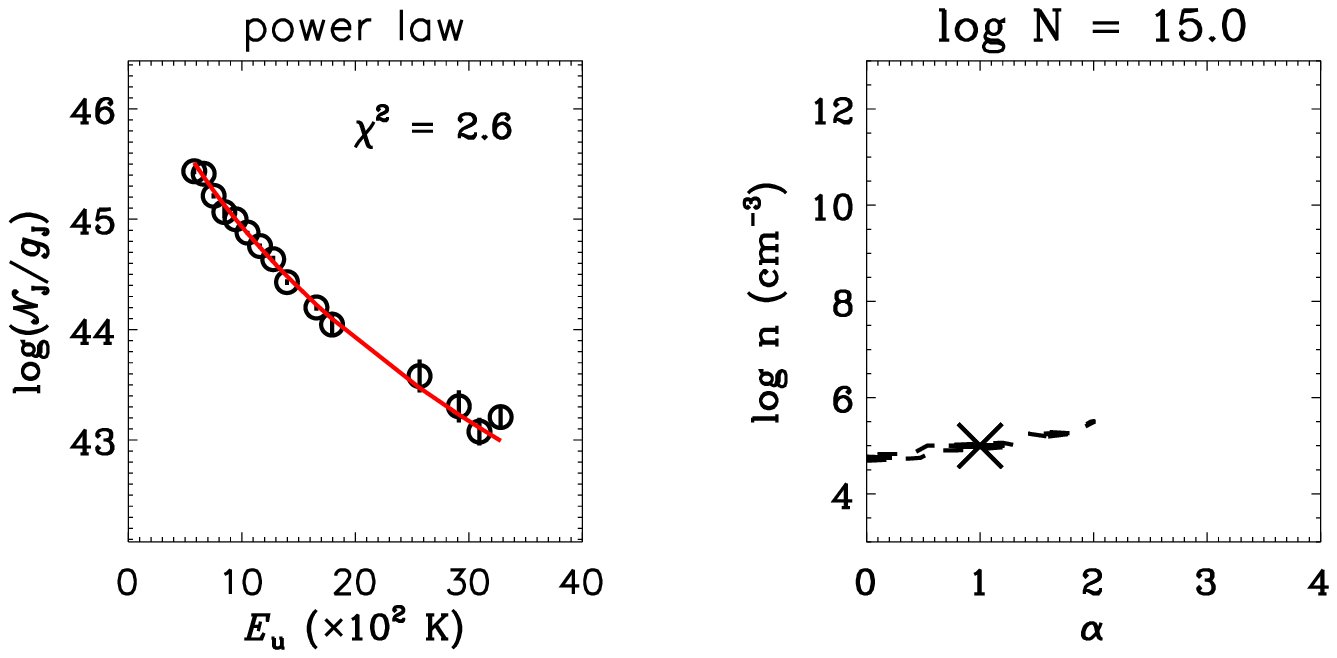}
\caption{The same as Fig.~\ref{tmc1_lvgco} but for L1489.
}
\label{l1489_lvgco}
\end{figure}

\begin{figure}
\epsscale{0.8}
\plotone{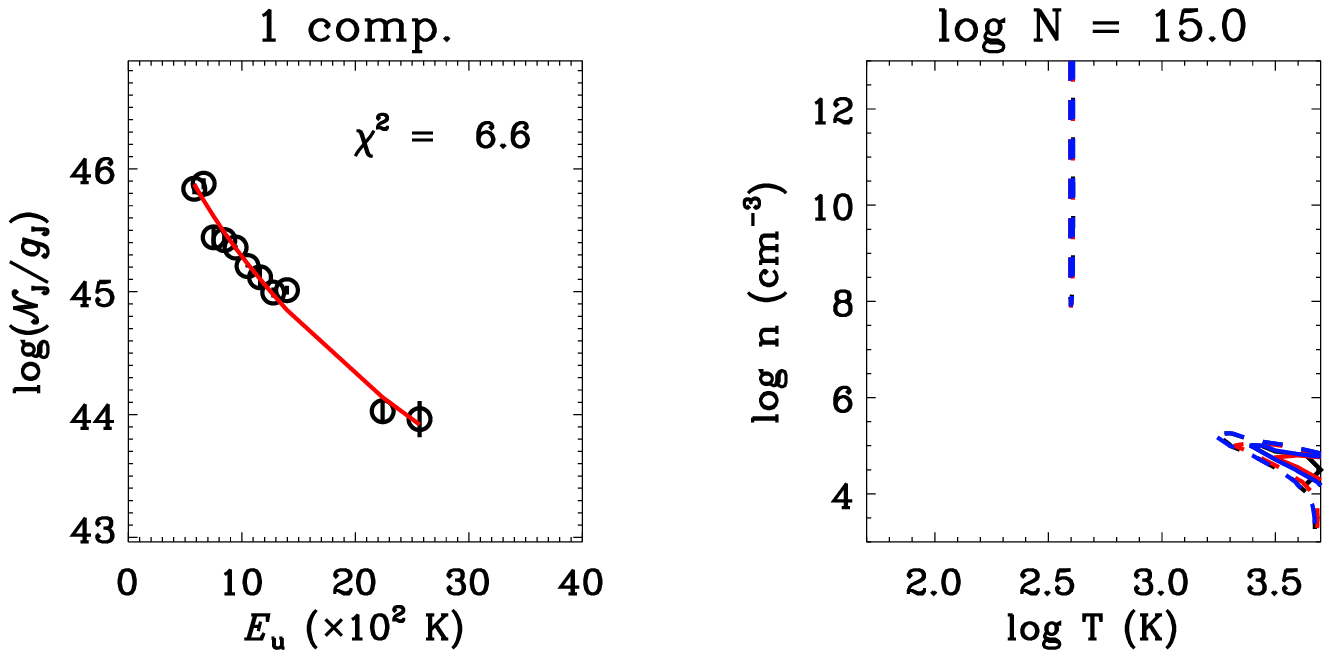}
\plotone{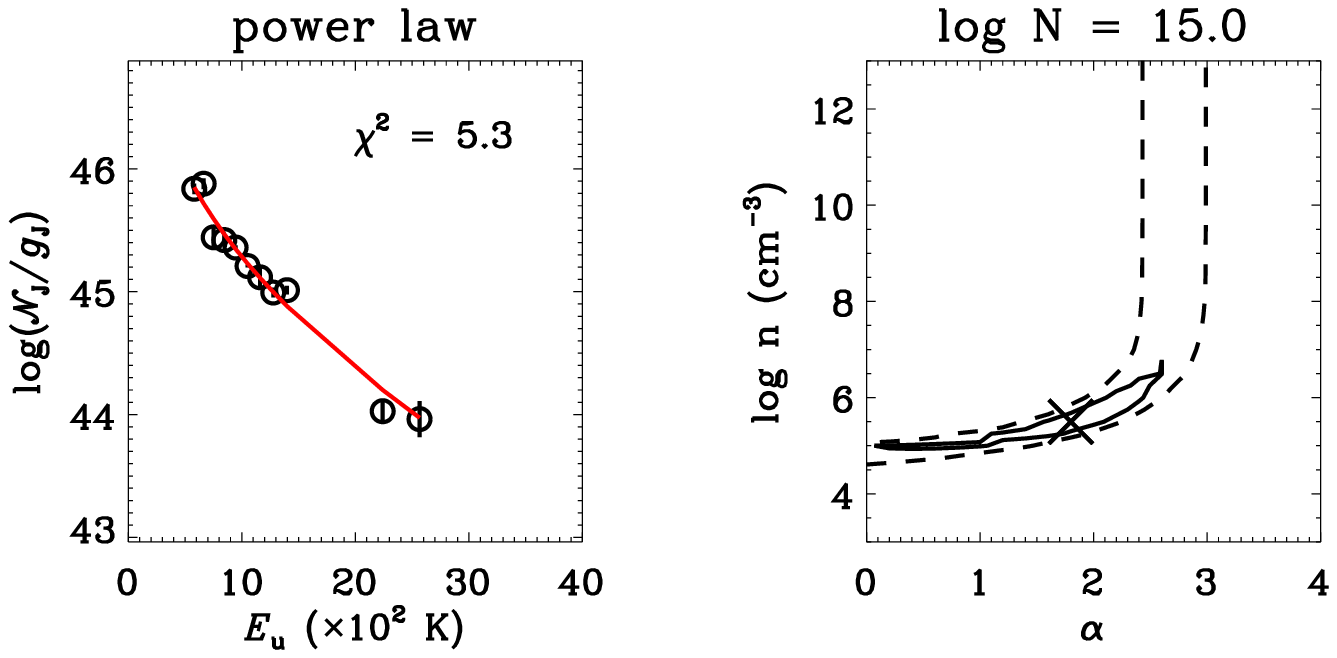}
\caption{The same as Fig.~\ref{tmc1_lvgco} but for L1551-IRS5.
}
\label{l1551_lvgco}
\end{figure}

\begin{figure}
\epsscale{0.8}
\plotone{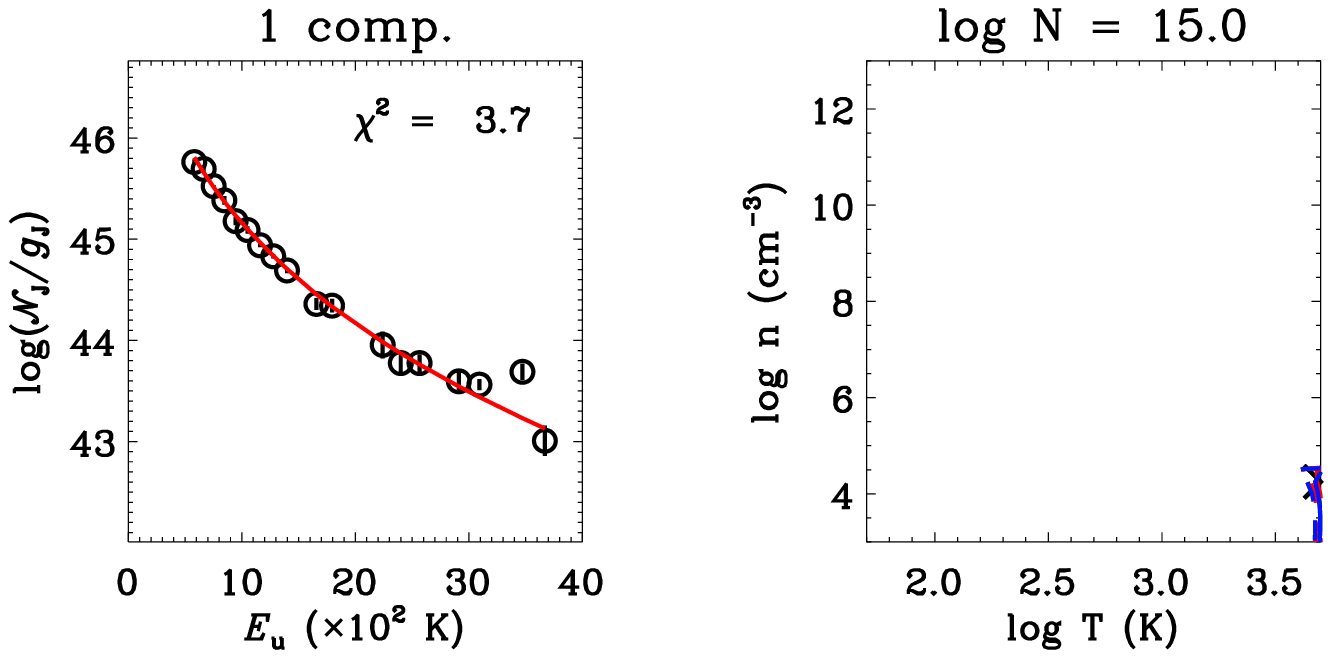}
\plotone{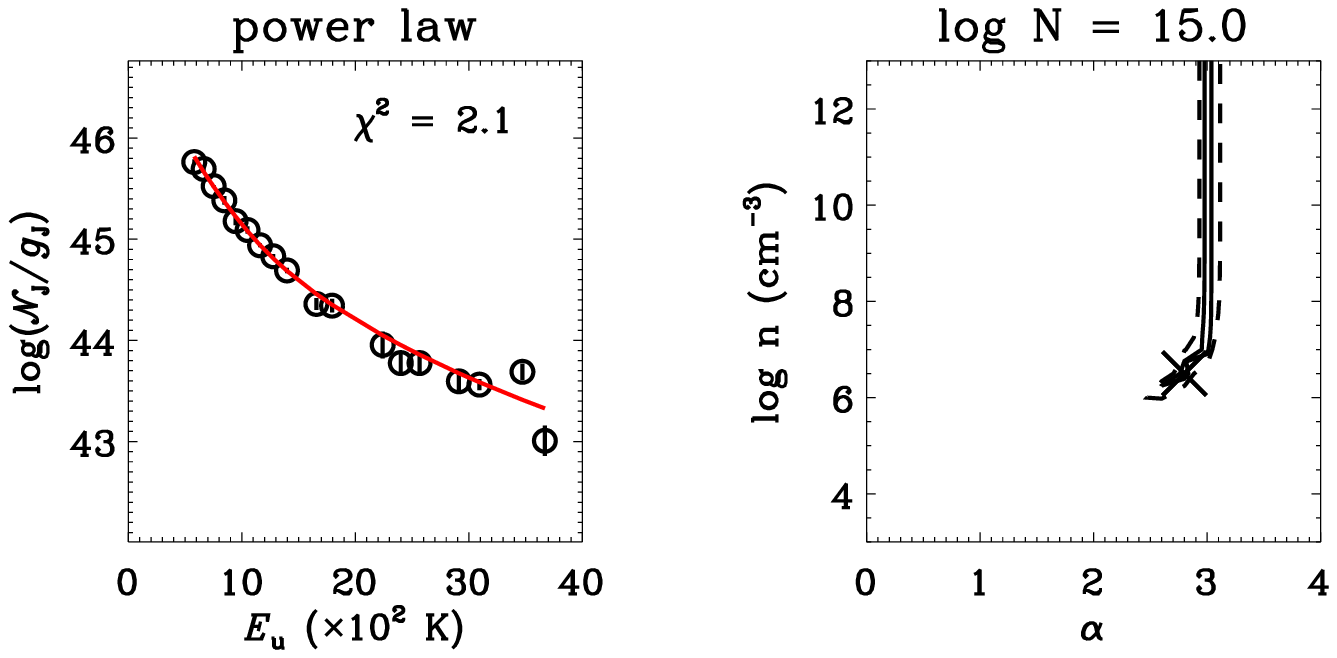}
\caption{The same as Fig.~\ref{tmc1_lvgco} but for TMR1.
}
\label{tmr1_lvgco}
\end{figure}

\begin{figure}
\epsscale{0.8}
\plotone{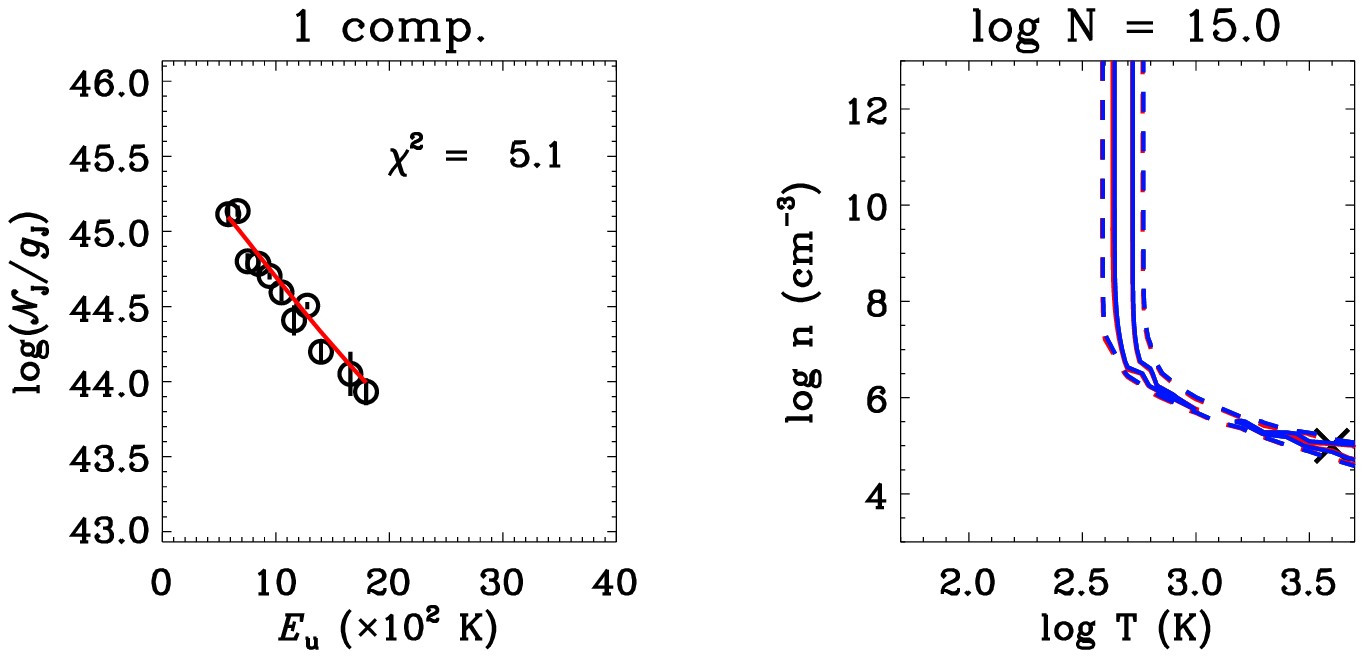}
\plotone{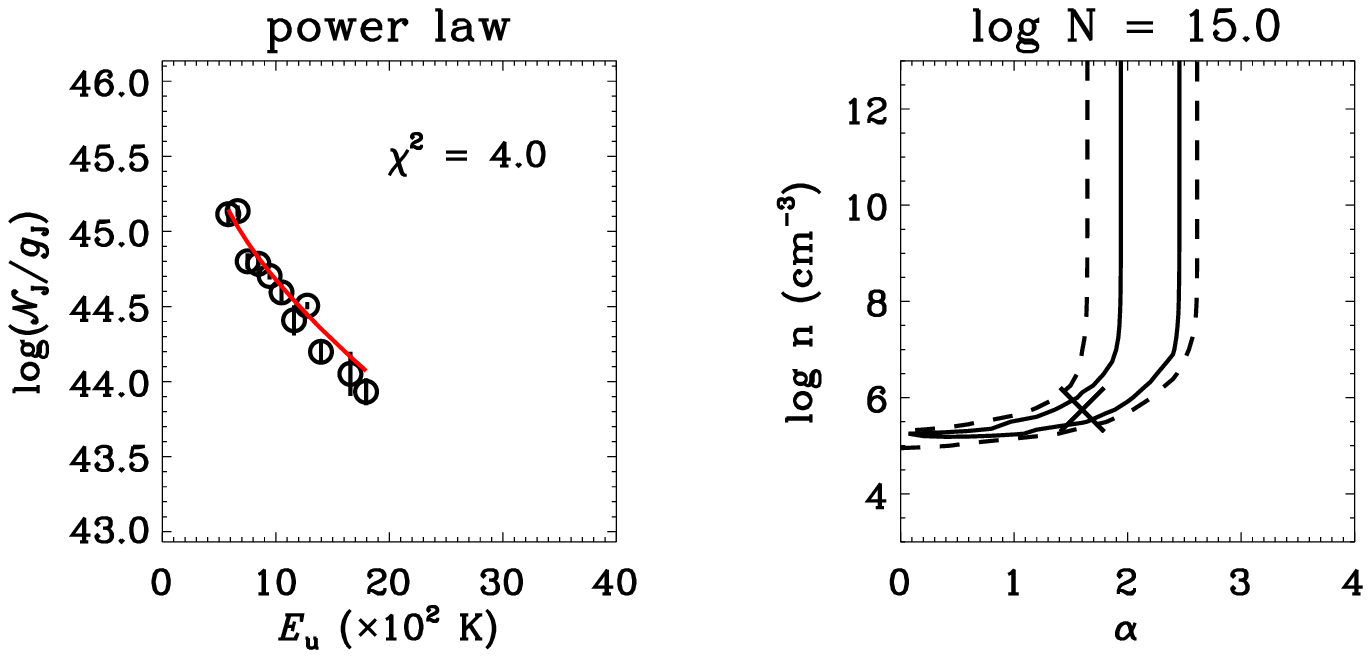}
\caption{The same as Fig.~\ref{tmc1_lvgco} but for TMC1-A.
}
\label{tmc1a_lvgco}
\end{figure}

\begin{figure}
\epsscale{0.8}
\plotone{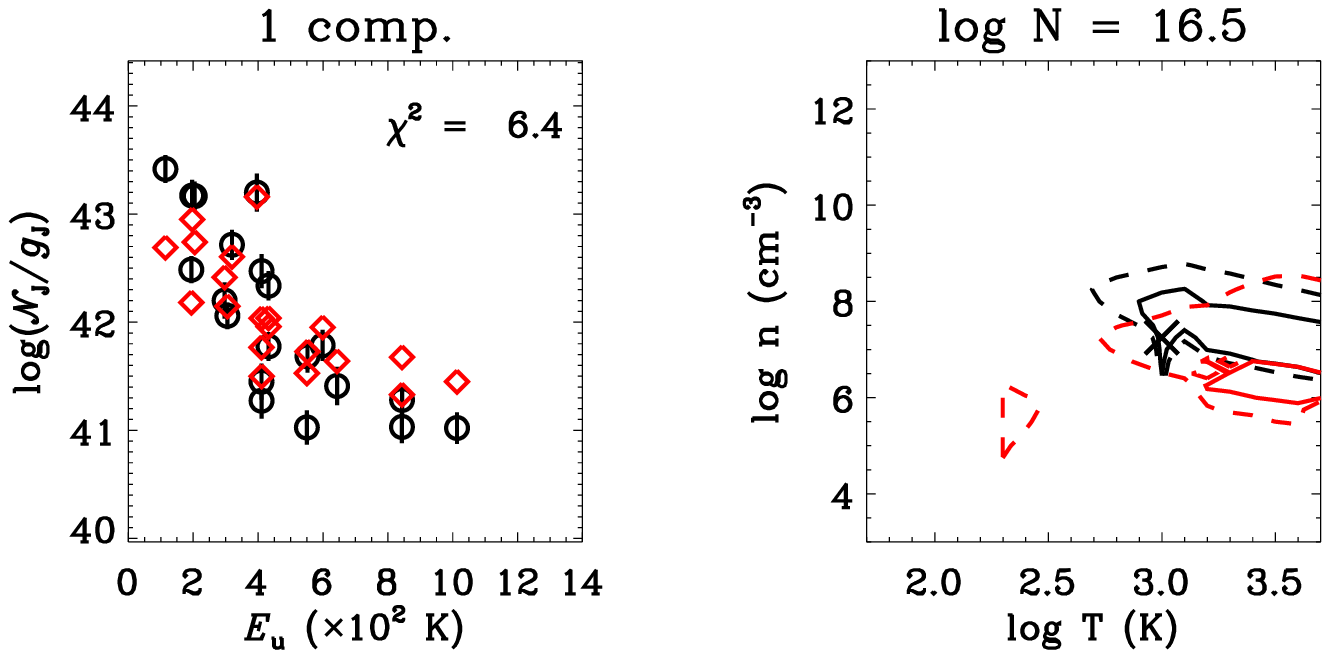}
\plotone{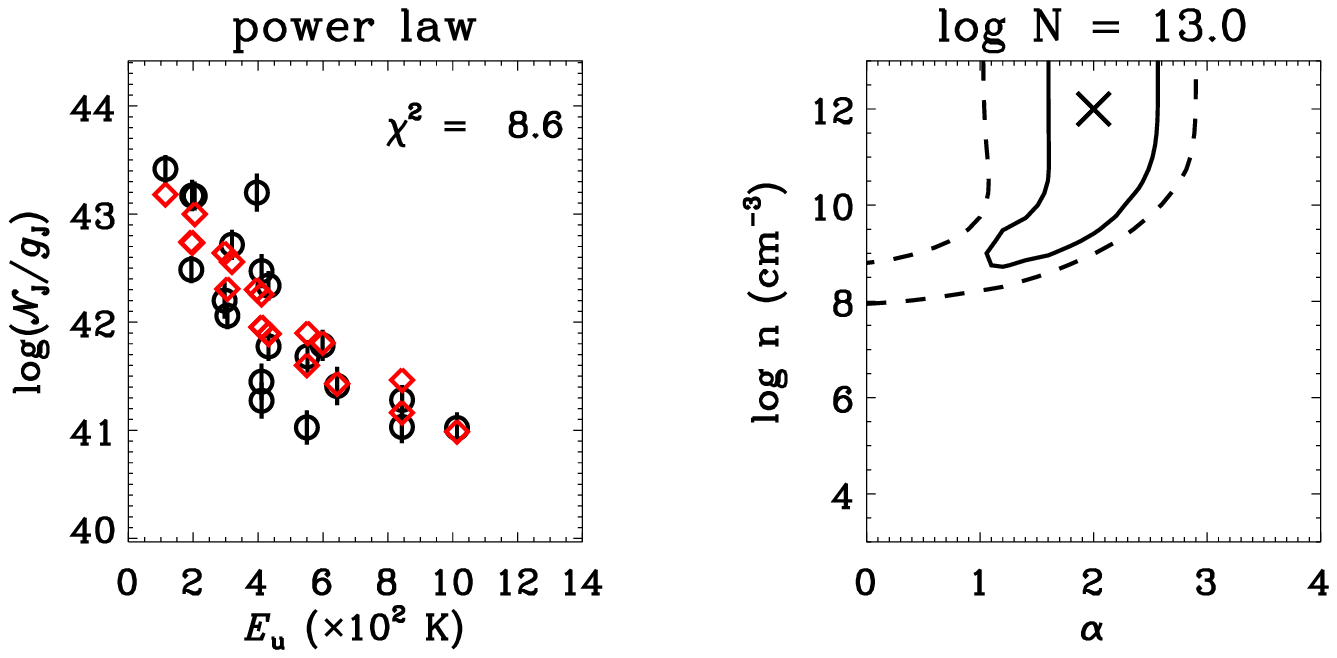}
\caption{The same as Fig.~\ref{fig_lvgh2o} but for L1489.
}
\label{tmc1a_lvgh2o}
\end{figure}

\begin{figure}
\epsscale{0.8}
\plotone{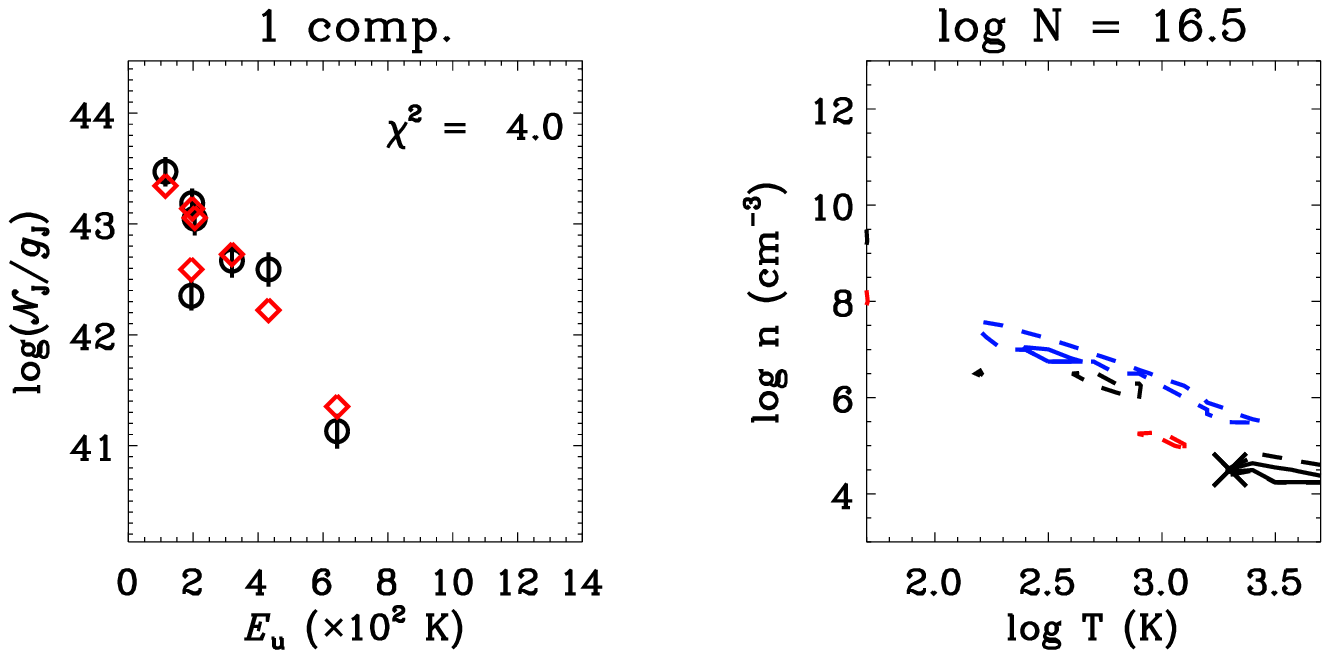}
\plotone{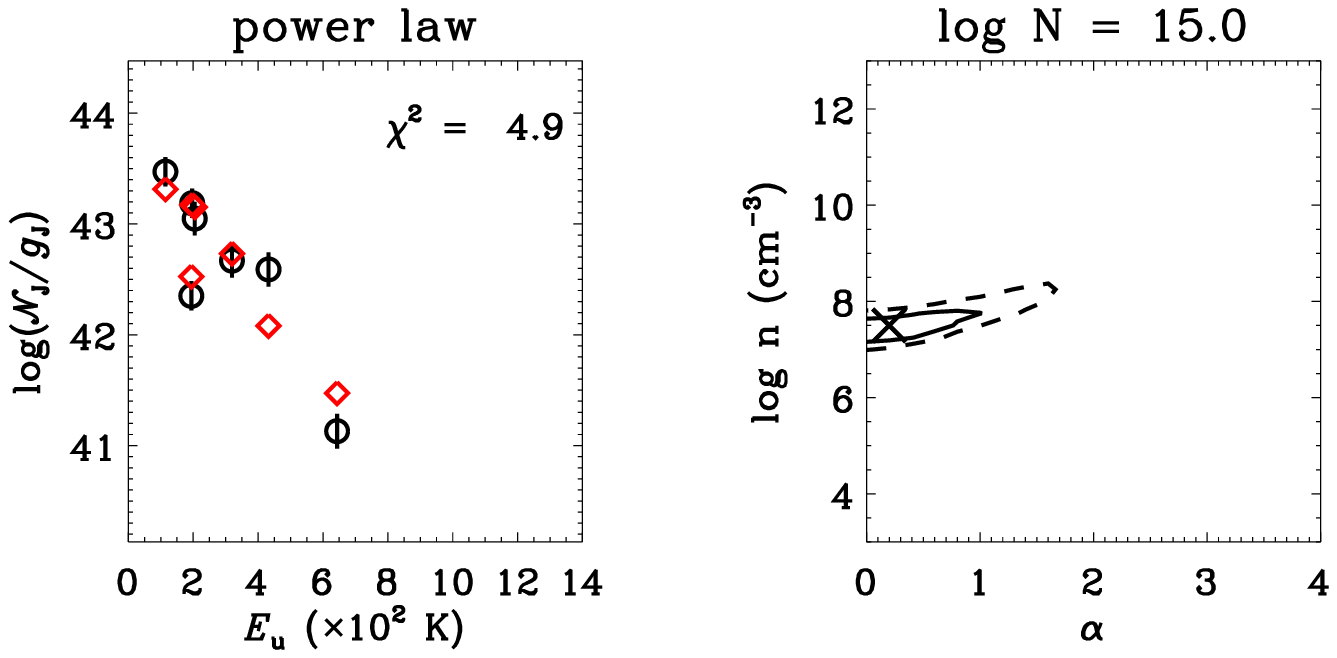}
\caption{The same as Fig.~\ref{fig_lvgh2o} but for L1527.
}
\label{l1527_lvgh2o}
\end{figure}

\begin{figure}
\epsscale{0.8}
\plotone{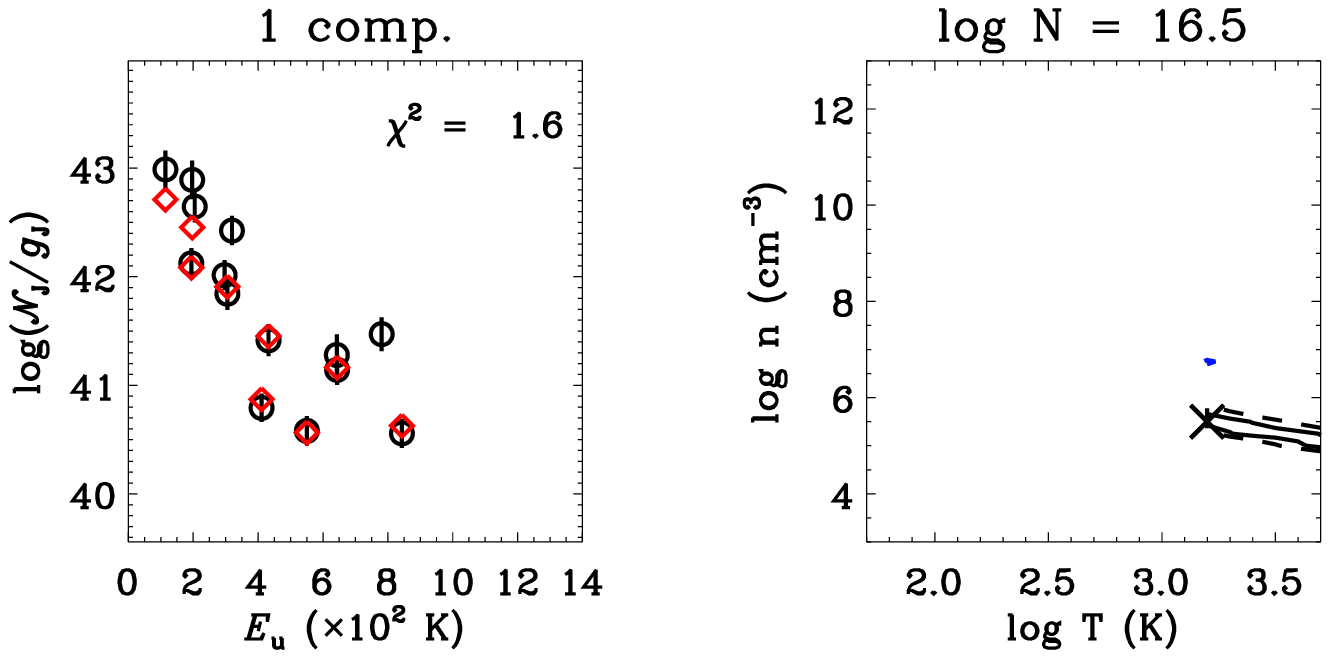}
\plotone{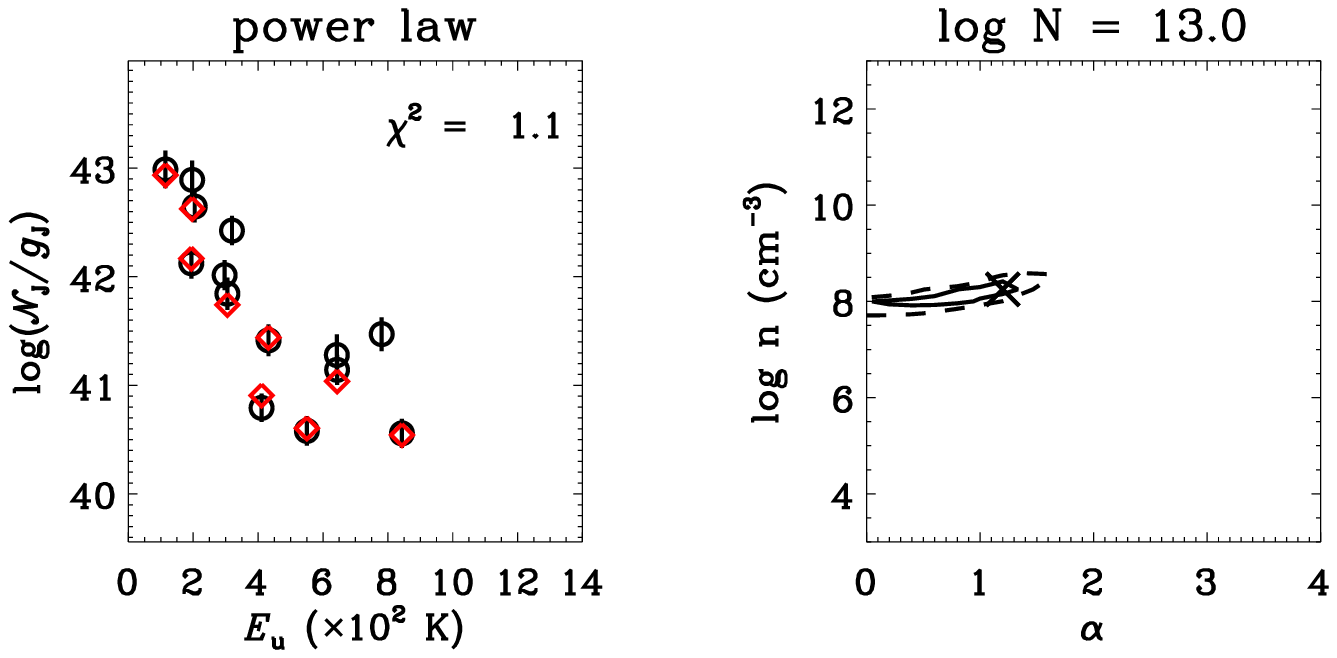}
\caption{The same as Fig.~\ref{fig_lvgh2o} but for TMC1.
}
\label{tmc1_lvgh2o}
\end{figure}

\clearpage

\begin{figure}
\epsscale{1.0}
\plottwo{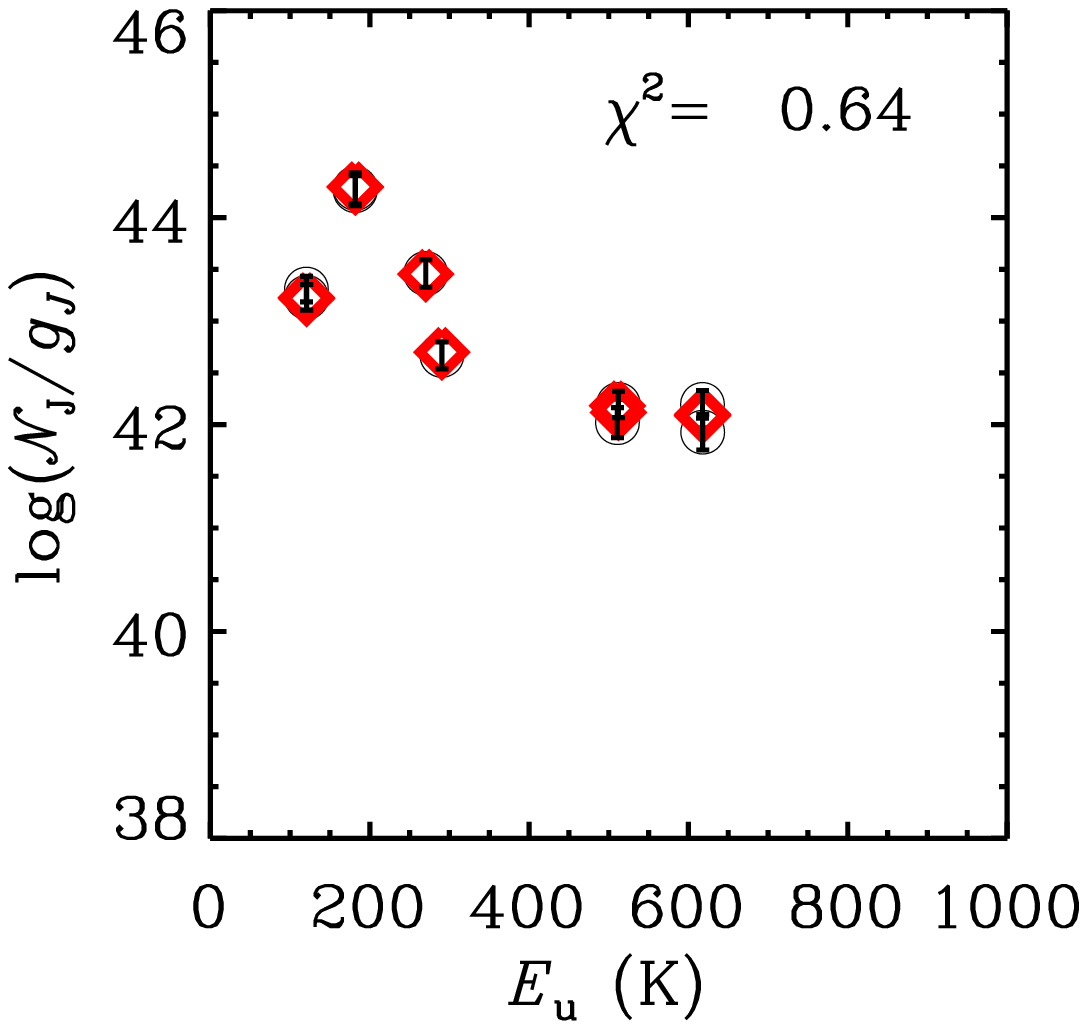}{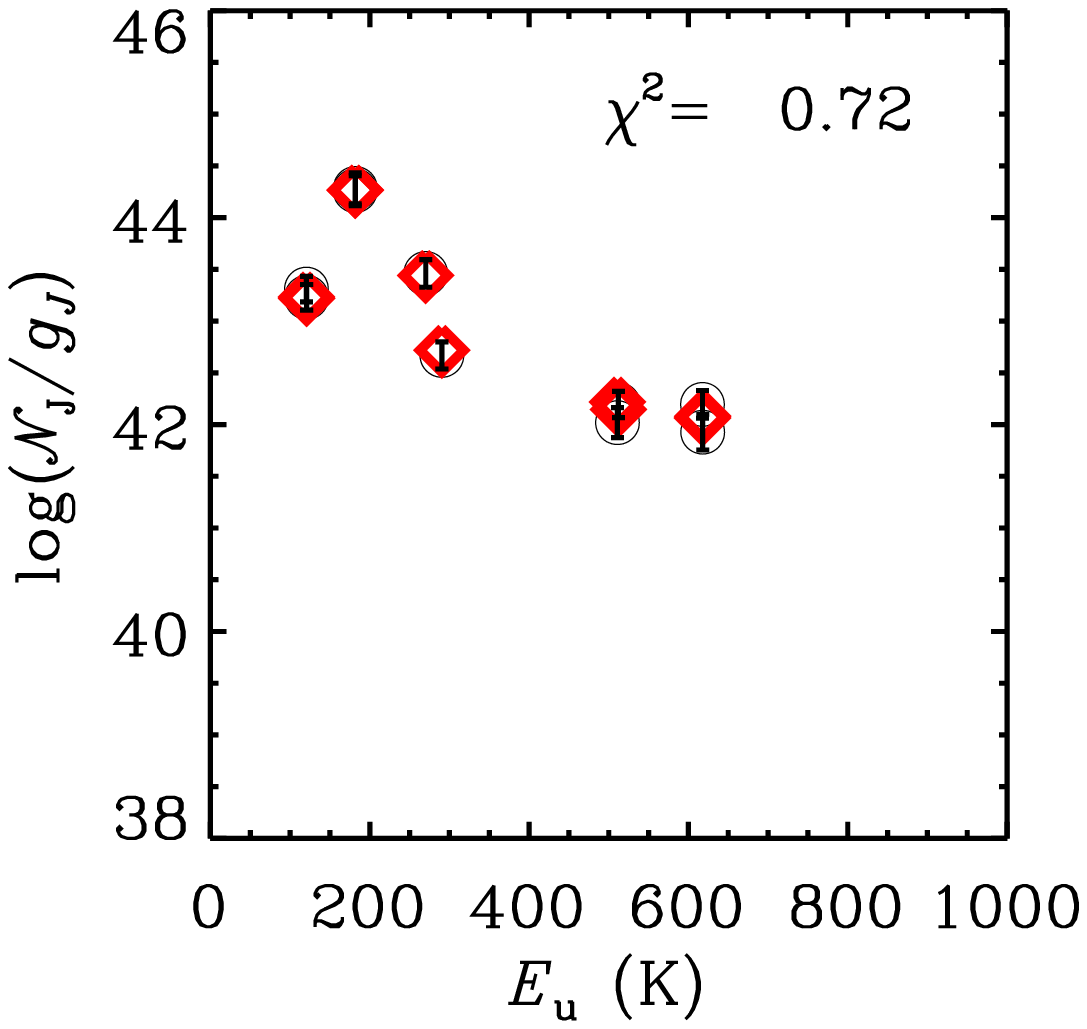}
\caption{The same as Fig.~\ref{tmc1_lvgoh} but for TMR1}
\label{tmr1_lvgoh}
\end{figure}

\begin{figure}
\epsscale{1.0}
\plottwo{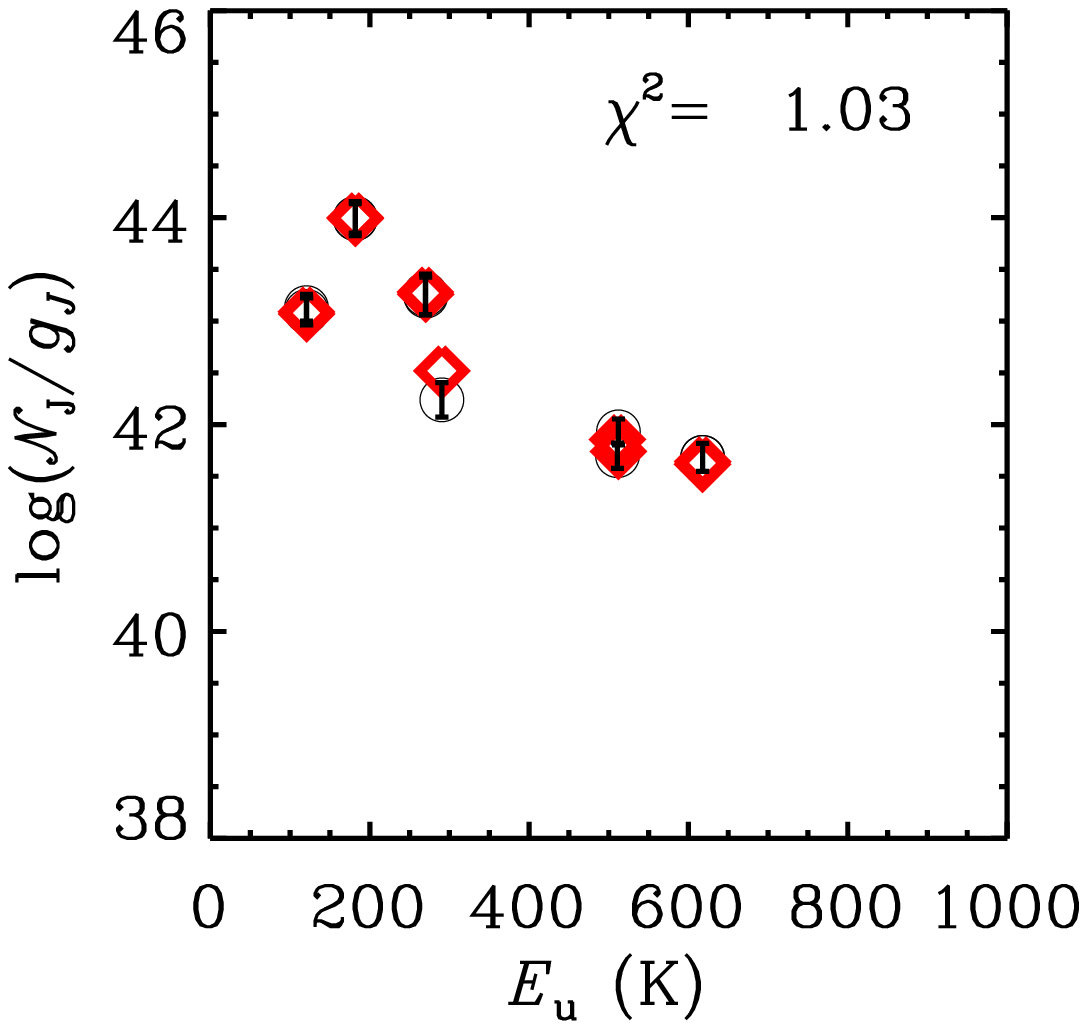}{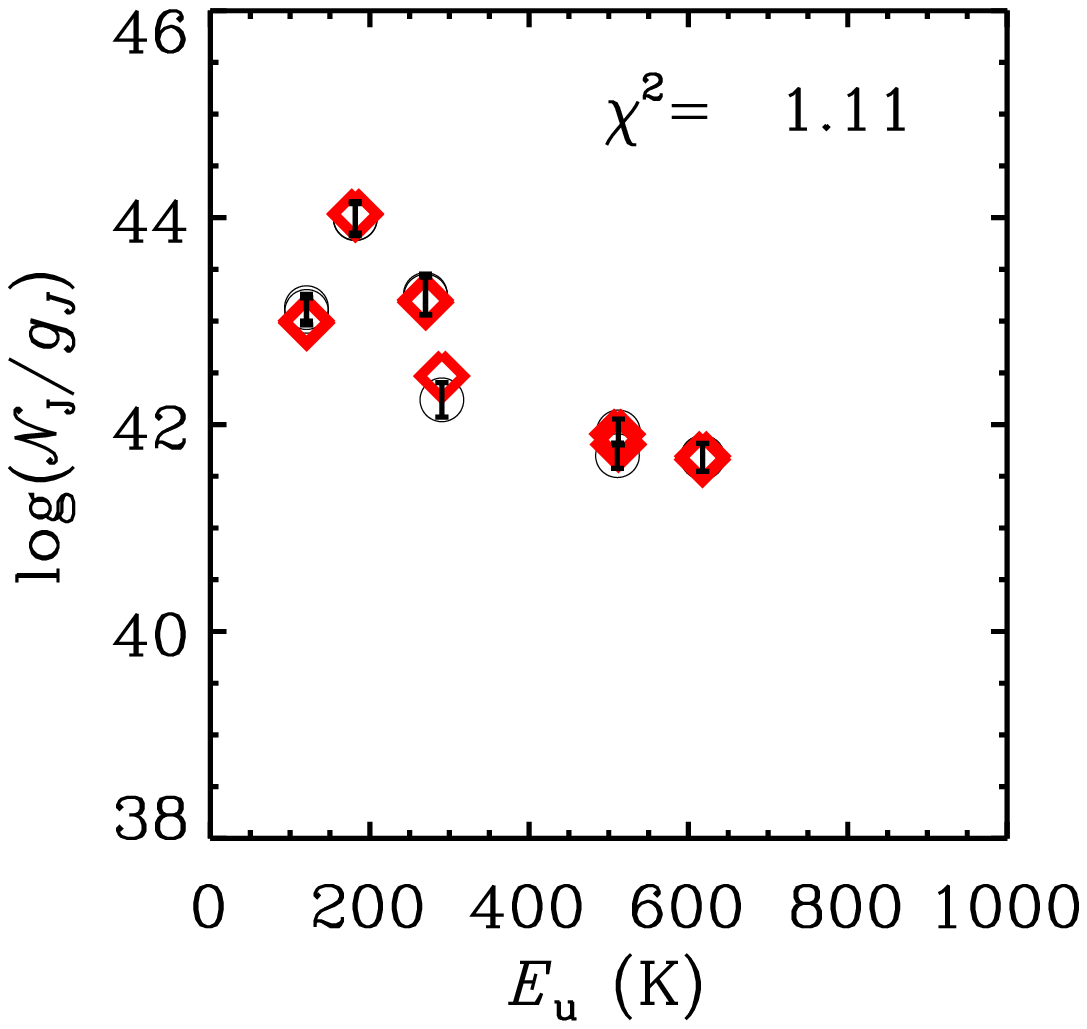}
\caption{The same as Fig.~\ref{tmc1_lvgoh} but for L1489}
\label{l1489_lvgoh}
\end{figure}

\end{document}